\documentclass[twocolumn]{aastex7}
\usepackage{xcolor}

\newcommand{\editone}[1]{\textcolor[HTML]{000000}{#1}}
\newcommand{\edittwo}[1]{\textcolor[HTML]{000000}{#1}}
\newcommand{\editthree}[1]{\textcolor[HTML]{000000}{#1}}
\newcommand{\editfour}[1]{\textcolor[HTML]{000000}{#1}}

\begin{document}

\title{Connecting Environment, Star Formation History, and Morphology of Massive Quiescent Galaxies at $3<z<4$ with JWST}

\correspondingauthor{Lalitwadee Kawinwanichakij}
\email[show]{lkawinwanichakij@swin.edu.au}

\author[0000-0003-4032-2445]{Lalitwadee Kawinwanichakij}
\affiliation{Centre for Astrophysics and Supercomputing, Swinburne University of Technology, PO Box 218, Hawthorn, VIC 3122, Australia}
\affiliation{JWST Australian Data Centre (JADC), Swinburne Advanced Manufacturing and Design Centre (AMDC), John Street, Hawthorn, VIC 3122, Australia}
\email[show]{lkawinwanichakij@swin.edu.au}

\author[0000-0002-3254-9044]{Karl Glazebrook}
\affiliation{Centre for Astrophysics and Supercomputing, Swinburne University of Technology, PO Box 218, Hawthorn, VIC 3122, Australia}
\affiliation{JWST Australian Data Centre (JADC), Swinburne Advanced Manufacturing and Design Centre (AMDC), John Street, Hawthorn, VIC 3122, Australia}
\email{kglazebrook@swin.edu.au}

\author[0000-0003-2804-0648]{Themiya Nanayakkara}
\affiliation{Centre for Astrophysics and Supercomputing, Swinburne University of Technology, PO Box 218, Hawthorn, VIC 3122, Australia}
\affiliation{JWST Australian Data Centre (JADC), Swinburne Advanced Manufacturing and Design Centre (AMDC), John Street, Hawthorn, VIC 3122, Australia}
\email{wnanayakkara@swin.edu.au}

\author[0000-0003-1362-9302]{Glenn G. Kacprzak}
\affiliation{Centre for Astrophysics and Supercomputing, Swinburne University of Technology, PO Box 218, Hawthorn, VIC 3122, Australia}
\email{gkacprzak@astro.swin.edu.au}

\author[0000-0001-5856-8713]{Harry George Chittenden}
\affiliation{Centre for Astrophysics and Supercomputing, Swinburne University of Technology, PO Box 218, Hawthorn, VIC 3122, Australia}
\affiliation{JWST Australian Data Centre (JADC), Swinburne Advanced Manufacturing and Design Centre (AMDC), John Street, Hawthorn, VIC 3122, Australia}
\email{hchittenden@swin.edu.au}

\author[0000-0003-4239-4055]{Colin Jacobs}
\affiliation{Centre for Astrophysics and Supercomputing, Swinburne University of Technology, PO Box 218, Hawthorn, VIC 3122, Australia}
\affiliation{JWST Australian Data Centre (JADC), Swinburne Advanced Manufacturing and Design Centre (AMDC), John Street, Hawthorn, VIC 3122, Australia}
\email{colinjacobs@swin.edu.au}

\author[0009-0004-1163-0160]{Ángel Chandro-Gómez}
\affiliation{International Centre for Radio Astronomy Research, The University of Western Australia, 35 Stirling Highway, Crawley, Western Australia 6009, Australia}
\affiliation{ARC Centre for Excellence in All-Sky Astrophysics in 3D, Australia}
\email{angel.chandrogomez@research.uwa.edu.au}

\author[0000-0003-3021-8564]{Claudia Lagos}
\affiliation{Cosmic DAWN Center, Niels Bohr Institute, University of Copenhagen, Jagtvej 128, Copenhagen N, DK-2200, Denmark}
\affiliation{ARC Centre for Excellence in All-Sky Astrophysics in 3D, Australia}
\affiliation{International Centre for Radio Astronomy Research, The University of Western Australia, 35 Stirling Highway, Crawley, Western Australia 6009, Australia}
\email{claudia.lagos@uwa.edu.au}

\author[0000-0001-9002-3502]{Danilo Marchesini}
\affiliation{Tufts University, Physics and Astronomy Department, 574 Boston Avenue, Medford, MA 02155, USA}
\email{danilo.marchesini@tufts.edu}

\author[0000-0002-6701-1666]{M. Martìnez-Marìn}
\affiliation{Centre for Astrophysics and Supercomputing, Swinburne University of Technology, PO Box 218, Hawthorn, VIC 3122, Australia}
\email{mmartinezmarin@swin.edu.au}

\author[0000-0001-5851-6649]{Pascal A. Oesch}
\affiliation{Cosmic DAWN Center, Niels Bohr Institute, University of Copenhagen, Jagtvej 128, Copenhagen N, DK-2200, Denmark}
\affiliation{Department of Astronomy, University of Geneva, Chemin Pegasi 51, Versoix, CH-1290, Switzerland}
\email{pascal.oesch@unige.ch}

\author[0009-0008-9260-7278]{Rhea-Silvia Remus}
\affiliation{University Observatory Munich, Faculty of Physics, Ludwig-Maximilians-University, Scheinerstrasse 1, 81679, Munich, Germany}
\email{rhea@usm.uni-muenchen.de}


\begin{abstract}
We present the morphological properties of 17 spectroscopically confirmed massive quiescent galaxies ($10.2 < \log(M_{\ast}/M_{\odot}) < 11.2$) at $3.0 < z < 4.3$, observed with JWST/NIRSpec and NIRCam. Using Sérsic profile fits to F277W and F444W imaging, we derive the size–mass relation and find typical sizes of $\sim$0.6–0.8 kpc at $M_{\ast} = 5 \times 10^{10}~M_{\odot}$, consistent with $\sim$7× growth from $z \sim 4$ to the present, including $\sim$2× by $z \sim 2$. \editfour{We find tentative evidence that formation history and morphology jointly influence galaxy sizes: late-forming bulge-dominated galaxies appear more compact by $\sim$0.2–0.3 dex relative to the expected relation, while late-forming disk-dominated galaxies are larger.} Using a random forest regressor, we identify local environmental density, quantified by $\log(1+\delta^{\prime}_{3})$ from the three nearest neighbors, as the strongest predictor of bulge-to-total ratio ($B/T$), which spans 0.25–1. In the \textsc{IllustrisTNG} simulation, the ex-situ stellar mass fraction ($f_{\ast,\mathrm{ex\text{-}situ}}$)—a proxy for mergers—is instead the dominant predictor of $B/T$. Galaxies with high $B/T$ in dense environments show bursty star formation and short quenching timescales ($\lesssim0.4$ Gyr), consistent with bulge growth through merger-driven starbursts; in simulations, such systems exhibit elevated ex-situ fractions ($\sim$20–30\%). In contrast, some high-$B/T$ galaxies in intermediate-density environments have low ex-situ fractions, suggesting that additional processes—such as violent disk instabilities—also contribute. These results point to multiple bulge growth pathways at high redshift, unified by rapid gas accretion, central starbursts, and AGN feedback, as predicted by cosmological simulations.
\end{abstract}

\section{Introduction} \label{sec:intro}

\par One of the remarkable findings from the Hubble Space Telescope (HST) is that many of the most massive galaxies at high redshifts ($z\approx2-3$)  exhibit compact structures, with small effective radii of  $\sim1-2$ kpc \citep[e.g.,][]{Buitrago2008,vanDokkum2008,vanderWel2014,Straatman2015, Belli2017,Stockman2020,Mowla2019,Marsan2019,Lustig2021,Esdaile2021}. This observation supports the interpretation that massive galaxies formed in an inside-out manner: their dense cores were already largely in place by  $z\approx2$, while subsequent growth, accounting for approximately half of their present-day mass, occurred later through minor mergers \citep[e.g.,][]{Oser2010,Patel2013,vandeSande2013}.

\par Several quenching mechanisms for massive galaxies have been proposed \citep[e.g., the review by][and references therein]{Man2018}.  The most commonly discussed mechanism related to the quenching of massive galaxies is active galactic nucleus (AGN) feedback and accretion shock heating \citep[][]{Dekel2006, Bluck2023}. This process involves gas accretion onto a galaxy, which fuels an AGN. Consequently, the AGN either ejects the gas reservoir from the galaxy or prevents further gas from being accreted, depending on the strength of the feedback. At high redshifts, strong AGN-driven outflows, sometimes referred to as quasar-mode feedback, have been commonly observed \citep[e.g.,][]{Maiolino2012, Forster-Schreiber2014}, and it has been proposed that major-merger events trigger strong AGN-driven outflows (e.g., Hopkins et al. 2006). In addition, low-accretion state AGN feedback, also referred to as radio-mode or jet-mode feedback, has been proposed to prevent gas cooling in the galaxy further and consequently make a galaxy remain quiescent 
\citep[e.g.,][]{Croton2006,Bower2006,Lagos2008,Somerville2008}.

\par The quenching mechanisms discussed above could plausibly leave an imprint on the morphology of galaxies. Previous studies of the relation between star formation activity (or absence thereof) and structural properties of galaxies utilizing a sample of nearby galaxies have found that quiescence is closely linked with morphological parameters, such as S\'{e}rsic index and central stellar mass density \citep[e.g.,][]{Kauffmann2003,Kauffmann2007,Schiminovich2007,Bell2008,Cheung2012,Fang2013,Lang2014,Whitaker2017,Abdurrouf2018,Spilker2019,Dimauro2022}. These studies consistently connect a galaxy's quenching with its high central stellar mass surface density, in line with a scenario typically referred to as inside-out quenching, in which galaxies start to quench first in the central core and then later in the outskirts.

\par At high redshifts ($z\gtrsim2$), however, the relationship between quenching and morphology may be more complex, with different evolutionary pathways contributing to the observed structural properties of quiescent galaxies. Massive quiescent galaxies at these redshifts are significantly more compact (approximately 2–4 times) than galaxies of similar stellar mass in the local universe \citep[e.g.,][]{Daddi2005, Trujillo2007,vanDokkum2008,Buitrago2008,Damjanov2009,McLure2013,Poggianti2013}. Cosmological simulations have shown multiple pathways to form compact galaxies, which can be broadly divided into two main mechanisms. The first involves intense, centrally concentrated starbursts, generally triggered by gas-rich major mergers between $z\sim2$–4, which reduce the half-mass radii of galaxies by a factor of a few to below 2 kpc \citep[e.g.,][]{Dekel2009Apj,Dekel2014,Zolotov2015,Tacchella2016a,Lapiner2023,Cenci2024,McClymont2025arXiv}. The second mechanism is the assembly of galaxies at very early cosmic times, when the universe was much denser; in this scenario, galaxies form compact and remain so until $z\sim2$ \citep{Wellons2015}. 
\par Recently, \citet{Ji2024} used JWST/NIRCam imaging to study galaxies at $3 < z < 4.5$ with $M_{\ast}\sim10^{10}~M_{\odot}$ and found that galaxies lying $0.2$–$0.7$ dex below the star-forming main sequence have dense central cores with stellar populations \edittwo{that are similar in age} or younger than their extended components, supporting the gas-rich compaction model for core growth. This suggests that, unlike at low redshift ($z\lesssim2$), where quenching is often associated with inside-out mechanisms, compact high-redshift galaxies may experience compaction-driven star formation and quenching, leading to a different spatial distribution of stellar ages.

\editthree{Consistent with this picture, recent JWST studies have confirmed that quiescent galaxies at $z>3$ are extremely compact, often exhibiting effective radii below 1 kpc. \citet{Wright2024} found that most quiescent candidates at $3<z<5$ are approximately 40\% smaller than expected from lower-redshift trends, with the most massive systems ($M_{\ast} \sim 10^{11}~M_{\odot}$) reaching sizes as small as $\sim$0.7 kpc. Similarly, \citet{Ito2024} reported that quiescent galaxies at rest-frame 0.5 $\mu$m have typical sizes of $\sim$0.6 kpc at $M_{\ast} = 5 \times 10^{10}~M_{\odot}$—significantly more compact than their low-redshift counterparts at similar mass. \citet{Cutler2024} further highlighted a sharp transition in the quiescent size–mass relation at $1<z<3$, with a break at $\log(M_{\ast}/M_{\odot}) \sim 10.3$: massive quiescent galaxies tend to have spheroidal morphologies and older stellar populations, while lower-mass systems are generally younger and more disky—consistent with mass-driven quenching at the high-mass end.}


\par HST and the James Webb Space Telescope (JWST) have sufficient spatial resolution ($\sim1$ kpc) and stable point spread functions (PSFs) to robustly quantify the light profiles of high-redshift galaxies. \edittwo{JWST/NIRCam imaging, with approximately twice the angular resolution of HST/WFC3 ($\sim0\farcs07$ vs. $\sim0\farcs15$), provides even finer detail, corresponding to scales of $\sim0.5$ kpc at $z=3-4$.} JWST allows us to probe the rest-frame near-IR properties of high-redshift galaxies. Hence, the size measurement in the rest-frame near-IR wavelength is less prone to dust attenuation and the outshining effect by the young stellar population \citep[e.g.,][]{Sawicki1998,Papovich2001,Shapley2001,Trager2008,Graves2010,Maraston2010,Pforr2013,Sorba2015}.

\par The interpretation for the correlation between central stellar mass surface density at $z=1-2$ and quenching \citep[e.g.,][]{Estrada-Carpenter2020} can be complicated as post-quenching processes can alter structural properties of galaxies \citep[e.g.,][]{Lagos2017,Lagos2018}. For instance, it is challenging to determine if the observed correlation is actually due to a causal connection between the development of dense cores and quenching, or is a consequence of another physical processes. Studying massive quiescent galaxies at higher redshifts of $z>3$, i.e., closer to the key epoch of the universe when quenching becomes a significant process \citep[e.g.,][]{Muzzin2013, Schreiber2018, Merlin2019,Shahidi2020,Carnall2023,Carnall2023Nature,Glazebrook2024Nature,deGraaff2024, Nanayakkara2024Nature,Nanayakkara2025} allows us to directly investigate the link between structural transformation and quenching, and therefore provides more accurate constraint on the physical connection of the two phenomena.

\par In this work, we investigate structural properties of a spectroscopically confirmed sample of 17 quiescent galaxies at $3.0<z<4.3$ with stellar masses of $10.2<\log(M_{\ast}/M_{\odot})<11.2$. \edittwo{This study leverages the unprecedented imaging and spectroscopic capabilities of JWST, utilizing targeted NIRCam imaging and NIRSpec observations presented in \cite{Glazebrook2024Nature,Nanayakkara2024Nature,Nanayakkara2025}. These galaxies were originally identified in the sample of \cite{Schreiber2018}, based on pre-JWST rest-frame $U-V$ vs. $V-J$ color selection \citep[e.g.,][]{Williams2009}, and were previously observed with the MOSFIRE spectrograph on the 10 m Keck telescope \citep{McLean2012}}.

\par \editone{The structure of this paper is as follows. Section~\ref{sec:data} describes the data used in this study. Section~\ref{sec:measureoverdensity} details the methodology for estimating the environmental densities of our sample. In Section~\ref{sec:morpho_measure}, we outline our measurements of galaxy morphologies, encompassing single-component Sérsic fits and bulge$+$disk decomposition. Section~\ref{sec:smr_measure} explains the derivation of the size-mass relation of quiescent galaxies.
Section~\ref{sec:result_morphology} presents the results on morphological properties, such as S\'{e}rsic index and axis ratio, and includes a discussion of these findings. In Section~\ref{sec:result_smr_sfh}, we examine the connection between galaxy sizes and their star formation histories. Section~\ref{sec:result_bulgediskdecomp} reports on bulge$+$disk decomposition, focusing on bulge-to-total ratios and the size-mass relations of bulges and disks. Section~\ref{sec:randomforest_analysis} describes our Random Forest (RF) analysis to identify the galaxy properties that best predict morphological characteristics, followed by the main RF results. Section~\ref{sec:discussion} compares our findings with previous studies and predictions from cosmological simulations, discussing their broader implications for understanding diverse morphologies of massive quiescent galaxies at $3<z<4$. Finally, Section~\ref{sec:conclusion} summarizes the key results and conclusions of this work. The Appendices include a summary of morphological parameter measurements for both single-component Sérsic fits and bulge$+$disk decomposition, as well as a detailed description of the methodology used to measure the environmental density of massive quiescent galaxies and validation tests using a mock galaxy catalog.} Throughout this paper, we adopt a $\Lambda$CDM cosmology with $H_{0} = 70~\mathrm{km} \mathrm{s}^{-1} \mathrm{Mpc}^{-1}$, $\Omega_{M} = 0.7$, and $\Omega_{\Lambda} = 0.3$. Magnitudes are reported in the AB system \citep{Oke1983}. \editone{For reference, $1^{\prime\prime}$ corresponds to approximately $7.5$ kpc at $z=3.5$.}

\section{Data and Sample Selection\label{sec:data}} 

\par \editone{The data sample consists of 17 spectroscopically confirmed massive quiescent galaxies at $3 < z_{\text{spec}} < 4.30$ with stellar masses in the range $\log(M_{\ast}/M_{\odot}) = 10.18 - 11.24$. These galaxies were observed as part of the JWST Cycle 1 program \textit{``How Many Quiescent Galaxies Are There at $3 < z < 4$ Really?"} (GO-2565, PI Glazebrook) using the Near Infrared Spectrograph \citep[NIRSpec; $0.6 - 5.3 \, \mu\text{m}$;][]{Jakobsen2022}. This program completes the spectroscopic confirmation of a pre-selected sample of 24 massive quiescent galaxy candidates from \citet{Schreiber2018}, which were identified using the following criteria:  $K < 24.5$, $M_{\ast} \geq 10^{10} \, M_{\odot}$, $z_{\text{phot}} > 2.8$, 
and the $UVJ$-based quiescent color selection \citep{Williams2009}. Spectroscopic redshifts were determined using \textsc{slinefit} \citep{Schreiber2018}, which fits stellar and emission line templates to the low-resolution ($50 < R < 500$) PRISM spectra \citep{Nanayakkara2024Nature, Glazebrook2024Nature}. \citet{Glazebrook2024Nature} demonstrated with simulated NIRSpec spectra that redshift recovery using \textsc{slinefit} achieves a precision of $\pm 0.005$ over $z = 3.19 - 3.23$.}

\par \editone{Our sample exhibits a diverse range of formation and quenching timescales. A majority (15 out of 17) shows a prominent Balmer break at $3640 \, \text{\AA}$, confirming their post-starburst nature \citep{Glazebrook2017, Nanayakkara2024Nature}. These galaxies formed half of their stellar mass at $z_{50}=3.4-5.9$, approximately within $\lesssim 1 \, \text{Gyr}$ before their observed epoch at $z \sim 3.5$ \citep{Nanayakkara2025}.}

\par \editone{In contrast, one galaxy, ZF-UDS-7329 at $z = 3.205$, analyzed in detail by \citet{Glazebrook2024Nature}, exhibits clear absorption features such as the $4000 \, \text{\AA}$ break and Mgb ($5174 \, \text{\AA}$), characteristic of an older stellar population ($> 1 \, \text{Gyr}$). Spectral fitting and star formation history modeling reveal a stellar mass of $1.24 \times 10^{11} \, M_{\odot}$, with most of its mass formed at 
$z_{\mathrm{50}} = 10.4^{+4.0}_{-2.2}$, approximately 400 Myr after the Big Bang. This makes ZF-UDS-7329 significantly older than most other quiescent galaxies at this epoch \citep[see also][]{Carnall2024, Nanayakkara2025}.}

\par Additionally, $\sim10$ galaxies in our sample exhibit strong NaD absorption features, indicative of Na enhancement and possible AGN-driven outflows. Among them, six galaxies in our sample -- ZF-UDS-6496, ZF-UDS-3651, ZF-COS-18842, ZF-UDS-8197, and 3D-EGS-18996 -- show significant detections of H$\alpha$ and/or $\mathrm{[OIII]}\lambda5007$ emission lines, with $S/N>5$ and line equivalent widths (EW) $>10 \, \text{\AA}$. Visual inspection of their spectra confirms that these emission lines are prominent over the local continuum level. Such features suggest possible AGN activity \citep{Nanayakkara2025,Carnall2024}, \edittwo{with AGN-driven outflows of neutral gas observed in similar systems \citep{Belli2024,Davies2024}. However, despite these detections, their overall star formation rates remain well below the star-forming main sequence, consistent with their classification as quiescent. }

\par
\begin{figure}
    \centering
    \includegraphics[width=0.45\textwidth]{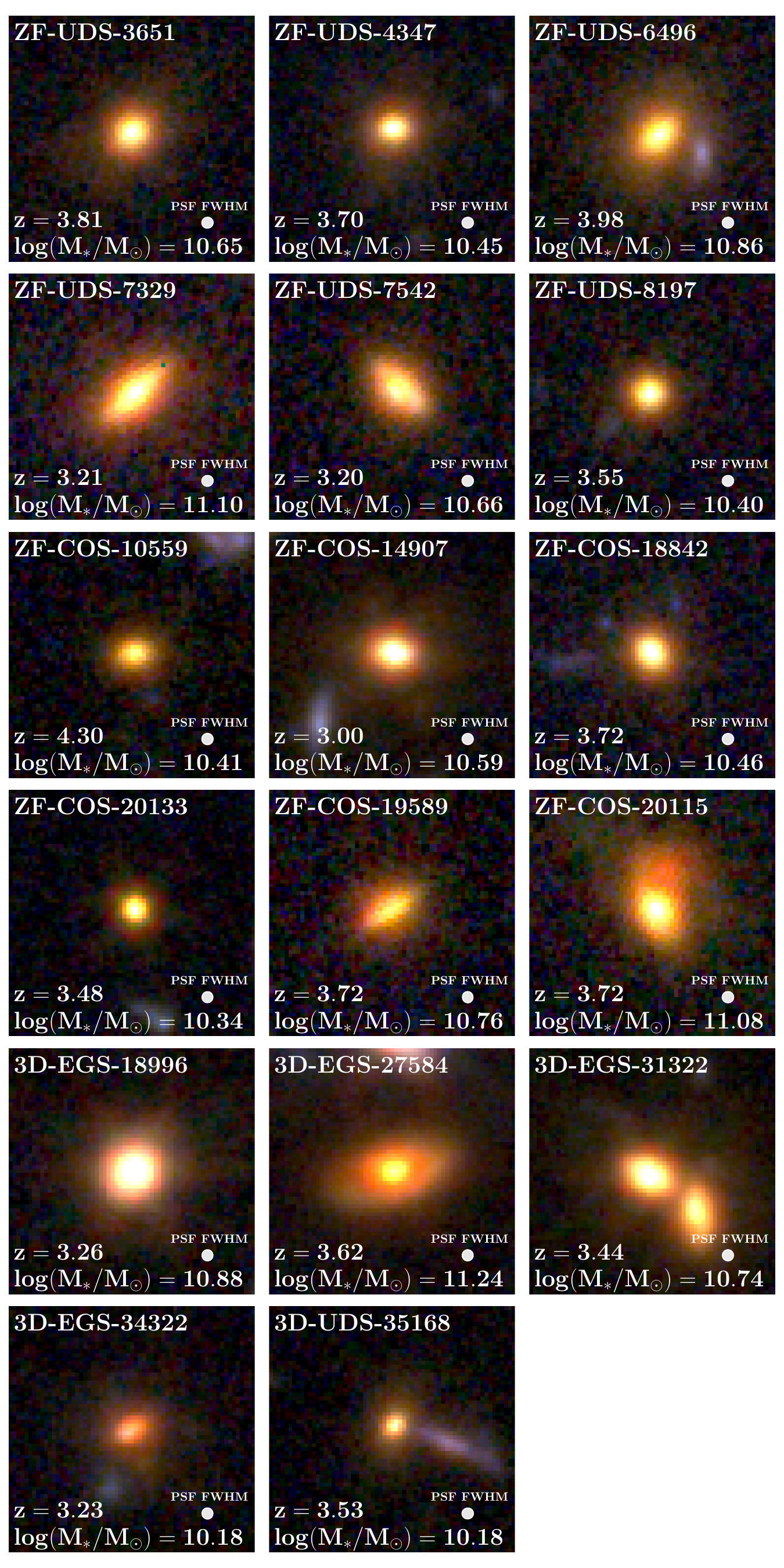}
    \caption{Our sample of spectroscopically confirmed massive quiescent galaxies at $z=3-4$ shown in color images (RGB). F150W, F277W, and F444W images are used for blue, green, and red, respectively. All images are $2\farcs5\times2\farcs5$ centered on each quiescent galaxy. The circle symbol indicates the PSF size of the F277W image (FWHM$=0\farcs11$). }
    \label{fig:RGBs}
\end{figure}

 \editone{Our massive quiescent galaxies are located in three HST legacy fields: the All-Wavelength Extended Groth Strip International Survey \citep[EGS/AEGIS;][]{Skelton2014}, the Cosmic Evolution Survey \citep[COSMOS;][]{Straatman2016}, and the UKIDSS Ultra-Deep Survey \citep[UDS;][]{Skelton2014, Straatman2016}. NIRCam imaging in the UDS and AEGIS fields (F150W, F277W, and F444W bands) was obtained through the PRIMER survey (GO-1837, PI Dunlop) and the CEERS survey \citep[DD-ERS-1345, PI Finkelstein;][]{Finkelstein2023}, respectively. Mosaicked images and weight maps were \edittwo{downloaded from the DAWN JWST Archive} (Brammer et al., in preparation), which were created following the procedures also described in  \citet{Valentino2023}, using the public \textsc{grizli} software package \citep{Brammer2021}}. \edittwo{Figure~\ref{fig:RGBs} presents color composite images (RGB) of our sample, illustrating their morphologies and the spatial resolution achieved with JWST/NIRCam imaging.}

\begin{figure*}
    \centering
    \includegraphics[width=\textwidth]{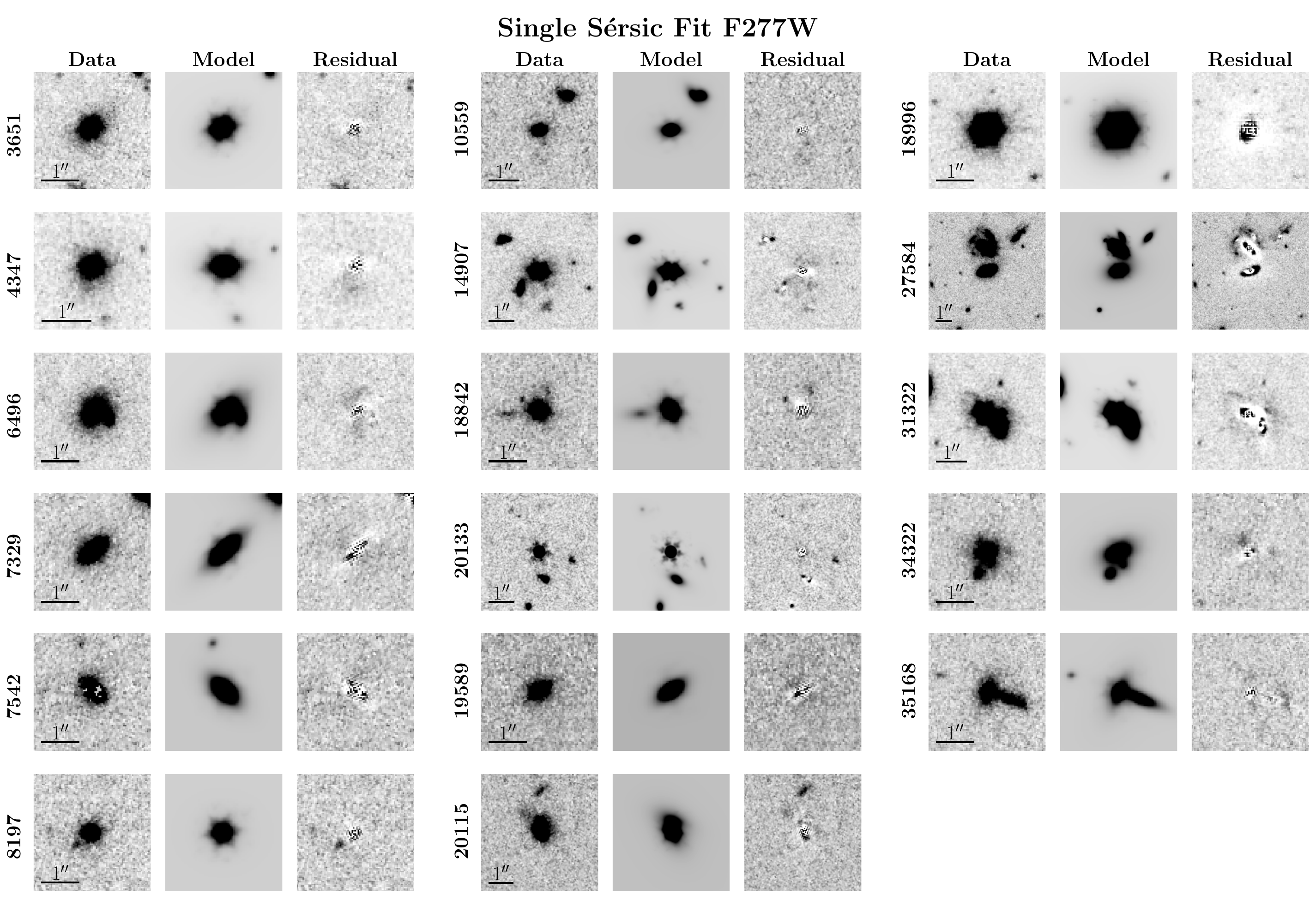}
    \caption{   Results of single S\'{e}rsic profile fitting for our sample of massive quiescent galaxies at $z=3-4$. Each set of three columns shows: (left) postage stamps in the JWST/F277W filter, (middle) the best-fitting S\'{e}rsic models, and (right) the residual images (data minus models). The data, models, and residuals are displayed \edittwo{with the same adaptive linear intensity scaling for each galay}. The black bar in the lower-left corner indicates an angular scale of 1 arcsecond. Our fitting procedure accurately reproduces galaxy cutouts in both isolated and crowded fields but struggles to capture complex morphologies, such as spiral arms. On average, the percentage residual is $3.5\%\pm0.5\%$ relative to the total light, \edittwo{with the values ranging from $1.8\%-7.7\%$.}}   
    \label{fig:singlesersic_f277w_galightfitting}
\end{figure*}

\begin{figure*}
    \centering
    \includegraphics[width=\textwidth]{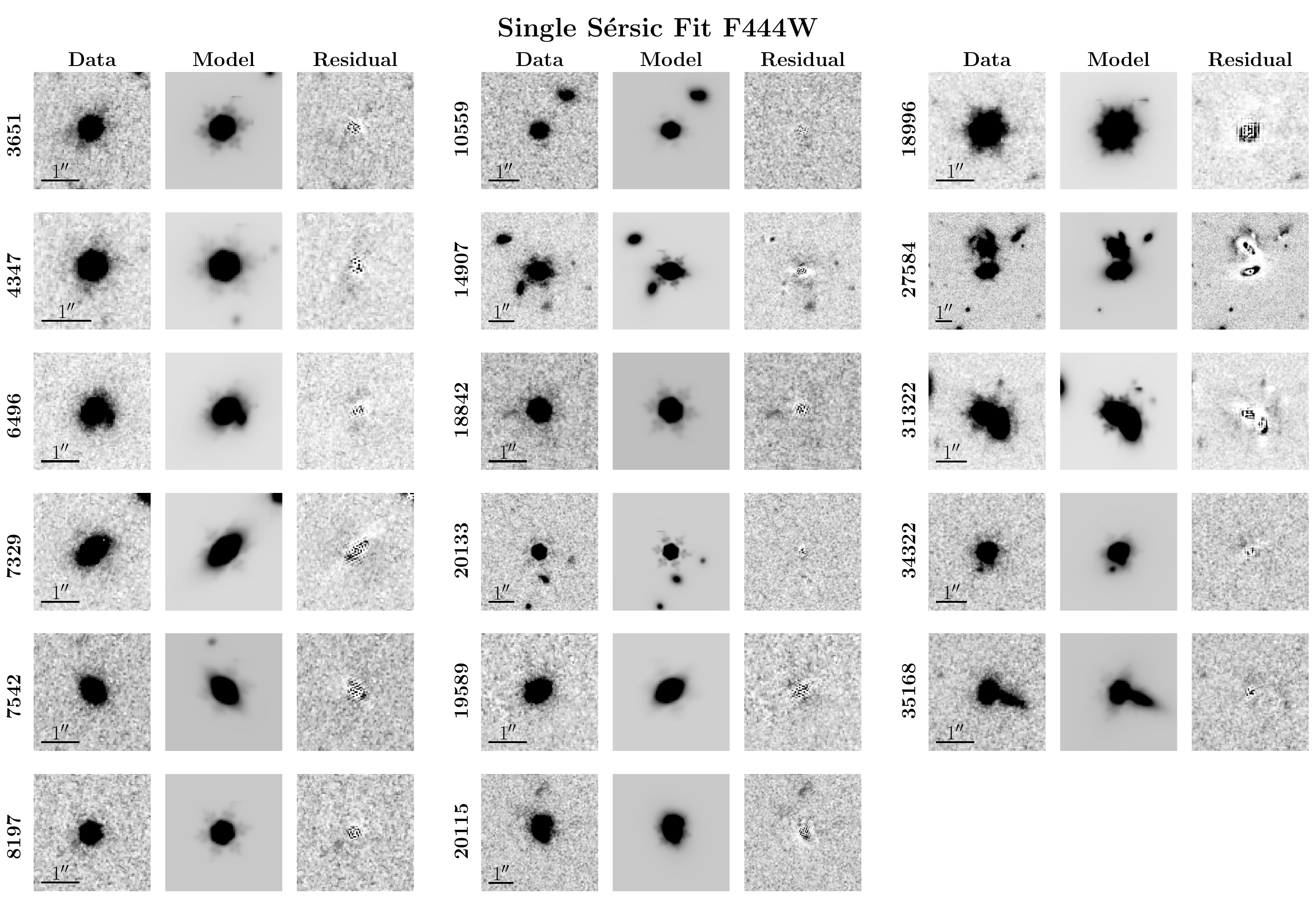}
    \caption{Similar to~\ref{fig:singlesersic_f277w_galightfitting} but for the JWST/F444W. On average, the percentage residual is $2.7\%\pm0.5\%$ relative to the total light, \edittwo{with the values ranging from $0.6\%-5.4\%$.}}
    \label{fig:singlesersic_f444w_galightfitting}
\end{figure*}

\section{Measurement of Environmental Overdensities}
\label{sec:measureoverdensity}
\par \editone{In this work, we quantify the environmental densities of massive quiescent galaxies at $z=3-4$ using the Bayesian-motivated estimator of local surface density, $\Sigma^{\prime}_{N}$. This approach builds upon the traditional $N$th nearest-neighbors estimator, which defines the local surface density of a galaxy as $\Sigma_{N} = N/(\pi d^{2}_{N})$, where $d_{N}$ is the projected distance to the $N$th nearest neighbor. The choice of $N$ typically ranges from 3 to 10 in the literature \citep[e.g.,][]{Dressler1980, Baldry2006, Muldrew2012, Kawinwanichakij2017,Papovich2018, Gu2021,Chang2022}. This traditional method reliably measure ``local density'' on scales internal to galaxy group halos, particularly for small $N$ \citep{Muldrew2012}.  A detailed description of the methodology, including the mathematical framework and validation tests, is provided in Appendix~\ref{appendix:measure_overdensity}. In this section, we briefly summarize the approach.}

\par \editone{We improve our measurement of environmental density by adopting a Bayesian-motivated estimator of the local surface density of galaxies (Equation~\ref{eq:bayesian_density_cowan}), which uses the distances to all $N$th nearest neighbors to improve accuracy and mitigate projection effects compared to the traditional $N$th nearest-neighbor estimator, as demonstrated by  \citep[]{Ivezic2005,Cowan2008} (see Appendix B of \citeauthor{Ivezic2005} \citeyear{Ivezic2005} and also Appendix A of \citeauthor{Kawinwanichakij2017} \citeyear{Kawinwanichakij2017}). Specifically, we define the environmental density of a galaxy using the dimensionless overdensity, $(1+\delta^{\prime}_{N})$, where the local surface density $\Sigma^{\prime}_{N}$ is normalized by the density expected for a uniform galaxy distribution. Thus, $\log(1+\delta^{\prime}_{N})>0$ indicates that a galaxy resides in an overdense environment, while $\log(1+\delta^{\prime}_{N})<0$ corresponds to an underdense environment.}

\par \editone{To measure $\Sigma^{\prime}_{N}$, we use photometric redshifts ($z_{\mathrm{phot}}$) from the photometric redshifts catalogs (v7.2) from the DAWN JWST Archive for CEERS and PRIMER surveys\footnote{\url{https://dawn-cph.github.io/dja/imaging/v7/}}. The photometric redshifts are derived using \textsc{EAZY-py}\citep{Brammer2023}, which is based on the EAZYcode\footnote{\url{https://github.com/gbrammer/eazy-py}}\citep{Brammer2008}. Photometry from all NIRCam ($0.9\mu\mathrm{m}-4.4\mu\mathrm{m}$) and HST ($0.44\mu\mathrm{m}-1.6\mu\mathrm{m}$) observations are used to constraint $z_{\mathrm{phot}}$, and we refer the reader to \cite{Valentino2023} for details on the SED modeling.  The typical photometric redshift uncertainty is $\sigma_{z} = \Delta z/(1+z_{\mathrm{spec}}) = 0.021$, with outlier fractions of $4.8\%-14.6\%$, depending on the survey field (Brammer et al., in preparation). \edittwo{Focusing on the sample at $3<z<4$, relevant for the environmental density analysis of our massive quiescent galaxies, we find that $\sigma_{z}$ increases to 0.026, while the outlier fractions decrease slightly, ranging from $3.8\%-12.2\%$, depending on the field.} Following \cite{Valentino2023}, we apply selection criteria to exclude galaxies with unreliable photometric modeling, requiring the number of available filters of $N_{\mathrm{filt}}\ge6$ and $\chi^{2}/N_{\mathrm{filt}}\le8$. Neighboring galaxies are defined as those satisfying these criteria and having stellar masses of $\log(M_{\ast}/M_{\odot})\ge8.5$. This stellar mass limit corresponds to the 90th percentile of the $M_{\ast}$ distribution for $UVJ$-selected red galaxies at $3<z<6.5$ in the CEERS and PRIMER catalogs \citep[][see their Figure C1]{Valentino2023}}.

\par \editone{In this work, we calculate the local surface density for each massive quiescent galaxy using distances to the three nearest neighbors ($\Sigma^{\prime}_{3}$). Neighboring galaxies are selected within a photometric-redshift separation of $| \Delta z | < 2\times \sigma_{z} (1+z_{\mathrm{spec}})$, where $z_{\mathrm{spec}}$ is the spectroscopic redshift of our massive quiescent galaxy. To mitigate contamination from foreground and background galaxies, we also test a narrower width of $\pm1.5\sigma_{z}\times(1+z)$ and find no significant impact on our conclusions.} 
\par \editone{In Appendix~\ref{appendix:mocktest_overdensity}, we validate this Bayesian-motivated estimator, applied to the data with the photometric redshift accuracy of JWST PRIMER and CEERS survey, using a mock galaxy catalog. For $N=3$, our Bayesian density metric accurately recovers the true projected overdensity, effectively identifying galaxies in the highest and lowest density quartiles with minimal contamination, particularly for small $N$. Our test also demonstrated that $N=3$ provides a good compromise between the accuracy of the measured environmental density and the ability studying the effects of local environment on the morphology of massive quiescent galaxies at $z\sim3-4$. }

\par \edittwo{While the mock tests in Appendix~\ref{appendix:mocktest_overdensity} validate the accuracy of our Bayesian-motivated estimator in recovering overdensities, they do not account for observational selection effects, such as detection completeness, which could bias our density measurements. Unlike those tests, which used mock catalogs with intrinsic properties, evaluating completeness requires a dataset with realistic photometry. To address this, we performed an additional test using the JAGUAR \citep[JWST Extragalactic Mock Catalog;][]{Williams2018}\footnote{\url{https://fenrir.as.arizona.edu/jwstmock/index.html}}, which includes synthetic photometry for both HST and NIRCam.}

\par \edittwo{ For each NIRCam and HST band, we adjusted the magnitude distributions in the JAGUAR mock to match those in the photometric catalogs of each survey field by probabilistically assigning non-detections to sources in each magnitude bin until the distributions aligned. We then applied the same stellar mass selection ($\log(M_{\ast}/M_{\odot})\ge8.5$) and required sources to be detected in at least six photometric bands ($N_{\mathrm{filt}}\ge6$), mirroring our observational selection.}

\par \edittwo{While the total number of galaxies in the completeness-adjusted mock sample decreased by $\sim50\%$, this reduction was almost entirely due to sources with $\log(M_{\ast}/M_{\odot})<8.5$, which are already excluded in our analysis.  Crucially, the number of galaxies satisfying the mass selection ($\log(M_{\ast}/M_{\odot})\ge8.5$) remained unchanged after applying the $N_{\mathrm{filt}}$ cut, confirming that our sample of neighboring galaxies is not significantly affected and that requiring $N_{\mathrm{filt}}\ge6$ does not introduce systematic biases in our environmental density estimates.}

\par Table~\ref{tab:overdensity_measurements} in Appendix~\ref{appendix:measure_overdensity} provides the environmental density measurements for massive quiescent galaxies at $z=3$–4. Due to their positions at the edge of the PRIMER COSMOS mosaic, ZF-COS-20115 and ZF-COS-19589 have unreliable density estimates and are excluded from the analysis of environmental impacts on morphological properties (Section~\ref{sec:randomforest_analysis}). 
\par \editthree{To place our sample in the broader context of known massive quiescent galaxies at similar or higher redshifts, we also compute environmental density estimates for RUBIES-EGS-QG1 ($z=4.886$, $\log(M_{\ast}/M_{\odot})=10.9$; \citealt{deGraaff2024}) and GS-9209 ($z=4.658$, $\log(M_{\ast}/M_{\odot})=10.58$; \citealt{Carnall2023Nature}), both of which are spectroscopically confirmed and have JWST/NIRCam imaging. While our measurements do not reveal new overdensity features for these two galaxies—particularly for RUBIES-EGS-QG1, which is already known to reside in an overdense environment—they serve to validate the applicability of our density estimator at $z=3-4$. We note that additional quiescent galaxies at these redshifts, including some with spectroscopic confirmation, have been identified in recent studies \citep[e.g.,][]{Carnall2024,Barrufet2025,Baker2025}, but incorporating them into our analysis would require additional bulge–disk decomposition and structural characterization, which is beyond the scope of this work.}


\par We note that two galaxies in our sample, ZF-UDS-7329 at $z=3.21$ and ZF-UDS-7542 at $z=3.20$, are within a projected distance of 12 arcseconds (corresponding to $\sim93$ physical kpc). ZF-UDS-7329 is also in close proximity to the star-forming galaxy RUBIES-UDS-46261 \citep{deGraaff2024}, with a projected separation of 6.5 arcseconds \citep{Turner2024}. \edittwo{These close projected separations raise the possibility that these galaxies may reside within a common overdense region or proto-group. Notably, \cite{Carnall2024} report an excess of massive quiescent galaxies at $z \sim 3.2$ in the same field, visible as a clear spike in the redshift distribution (their Figure 2), which further supports the presence of a significant overdensity at this epoch. Although spectroscopic redshifts alone cannot confirm physical association without velocity information, their proximity suggests potential environmental influence or a shared evolutionary context, which may be relevant for interpreting their morphologies.}

\par \edittwo{Moreover, one of our galaxy, 3D-EGS-31322 is located in the densest region of the large-scale structure Cosmic Vine at $z=3.44$ \citep{Jin2024}. This structure consists of 20 spectroscopically confirmed galaxies at $3.43<z_{\mathrm{spec}}<3.45$ and six galaxy overdensities at $4-7\sigma$ significance over the field level. The identification of 3D-EGS-31322 in the large-scale structure reinforces our enviromental density measurement using photometric redshifts, where we find $\log(1+\delta^{\prime}_{3})=0.95$, ranking it among the top three galaxies in the highest-density environments in our sample. This independent confirmation further supports the reliability of our environmental density estimator.}

\section{Morphological parameter measurement} \label{sec:morpho_measure}
We perform two-dimensional fits to the surface brightness distributions of the NIRCam F277W and F444W galaxy images using a Python package, \textsc{GaLight} \citep{Ding2020,Ding2022}. With \textsc{GaLight}, we prepare a cutout galaxy image and detect neighboring sources in the field of view; \textsc{GaLight} allows for the simultaneous fitting of as many neighboring objects as needed to avoid source blending.  The quality of these fits benefits from the high signal-to-noise ratios (S/N) of the images. The median integrated S/N for F277W images of our 17 galaxies is $\mathrm{S/N}= 319 \pm 52$, with values ranging from S/N$=$101 to 1115. For F444W images, the median integrated SNR for F277W images is $\mathrm{S/N}= 301 \pm 44$, with values ranging from S/N$=$134 to 1592. The integrated S/N is computed over all pixels within the galaxy cutout where the segmentation map identifies the galaxy, ensuring that pixels associated with the galaxy and any simultaneously fit neighboring sources are included, while background regions are excluded. \editthree{In total, five galaxies in our sample—ZF-UDS-6496, ZF-COS-20115, 3D-EGS-31322, 3D-EGS-34322, and 3D-UDS-35168—are blended with a nearby source in the NIRCam cutouts. For these systems, we simultaneously fit multiple Sérsic components using \textsc{GaLight}, and the morphological parameters reported in this paper correspond specifically to the best-fit model of the primary (target) galaxy. These target galaxies have also been observed spectroscopically with JWST/NIRSpec, as presented in \citet{Nanayakkara2025}.}


\par The PSF can vary spatially and temporally across the detector due to aberration and breathing effects \citep[e.g.,][]{Bely1993,Rhodes2006,Rhodes2007}. To account for this, we construct a PSF library for each filter using isolated, unsaturated stars with high signal-to-noise ratios selected from the PRIMER and CEERS photometric catalogs. \editthree{We first identify candidate stars based on SExtractor parameters, requiring $\texttt{SNR\_WIN} > 100$, $\texttt{FLAGS} < 2$, $\texttt{ELONGATION} < 1.5$, $\texttt{CLASS\_STAR} > 0.8$, and magnitude $< 22.5$. This yields an initial sample of 15 (20), 44 (38), and 15 (16) stars in the F277W (F444W) images for the PRIMER/UDS, PRIMER/COSMOS, and CEERS/EGS fields, respectively.} We then visually inspect the candidates and manually select PSF-like stars based on compactness, central symmetry, and the absence of nearby contaminants. The final PSF libraries consist of 12 (14), 15 (23), and 9 (14) stars in F277W (F444W) for PRIMER/UDS, PRIMER/COSMOS, and CEERS/EGS, respectively. Many bright stars are clipped or saturated at the core in the mosaics, which limits the number of usable stars. \editthree{While we do not explicitly model spatial variation of the PSF across the field, the selected stars are distributed throughout each mosaic. The dispersion in PSF profiles within the library captures both spatial and temporal variations and is used to estimate the uncertainties in our morphological measurements.} The median FWHMs of our PSFs are $0\farcs06$, $0\farcs11$, and $0\farcs15$ for the F150W, F277W, and F444W filters, respectively.

\subsection{Single S\'{e}rsic Fit\label{sec:sersicfit}}
We first fit a single-component S\'{e}rsic profile \citep{Sersic1968} to the two-dimensional surface brightness distributions of all detected objects using galaxy imaging data, noise level maps, and PSF models. The fit parameters are total magnitudes ($M$), the effective radius ($R_{e}$), S\'{e}rsic index ($n$), and axis ratio ($b/a$). \editone{Following the literature \citep[e.g.,][]{vanderWel2014,Ito2024}, we} define the effective radius, $R_{e}$, as the semi-major axis of the ellipse that contains half of the total flux of the best-fitting S\'{e}rsic model. With the input ingredients, we utilize \textsc{GaLight}  to convolve the theoretical model (i.e., S\'{e}rsic profile) with each PSF in the library before fitting it to the galaxy image. We avoid any unphysical results by setting upper and lower limits on the parameters: effective radius $R_{e}\in [0\farcs1,1\farcs0]$ and S\'{e}rsic index $n\in [0.3,9]$. Finally, \textsc{GaLight} finds the maximum likelihood of the parameter space by adopting the Particle Swarm Optimizer \citep[PSO;][]{Kennedy1995}. 

\par Following the procedure of \cite{Ding2020}, we weight each PSF in the library based on the relative goodness of fit while ensuring that at least five PSFs are used to capture the range of systematic uncertainties. Specifically, we rank the performance of each PSF based on the $\chi^{2}$ value and select the top five PSFs as representative of the best-fit PSFs. Then, we determine the S\'{e}rsic parameters (i.e., flux, $R_{e}$, S\'{e}rsic index, and the axis ratio) using a weighted arithmetic mean, which is calculated using:
\begin{equation}
w_{i}=\exp\left ( -\alpha \frac{(\chi^{2}_{i}-\chi^{2}_{\mathrm{best}})}{2\chi^{2}_{\mathrm{best}}} \right ),
\end{equation}
\noindent for a PSF $i$ and $\alpha$ is an inflation parameter, defined such that when $i=5$, 
\begin{equation}
\alpha \frac{(\chi^{2}_{i=5}-\chi^{2}_{\mathrm{best}})}{2\chi^{2}_{\mathrm{best}}} =2.
\end{equation}
\noindent We then compute the morphological parameter $x$ (i.e., flux, $R_{e}$, $n$, $b/a$) and its uncertainty ($\sigma$) as
\begin{equation}
\overline{x} = \frac{\sum_{i=1}^{N}x_{i}w_{i}}{\sum w_{i}}, 
\end{equation}
\begin{equation}
\sigma = \sqrt{\frac{\sum_{i=1}^{N}(x_{i}-\overline{x})^{2} w_{i}}{\sum w_{i}}}
\end{equation}
\noindent where $N$ is the number of the ranking PSF, i.e., $N=5$. 

\par \edittwo{The systematic uncertainty on the morphological parameters is derived from the spread of the best-fit values across the top five PSFs, weighted by their relative $\chi^{2}$. Since the uncertainty calculation accounts for the variation among these best-fit solutions rather than the specific ranking order, small changes in the ranking of the PSFs should not significantly impact the final systematic errors. To verify this, we tested slight variations in the PSF ranking (e.g., selecting the top four or top six instead of five) and found that the resulting uncertainties remain consistent, indicating that our method is robust against minor changes in the PSF ranking.}

\par Figure~\ref{fig:singlesersic_f277w_galightfitting} and~\ref{fig:singlesersic_f444w_galightfitting} show our single-S\'{e}rsic fitting results in the F277W and F444W filters, \editthree{using stacked PSFs constructed from the top five ranked stars in each field.} These two bandpasses are chosen because at $z=3-4$, F277W probes the rest-frame optical ($\sim0.5\mu$m), enabling direct comparison with previous studies of galaxy structure at similar rest-frame wavelengths \citep[e.g.,][]{vanderWel2014,Kawinwanichakij2021,Forrest2022,Ito2024}. In contrast, F444W corresponds to the rest-frame near-IR ($\sim1\mu$m) and serves as a proxy for the underlying stellar mass distribution \citep[e.g.,][]{Suess2022, Boogaard2024,Costantin2024,Martorano2024}. \editthree{Additionally, we performed Sérsic fits in the F150W filter to enable direct comparison with archival HST/WFC3/F160W measurements. These F150W fits were derived using the same methodology and are discussed later in this section and in Appendix~\ref{sec:compare_size_nircam_wfc3}.}

\par \editthree{Morphological parameters in F277W and F444W were measured independently using the same fitting procedure described above. We further explore the wavelength dependence of S\'{e}rsic index and axis ratio in Appendix~\ref{appendix:compare_sersic_axisratio_f277w_f444w}, where we compare the structural parameters derived from each filter for individual galaxies.}


As shown in Figures~\ref{fig:singlesersic_f277w_galightfitting} and~\ref{fig:singlesersic_f444w_galightfitting}, \editfour{six galaxies—ZF-UDS-7329, ZF-UDS-7542, ZF-COS-19589, 3D-EGS-27584, 3D-EGS-34322, and 3D-EGS-31322—show a significant improvement in residuals, defined as a relative decrease of more than 10\% compared to the single Sérsic fit (see Sections~\ref{sec:compute_residual_singlesersic} and~\ref{sec:compute_residual_BD}, and Table~\ref{tab:residual_comparison_f277w}). For example, a drop from 1.0\% to 0.9\% residual meets this threshold. This indicates that a single Sérsic function does not adequately describe their light profiles. In some cases, such as 3D-EGS-27584 and 3D-EGS-31322, the need for a more complex model is further supported by visible substructures like clumps or spiral arms, which can reduce the reliability of structural parameters derived from single-component fits.}

\begin{figure*}
\centering
\includegraphics[width=\textwidth]{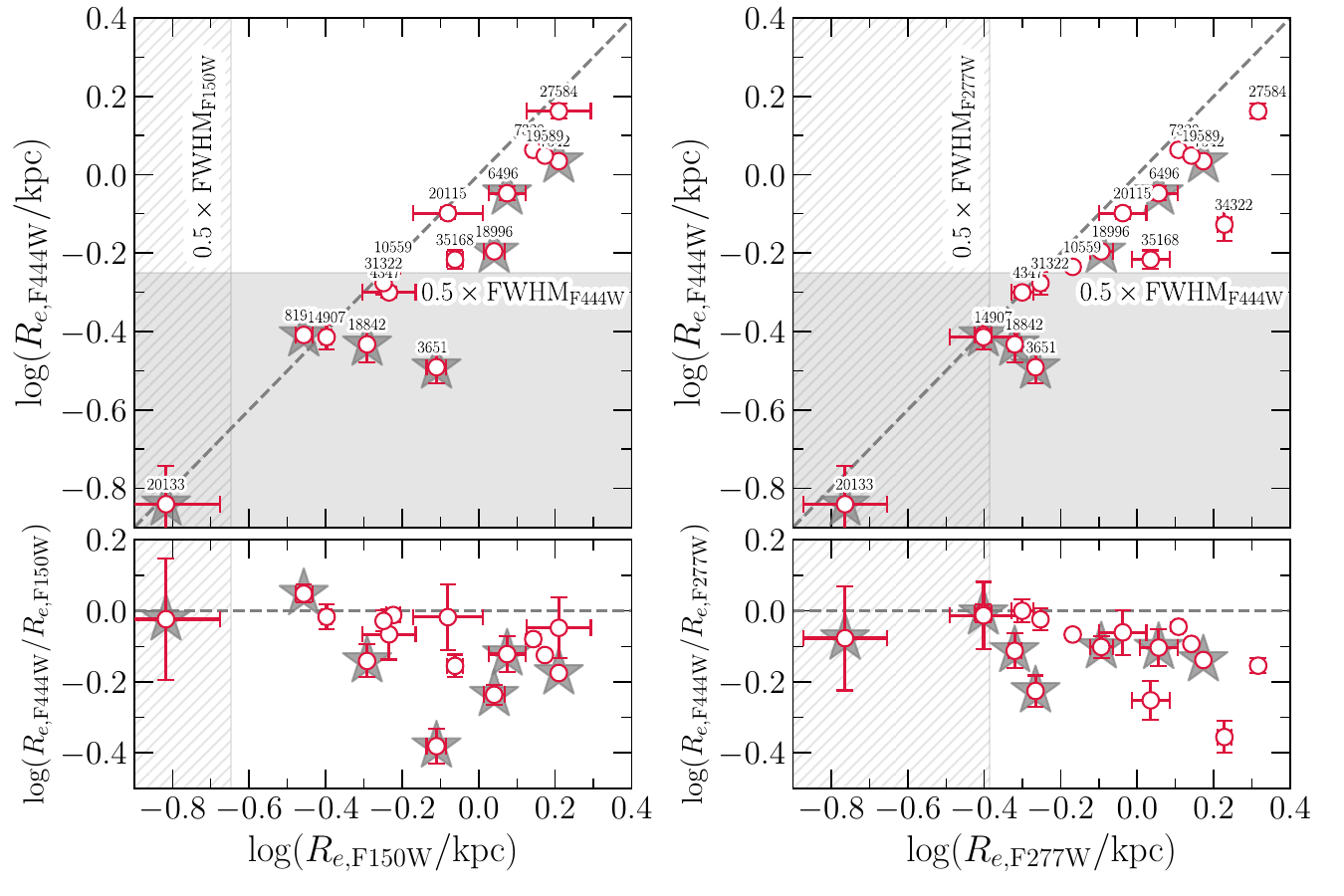}
\caption{Comparison between 4.4$\mu$m size of massive quiescent galaxies at $3<z<4$, which is based on a single S\'{e}rsic fit to the images of the galaxies in F444W filter, the longest wavelength filter of JWST/NIRCam used in this paper, with sizes measured in the shorter wavelength filters: F150W and F277W.  Left: 4.4$\mu$m size as a function of  1.5$\mu$m size. Right: 4.4$\mu$m size as a function of 2.7$\mu$m size. The bottom panels show the difference (in a logarithmic scale) between 1.5$\mu$m sizes and 2.7$\mu$m sizes. Star markers indicate galaxies identified as potential AGN based on significant detections of prominent rest-frame optical emission lines \citep{Nanayakkara2025}. \editone{The shaded region indicates sizes smaller than half of the FWHM of the F150W, F277W and F444W PSFs at $z=3.5$, where measurements are potentially challenging.} Massive quiescent galaxies are smaller in $4.4\mu$m than in the shorter wavelengths, and the size difference is most prominent when comparing the $4.4\mu$m sizes with $1.5\mu$m sizes, implying negative color gradients in galaxies (e.g., the outskirts of galaxies are bluer than the central regions). }
\label{fig:sizediff_threebands}
\end{figure*}
\par  In Figure~\ref{fig:sizediff_threebands}, we show a comparison between 4.4$\mu$m size of massive quiescent galaxies at $3<z<4$, which is based on a single S\'{e}rsic fit to the F444W images, to the 1.5$\mu$m size and 2.7$\mu$m size. Massive quiescent galaxies are smaller in $4.4\mu$m than in the shorter wavelengths, and the size difference is most prominent when comparing the $4.4\mu$m sizes with $1.5\mu$m sizes, consistent with previous works, 
\citep[e.g.,][]{Tortora2010,Wuyts2010,Guo2011,Szomoru2013, Mosleh2017, Suess2019,Suess2022}. These studies generally showed that the observed galaxy half-light radii measured in longer wavelength bands are smaller than those in short wavelength bands, implying negative color gradients in galaxies (e.g., the outskirts of galaxies are bluer than the central regions).

\par \editone{ZF-COS-20133 stands out in our sample due to its extremely small size ($R_{e} \sim 0\farcs02$), which is well below the resolution limit of the F444W PSF. Its measured S\'{e}rsic index ($n \sim 0.5$) is consistent with a flattened or diffuse light profile, typically associated with disk-like morphology. This unresolved or marginally resolved nature in the NIRCam F277W and F444W images suggests that the structural measurements are heavily influenced by the PSF and are therefore less reliable. Furthermore, the unusually low S\'{e}rsic index may reflect degeneracies in the fitting process, where the algorithm compensates for the unresolved structure by flattening the profile to match the observed light distribution under PSF constraints. The possible presence of an AGN, as proposed by \citet{Martinez2024}, could also contribute to its peculiar appearance. Consequently, caution is warranted when interpreting the structural parameters of this galaxy.}

\par \editone{Additionally, we compare our JWST/NIRCam $1.5\mu$m size measurements to HST/WFC3 $1.6\mu$m size measurements from  \cite{vanderWel2012}, \cite{Straatman2015}, and \cite{Esdaile2021}. Overall, we find good agreement between the two datasets, with a median logarithmic size ratio of $\log(R_{e,\mathrm{F160W}}/R_{e,\mathrm{F150W}})=0.04\pm0.16$ dex (Figure~\ref{fig:comparesize_jwst_hst}). However, discrepancies mainly arise for compact galaxies with sizes smaller than half the F160W PSF. Detailed investigations reveal that JWST's superior resolution and sensitivity allow for more precise size measurements, particularly for high-redshift compact galaxies, capturing faint extended structures that are less apparent in HST images. For a comprehensive discussion, including reprocessed measurements and comparisons of specific galaxies, see Appendix~\ref{sec:compare_size_nircam_wfc3}.}

\subsubsection{Residual of  Single S\'{e}rsic Fit\label{sec:compute_residual_singlesersic}}

\par \editone{We compute the percentage residual, $R_{\%}$, to quantify the fraction of total light not captured by the S\'{e}rsic model using:
\begin{equation}
R_{\%,\mathrm{single~S\acute{e}rsic}  } = \frac{\sum (D - M)}{\sum D} \times 100,
\end{equation}
\label{eq:percent_residual}
}
\noindent where $D$ and $M$ are the 2D data and model images, respectively. The  summation is over all pixels in the 2D array where the mask equals one, which is determined from a segmentation map generated using source detection output by \textsc{Galight}. This ensures that the residual calculation includes all pixels associated with the galaxy of interest and its neighboring sources while excluding background regions. We report the $R_{\%,\mathrm{single~S\acute{e}rsic}}$ for F277W and F444W for each galaxy in Tables~\ref{tab:residual_comparison_f277w} and ~\ref{tab:residual_comparison_f444w}, respectively (Appendix~\ref{app:morphotables}).
\par The median percentage residual for single S\'{e}rsic fit  to the F277W images is $3.5\% \pm 0.5\%$, with values ranging from $-2.1\%$ to $7.7\%$. For F444W, the median residual is $2.7\%\pm0.5\%$, with values ranging from $-0.75\%$ to $5.4\%$. The difference in the percentage residuals likely reflects wavelength-dependent galaxy structures or substructures (e.g., clumps or asymmetries) that are less pronounced at longer wavelengths. The smaller residuals in F444W suggest smoother galaxy light profiles in the near-IR, making them better approximated by a S\'{e}rsic model. These results underscore the influence of morphological complexity on fit quality, particularly in shorter-wavelength bands where finer structures dominate. \edittwo{However, despite these substructures, the majority of the stellar mass is still contained within the smooth S\'{e}rsic component, particularly at longer wavelengths where the near-IR light better traces the underlying stellar mass distribution.}

\begin{figure*}
    \centering
    \includegraphics[width=\textwidth]{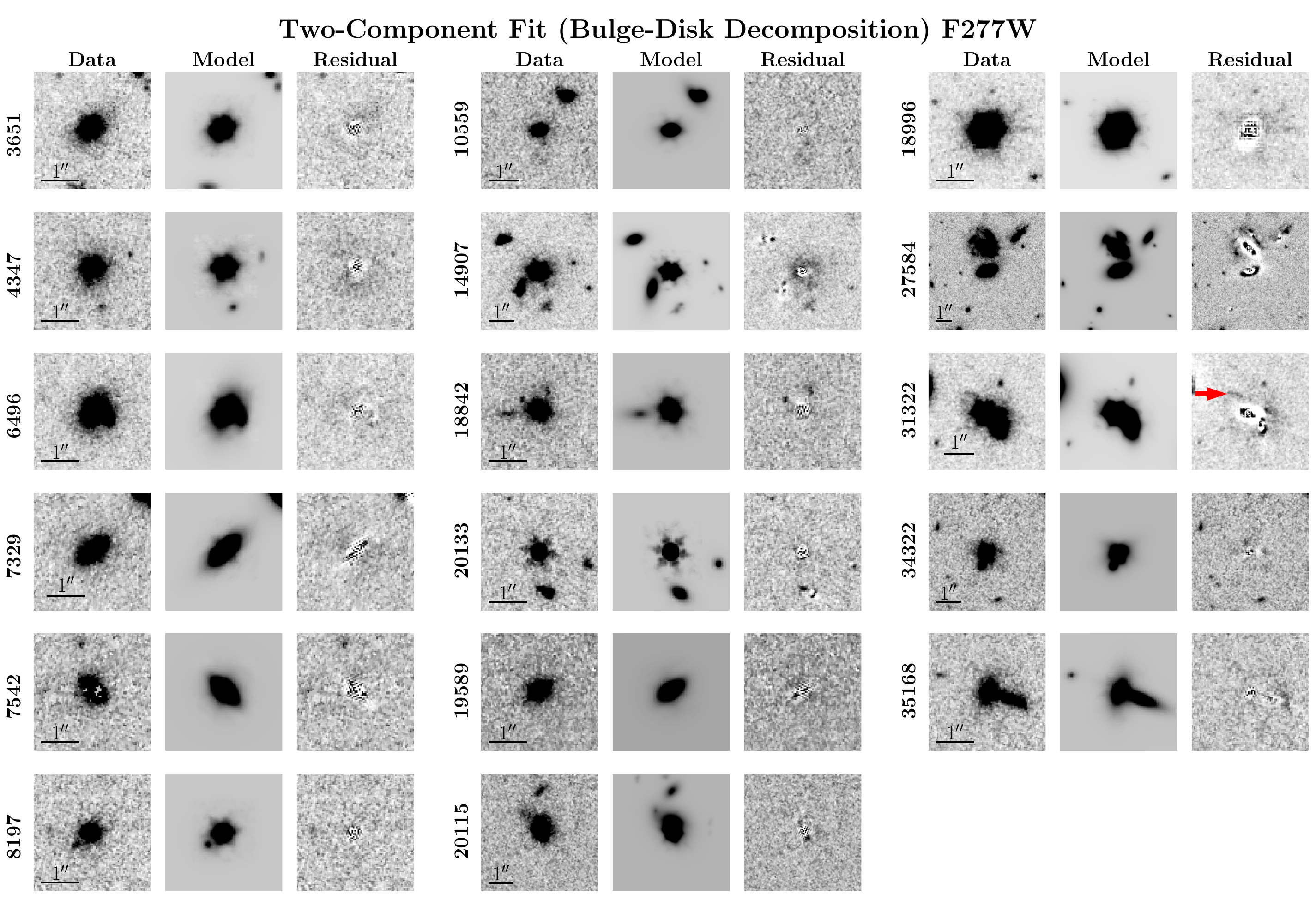}
    \caption{The bulge-disk decomposition of our sample of massive quiescent galaxies at $z=3-4$. In each set of three columns, the left column displays postage stamps in the JWST/F277W filter; the middle column displays the best-fitting models; the right column displays the residual images, i.e., the models subtracted from the data. The data, model, and residuals are on the same color scale for each galaxy. The black bars in the lower left corner of each plot set indicate an angular scale of 1 arcsecond. The median residual for F277W is $3.2\%\pm0.8\%$, with the values ranging from $1.5\%-9\%$. \edittwo{3D-EGS-31322 exhibits tidal tail and is indicated by arrow}.}
    \label{fig:BDdecomp_f277w_galightfitting}
\end{figure*}

\begin{figure*}
    \centering
    \includegraphics[width=\textwidth]{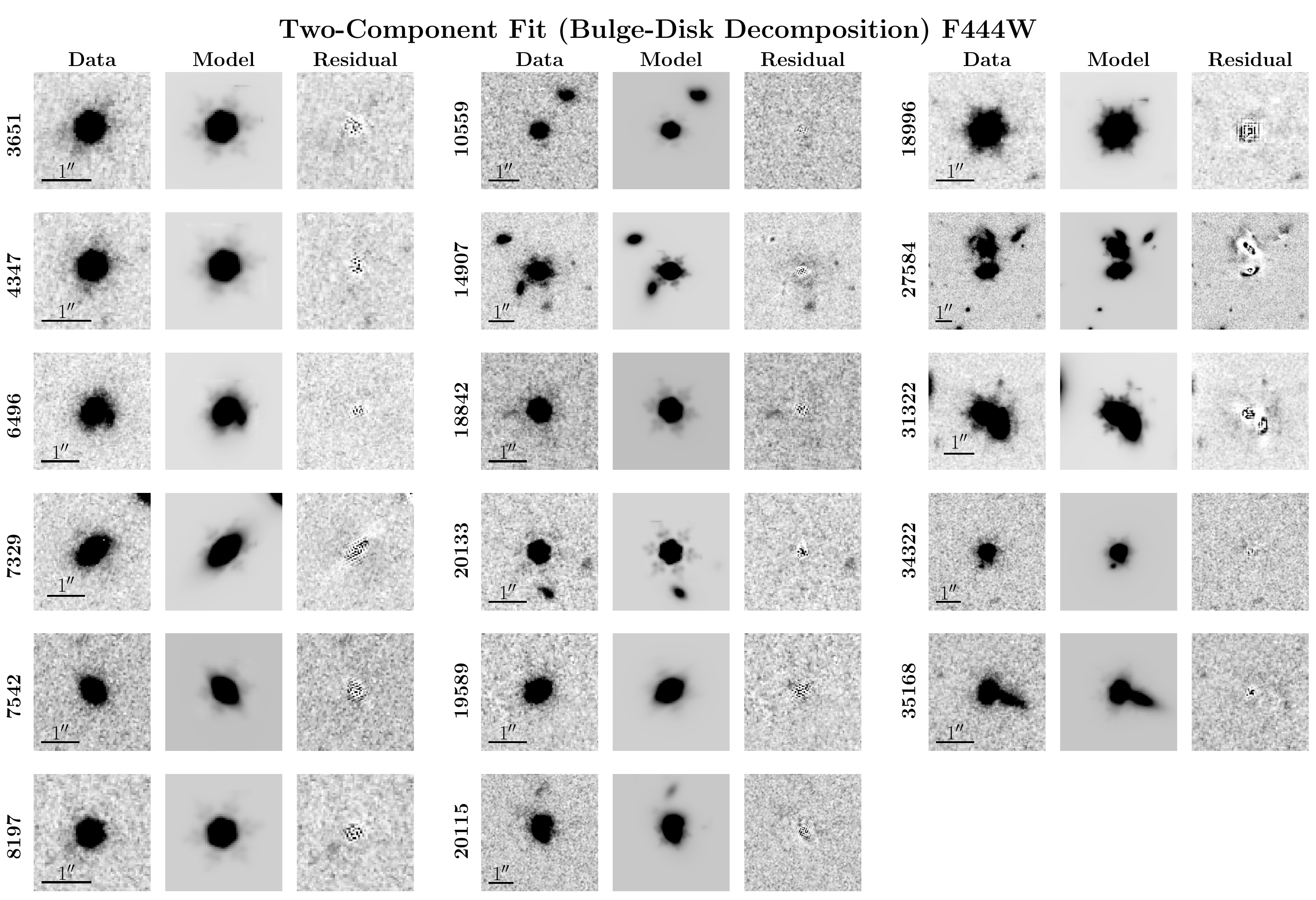}
    \caption{Similar to~\ref{fig:BDdecomp_f277w_galightfitting} but for the JWST/F444W. \editone{The median residual is $2.4\%\pm0.4\%$, with the values ranging from $0.4\%-5\%$.}}
    \label{fig:BDdecomp_f444w_galightfitting}
\end{figure*}

\subsection{Bulge and Disk Decomposition}
\par While the S\'{er}sic index $n$ is often used as a proxy of the contribution of the bulge component, it is important to note that the relationship between S\'{e}rsic index and the bulge-to-total ratio $B/T$ is not one-to-one \citep[e.g.,][]{Andredakis1995, Lang2014, Bruce2012, Bruce2014,Kennedy2016}: the increasing S\'{e}rsic index can arise from an increasing in $B/T$ or increasing in the size of the disk component relative to that of the bulge while leaving $B/T$ constant. For this reason, we further perform bulge-disk decompositions for massive quiescent galaxies at $z=3-4$ using \textsc{GaLight}. We adopted a procedure similar to that implemented in \cite{Bruce2012} and \cite{Lang2014}. Specifically, for each galaxy and each filter, we parametrize the bulge and disk light as a sum of two-component S\'{e}rsic profiles with $n_{\mathrm{bulge}}=4$ (a de Vaucouleurs profile) and $n_{\mathrm{disk}}=1$ (i.e., an exponential profile), respectively. We follow the same procedure described above to simultaneously fit the bulge component, disk component, and neighbor objects. \editthree{While the target galaxy is modeled with a two-component bulge+disk decomposition, all neighboring sources are modeled using single Sérsic profiles to reduce complexity and minimize parameter degeneracy in crowded fields.} We additionally constrain the center of both bulge and disk components to be the same, \editthree{ but allow other structural parameters—including position angle and axis ratio—to vary independently.} We define the effective radius of the bulge component, $R_{e,\mathrm{Bulge}}$, and disk component, $R_{e,\mathrm{Disk}}$, as the radius at which the component contains half of its light. The result of the surface brightness modeling of the disk with an exponential profile is a scale radius $h$; we, therefore, compute the effective radius of the disks as $R_{e,\mathrm{Disk}}=1.678\times h$ \citep{Graham2005}. 

\par We calculate the bulge-to-total flux ratio ($B/T$) for our massive quiescent galaxies as the ratio of the bulge flux to the total flux (sum of the bulge and disk components). In this work, we primarily focus on $B/T$ measurements obtained using the F444W filter, which corresponds to rest-frame near-infrared light at $\sim 1 \, \mu\mathrm{m}$. This wavelength is expected to serve as a reasonable proxy for the underlying stellar mass distribution \citep[e.g.,][]{Suess2022} and, therefore, provides a reliable tracer of the bulge-to-total mass ratio. We denote the F444W-based measurement as $(B/T)_{\mathrm{F444W}}$. For completeness, we also present $B/T$ measurements obtained with the F277W filter, which we denote as $(B/T)_{\mathrm{F277W}}$.

\par We present the results of bulge-disk decompositions in the F277W and F444W filters in Figures~\ref{fig:BDdecomp_f277w_galightfitting} and~\ref{fig:BDdecomp_f444w_galightfitting}, respectively. The best-fit morphological parameters for the two-component fitting are listed in Table~\ref{tab:morphomeas_BD}. 

\subsubsection{Residual of Two-Component Fit (Bulge–Disk Decomposition)\label{sec:compute_residual_BD}}

\par \editone{To evaluate the fit quality of our bulge–disk decompositions, we compute the percentage residual (Equation~\ref{eq:percent_residual}), $R_{\%, \mathrm{two-component}}$, for each galaxy. We quantify the improvement over the single S\'{e}rsic fit using the residual difference, $\Delta R_{\%} = |R_{\%, \mathrm{single~S\acute{e}rsic}}| - |R_{\%, \mathrm{two-component}}|$, and the relative percentage improvement, $\Delta R_{\%} / | R_{\%, \mathrm{single~S\acute{e}rsic}}| \times 100$. Results for the F277W and F444W bands are presented in Tables~\ref{tab:residual_comparison_f277w} and~\ref{tab:residual_comparison_f444w} (Appendix~\ref{app:morphotables}).}

\par For F277W, the median residual is $R_{\%} = 3.2\% \pm 0.8\%$, while for F444W it is slightly lower at $2.7\% \pm 0.6\%$. In F277W, 7 out of 17 galaxies show significant improvement with the two-component model, with relative differences ranging from $\sim10\%$ to $60\%$. Notably, 3D-EGS-27584 shows improvement in both filters, with residual reductions of $\sim60\%$ (F277W) and $35\%$ (F444W). Conversely, six galaxies exhibit increased residuals (by $\sim15\%$–$70\%$) relative to the single S\'{e}rsic fit. This suggests that the two-component model may struggle to represent galaxies with irregular morphologies, asymmetries, or deviations from idealized bulge–disk ($n=4$ and $n=1$) profiles—whereas a single S\'{e}rsic model can sometimes better capture the global light distribution.

\par \editfour{To further investigate the performance of the two-component fits, we examine whether residuals correlate with Sérsic index. We find a moderate positive correlation in F277W ($\rho = 0.49$, $p = 0.044$), but no significant trend in F444W ($\rho = 0.18$, $p = 0.501$). In particular, galaxies with high Sérsic indices ($n_{F277W} > 6$) tend to exhibit systematically larger residuals—by $\sim5$–$10\%$ on average—than those with lower values. This supports the idea that fixed-component bulge–disk models are less effective at modeling strongly peaked or extended profiles, which are more prominent in the rest-frame optical morphologies traced by F277W. A visual comparison of residual percentage versus Sérsic index for both bands is provided in Figure~\ref{fig:residual_vs_sersicindex} in Appendix~\ref{appendix:compare_bt_sersic}.}

\par Overall, the two-component model performs better in the F444W band, where smoother profiles and lower residuals suggest that longer-wavelength imaging more effectively captures the bulk of the stellar mass distribution. In contrast, higher residuals and scatter in F277W likely reflect the presence of finer-scale structures—such as clumps or asymmetries—that the parametric models cannot easily reproduce. These results underscore the wavelength dependence of morphological structure and highlight the limitations of both single S\'{e}rsic and bulge–disk models in fitting complex systems. More flexible or non-parametric methods may be needed to fully capture this diversity.

\par In addition to residual-based comparisons, we assess the reliability of single-component Sérsic fits as structural tracers by comparing them to bulge+disk decompositions across our sample. Specifically, we examine the relationship between the bulge-to-total flux ratio ($B/T$) and the single-component Sérsic index ($n$). As shown in Appendix~\ref{appendix:compare_bt_sersic}, we find a general correlation between $B/T$ and $n$, particularly for $n \lesssim 4$, but also observe considerable scatter across the full range of Sérsic indices. Some galaxies with similar $B/T$ values show substantial variation in $n$, driven by differences in the sizes and shapes of the bulge and disk components. Conversely, galaxies with similar $n$ may differ in $B/T$ depending on how the light is distributed spatially between components.

\editfour{We note that in our two-component bulge--disk decompositions, the bulge S\'ersic index is fixed to $n_{\mathrm{bulge}} = 4$. This choice can limit the dynamic range of fitted $B/T$ values and may contribute to the scatter seen at high $n$ in Figure~\ref{fig:bt_vs_sersicindex} (here $n$ refers to the \emph{single-component} S\'ersic index). At intermediate $n$ ($\approx 3$--$4$), we still find a non-negligible spread in $B/T$ from the two-component fits, even after excluding flagged or unreliable morphologies (e.g., ZF-COS-20133). Several of these $n \approx 3$--$4$ systems also show a significant relative improvement in residuals ($>10\%$ compared to the single-S\'ersic fit; see Sections~\ref{sec:compute_residual_singlesersic} and~\ref{sec:compute_residual_BD}), pointing to genuine differences in bulge/disk size ratios and light distributions that are not uniquely encoded in the single-S\'ersic index alone. This supports our conclusion that two-component decompositions capture structural diversity beyond what is obtained from single-component S\'ersic fits.}



\par \edittwo{Throughout this paper, we adopt the single S\'{e}rsic fit for structural parameters such as effective radius, Sérsic index, and axis ratio. This choice enables direct comparison with previous studies at intermediate redshifts ($z \sim 0.5$–3) and at similar redshifts ($z = 3$–4). For completeness, we also present bulge–disk decomposition results—including $B/T$ ratios and component sizes—in Section~\ref{sec:result_bulgediskdecomp}. In particular, the $B/T$ ratio serves as our primary structural indicator in the Random Forest analysis (Section~\ref{sec:randomforest_analysis}), where we identify the galaxy and environmental properties most predictive of morphology.}

\begin{figure*}
    \centering
    \includegraphics[width=1\textwidth]{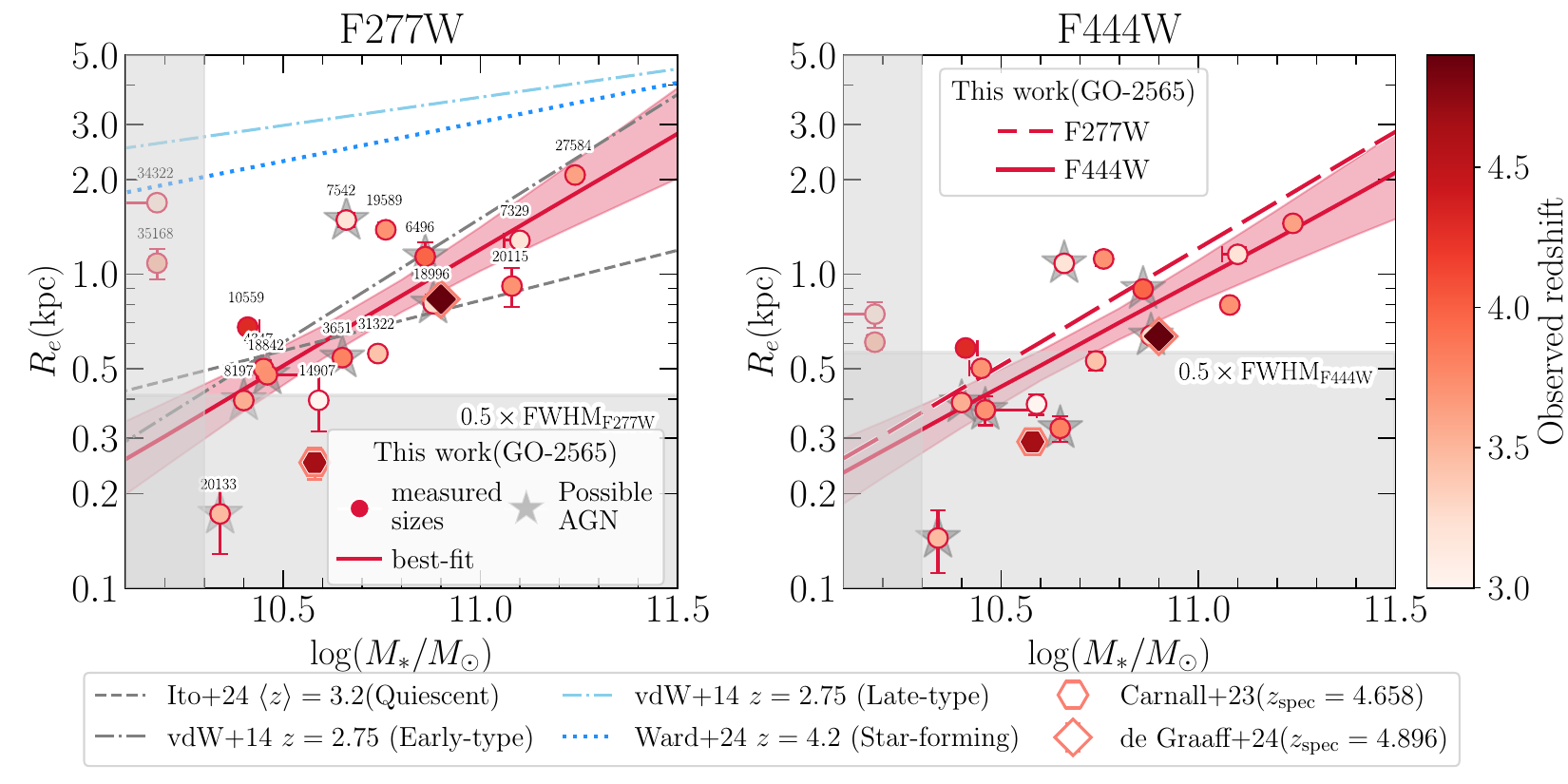}
    \caption{Size-mass relations of massive and quiescent galaxies at $3<z<4$ observed as part of JWST-GO-2565 program. The size measurements are based on the JWST F277W sizes (rest-frame $\sim0.6 \, \mu\mathrm{m}$; left) and F444W sizes (rest-frame $\sim1\, \mu\mathrm{m}$; right) sizes. Star symbols indicate that galaxies are potentially AGN based on significant detections of prominent rest-frame optical emission lines \citep{Nanayakkara2025}. The color scale of the symbols indicates their observed redshifts. \editthree{The data points located in the vertical gray shaded region ($\log(M_{\ast}/M_{\odot})<10.3$) are not used for the analytical fit.} The red line indicates the best-fit size-mass relation of the form $r_{e}/\mathrm{kpc}=A\times {m_{\star}}^{\alpha}$, where $m_{\star}\equiv M_{\ast}/5\times10^{10}M_{\odot}$. The best-fit slope and intercept of the relation in F277W are given by $\alpha={0.74}^{+0.17}_{-0.17}$ and $\log (A/\mathrm{kpc})={-0.14}^{+0.05}_{-0.05}$, respectively. The best-fit slope and intercept of the relation in F444W are given by  $\alpha={0.68}^{+0.15}_{-0.18}$ and $\log (A/\mathrm{kpc})={-0.22}^{+0.04}_{-0.05}$, respectively.\editone{Spectroscopically confirmed quiescent galaxies from the literature are included: GS-9209 ($\log(M_{\ast}/M_{\odot}) = 10.58, z = 4.658$; hexagon, \citealt{Carnall2023Nature}) and RUBIES-EGS-QG1 ($\log(M_{\ast}/M_{\odot}) = 10.9, z = 4.886$;  diamond, \citealt{deGraaff2024}).  The size measurements for these two galaxies in F277W and F444W are based on our analysis. The shaded region indicates sizes below half the FWHM of the F277W and F444W PSFs at $z=3.5$.} }
    \label{fig:size_mass_relation_f277w_f444w}
\end{figure*}

\section{Size-Mass Relation of Massive Galaxies\label{sec:smr_measure}}

We present the size-mass relation of our sample of massive galaxies in Figure~\ref{fig:size_mass_relation_f277w_f444w}. The figure shows the best-fit relation based on effective radii measured in both the F277W and F444W filters \edittwo{based on a Single S\'{e}rsic fit}, along with comparisons to previous studies. Below, we describe the method used to derive the size-mass relation and the resulting best-fit parameters.

\par To analytically model the size-mass relation, we follow methods in the literature \citep[e.g.,][]{Shen2003,vanderWel2014,Mowla2019,Kawinwanichakij2021,Nedkova2021,Ito2024,Allen2024}. \edittwo{Specifically, we model the intrinsic distribution of galaxy sizes by assuming a log-normal form, parameterized as $N(\log r,\sigma_{\log r})$, where $r$ represents the mean intrinsic effective radius, and $\sigma_{\log r}$ is the intrinsic scatter of the distribution.  This is distinct from the measured galaxy size, $R_{e}$, which is subjected to measurement uncertainties, $\delta \log R_{e}$.} We parameterize the mean intrinsic size $r$ as a function of stellar mass as 
\begin{equation}
\label{equation:singlepowerlaw}
\mathrm{kpc}=A\left( \frac{M_{\ast}}{5\times 10^{10}M_{\odot}}\right)^{\alpha},
\end{equation}
\noindent where $A$ is the effective radius at $M_{\ast} = 5\times10^{10}M_{\odot}$, and $\alpha$ is the slope of the size-mass relation. We additionally assume that $\sigma_{\log r}$ is independent of mass.
\par We then compute the total likelihood of three parameters ($A$,$\alpha$, and $\sigma_{\log r_{e}}$) as
\begin{equation}
 \mathcal{L} = \sum \ln \left[W \left< N (\log r, \sigma_{\log r}),N(\log R_{e},\delta \log R_{e}) \right>\right]
\end{equation} 
\edittwo{where $W$ is the weighting factor designed to prevent the fit from being dominated by the more numerous low-mass galaxies \citep{vanderWel2014}. We compute $W$ as the inverse of the stellar mass function (SMF) of quiescent galaxies at $3<z<3.5$, following \cite{Weaver2023}. The SMF is parameterized by a Schechter function \citep{Schechter1976}:
\begin{equation}
\Phi (M_{\ast}) \propto \left( \frac{M_{\ast}}{M_0} \right)^{\alpha+1} \exp \left( - \frac{M_{\ast}}{M_0} \right), 
\end{equation}
where we adopt $M_{0}=10^{10.41}~M_{\odot}$ and $\alpha=-1.41$. The weight is then defined as:
\begin{equation} 
W(M_{\ast}) = \frac{1}{\Phi (M_{\ast})}. 
\end{equation}
This ensures that all stellar mass bins contribute equally to the likelihood calculation, preventing bias toward the most abundant low-mass galaxies.}

\par Finally, we fit the observed $R_{e}-M_{\ast}$ relation and estimate posterior distribution functions of the parameters ($A$,$\alpha$, and $\sigma_{\log r_{e}}$) by using DYNESTY nested sampling \citep[][]{Skilling2004,Skilling2006,Speagle2020,Koposov2024}. We assume flat priors on all parameters: $\alpha=[0,1]$, $\log A=[-2,1]$, and $\sigma_{\log r_{e}}=[0.05,0.5]$.

\par \editthree{Several studies at lower redshifts have reported that the size-mass relation flattens below a pivot mass of $\log(M{\ast}/M_{\odot}) \sim 10.3$ \citep[e.g.,][]{Mowla2019a,Kawinwanichakij2021,Nedkova2021,Cutler2022}, implying that a single power-law form (Equation~\ref{equation:singlepowerlaw}) may not adequately describe the full stellar mass range. Motivated by this, we fit the size–mass relation using only galaxies with $\log(M_{\ast}/M_{\odot}) > 10.3$, excluding the two lowest-mass systems in our sample—3D-EGS-34322 and 3D-UDS-35168. We summarize our best-fit parameters of the size-mass relations for our sample of massive quiescent galaxies at $3<z<4$ in  F277W and F444W filters in Table~\ref{tab:size_mass_params}.}
\begin{deluxetable}{lccc}
\tablecaption{Best-fit Parameters of the Size-Mass Relation\label{tab:size_mass_params} of 17 massive quiescent galaxies at $3<z<4$}
\tablehead{
\colhead{Filter} & \colhead{Slope $\alpha$} & \colhead{$\log (A/\mathrm{kpc})$} & \colhead{Scatter $\sigma_{\log R_e}$}
}
\startdata
F277W & $0.74^{+0.17}_{-0.17}$ & $-0.14^{+0.05}_{-0.05}$ & $0.18^{+0.05}_{-0.03}$ \\
F444W & $0.68^{+0.15}_{-0.18}$ & $-0.22^{+0.04}_{-0.05}$ & $0.17^{+0.04}_{-0.03}$ \\
\enddata
\tablecomments{$\log (A/\mathrm{kpc})$ is the normalization at $M_{\ast} = 5 \times 10^{10} M_{\odot}$), and $\sigma_{\log R_e}$ is an intrinsic scatter of the relation.}
\end{deluxetable}

\par \editone{ A comparison of the best-fit parameters from the size-mass relations in F277W and F444W reveals a systematic difference. The effective radius at $M_{\ast}=5\times10^{10}M_{\odot}$ measured in F444W of  $R_{e}=0.60\pm0.07$ kpc is approximately a factor of $\sim1.2$ smaller than that in F277W ($R_{e}=0.72\pm0.08$ kpc). This difference between rest-frame near-IR and optical sizes suggests that stellar mass profiles are more compact than optical light profiles, consistent with previous studies \citep[e.g.,][]{Suess2022,Martoran2024}. The lower intrinsic scatter in the F444W-based relation is also expected, as this filter probes the rest-frame near-IR ($\sim1\mu$m), a better proxy for the stellar mass distribution.}

\par In Figure~\ref{fig:size_mass_relation_f277w_f444w}, we also show results from other studies which derived the size-mass relation for quiescent and star-forming population at similar redshift and based on the size measurement at rest-frame wavelength at $0.5\mu$m, similar to our measurement, namely those by \cite{vanderWel2014, Ito2024,Ward2024}. We also present our F277W and F444W size measurements for two spectroscopically confirmed quiescent galaxies from the literature including, GS-9209 ($\log(M_{\ast}/M_{\odot}) = 10.58, z = 4.658$) from \citealt{Carnall2023Nature}) and RUBIES-EGS-QG1 ($\log(M_{\ast}/M_{\odot}) = 10.9, z = 4.886$) from \citealt{deGraaff2024}. We defer a detailed comparison of our size measurements with those obtained by these authors to Section~\ref{sec:compare_otherworks}.

\begin{figure}
    \centering
    \includegraphics[width=0.48\textwidth]{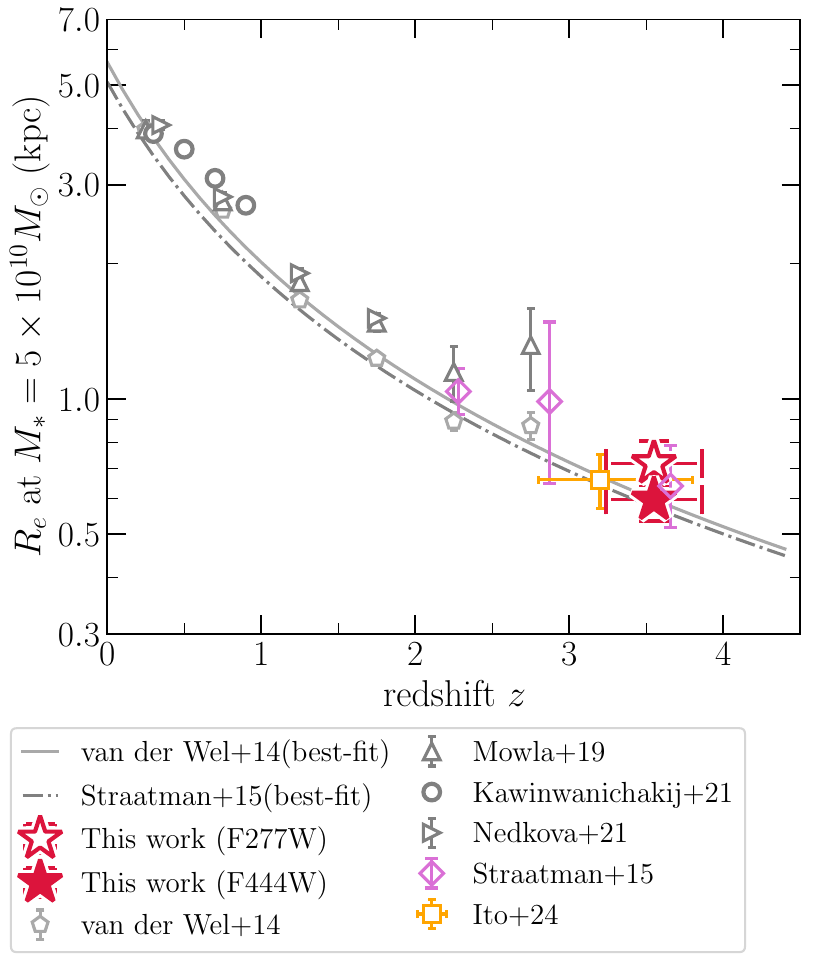}
\caption{Redshift evolution of the effective radius at $M_{\ast}=5\times10^{10}M_{\odot}$ based on the best-fit size-mass relation measured in both F277W and F444W. The large open red star shows the F277W-based size, corresponding approximately to rest-frame 0.5$\mu$m \editthree{and used for direct comparison with lower-redshift studies.} The filled red star shows the F444W-based size, \editthree{which corresponds to a longer rest-frame wavelength and is less affected by recent star formation, thus more representative of the underlying stellar mass profile.} The error bars reflect the uncertainty in the best-fit intercept of the size-mass relation. Grey markers represent lower-redshift measurements from \citet{vanderWel2014,Mowla2019,Kawinwanichakij2021,Nedkova2021}. The solid curve indicates the median size evolution of quiescent galaxies with $\log(M_{\ast}/M_{\odot})=10.5-11$ from \citet{vanderWel2014}, modeled as $r_{e}=5.62(1+z)^{-1.48}$ kpc. The dot-dashed curve shows the size evolution from \citet{Straatman2015} for a similar mass range. \editthree{Including both filters illustrates the potential impact of rest-frame wavelength on size measurements at high redshift.}  }
    \label{fig:size_evol}
\end{figure}
\subsection{Evolution of the Size-Mass Relation of Quiescent Galaxies from $z\sim4$}

\par We explore the redshift evolution of the effective radius at $M_{\ast}=5\times10^{10}M_{\odot}$ in Figure~\ref{fig:size_evol}. In the Figure, we compare our result with the median size evolution for quiescent galaxies with $\log(M_{\ast}/M_{\odot})=10.5-11$ from the 3D-HST+CANDELS \citep{vanderWel2014}, with the form of $r_{e}=5.62(1+z)^{-1.48}$ kpc, and the size evolution of quiescent galaxies at $3.4\le z <4.2$ and with $\log(M_{\ast}/M_{\odot})=10.5-11$ from the ZFOURGE+CANDELS dataset \citep{Straatman2015}, with the form of $r_{e}=5.08(1+z)^{-1.44}$ kpc. Our measurements of $R_{e}$ at $M_{\ast}=5\times10^{10}M_{\odot}$ of  $R_{e}=0.72\pm0.08$ kpc in F277W are overall in good agreement with the extrapolation of the redshift evolution of the effective radii presented in \citet{vanderWel2014} and \citet{Straatman2015}, implying a factor of $\sim7$ growth in size from $z\sim4$ to the present day at fixed stellar mass. The parameterized size evolution of galaxies also suggests a significant fraction of this size growth -- by a factor of $\sim2$ at fixed stellar mass -- occurs between $z\sim4$ to $z\sim2$, an epoch of peak cosmic star formation \citep[colloquially referred to as ``cosmic noon"; e.g., see][and reference therein]{ForsterSchreiber2020}.
\par \editthree{We note that the size measurements of \citet{vanderWel2014} and \citet{Straatman2015} differ in their rest-frame wavelengths: \citet{vanderWel2014} corrected all sizes to rest-frame $5000\AA$, while \citet{Straatman2015} used observed F160W imaging at $z\sim3.7$, corresponding to the rest-frame $\sim3400\AA$. In addition, \citeauthor{Straatman2015} emphasize that the smaller sizes in their $z\sim4$ quiescent sample are primarilly driven by the inclusion of galaxies with $\log(M_{\ast}/M_{\odot})\ge11$, which are systematically more compact. When restricting to lower-mass galaxies, their size evolution aligns more closely with the trend reported by \citeauthor{vanderWel2014}}
\begin{figure*}
    \centering
    \includegraphics[width=\textwidth]{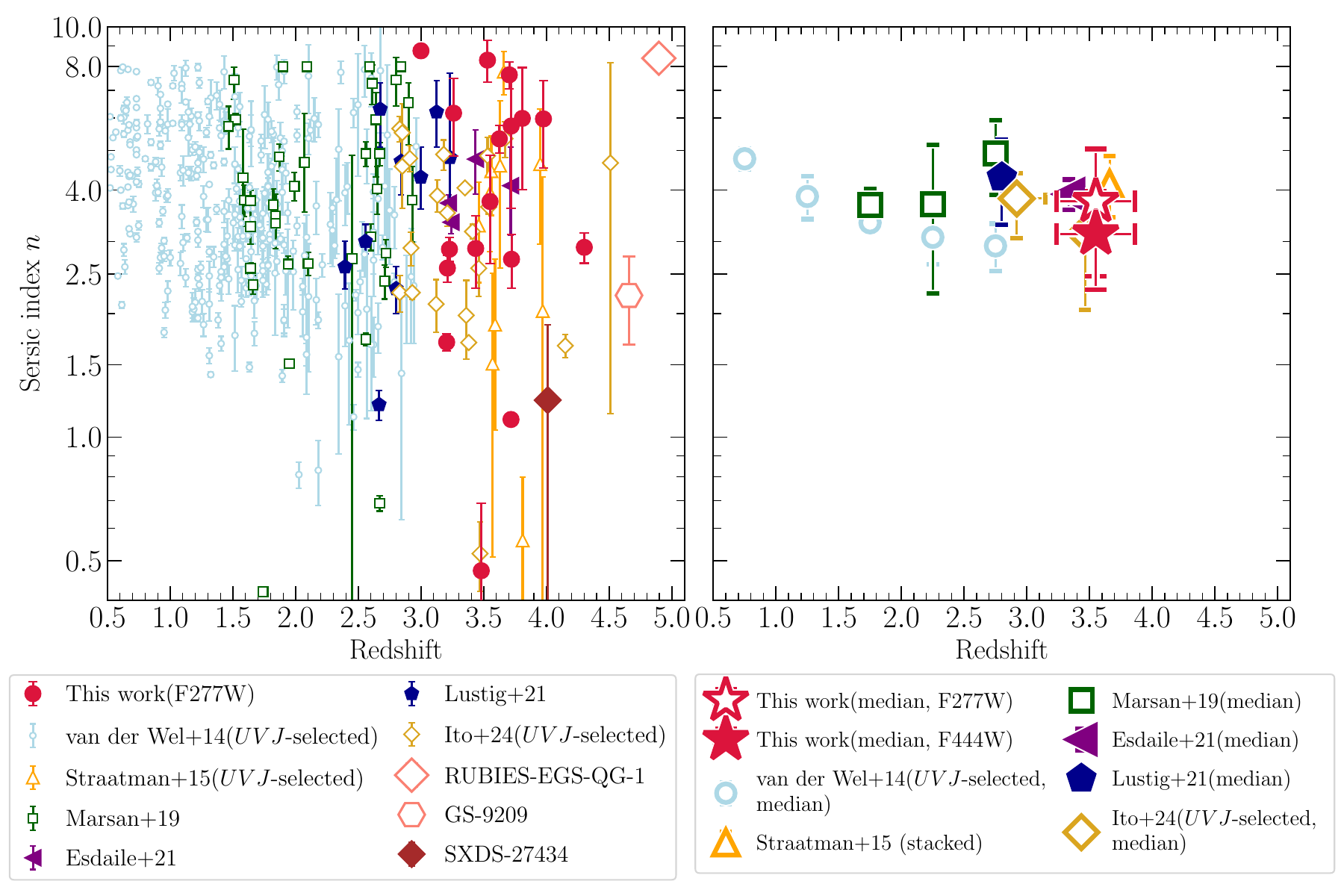}
\caption{\editone{The S\'{e}rsic index ($n$) of massive galaxies as a function of redshift. The \textbf{left panel} shows individual measurements from this work (crimson filled circles) and literature datasets, while the \textbf{right panel} highlights median values for each dataset. ZF-COS-20133, with a size below half the FWHM of the F444W PSF, has uncertain structural measurements and is shown with reduced opacity. Literature data include $UVJ$-selected quiescent galaxies (open symbols): \citet{Straatman2015} ($10.6 < \log(M_{\ast}/M_{\odot}) < 11.25$), \citet{Ito2024} ($9.8 < \log(M_{\ast}/M_{\odot}) < 11.4$), and \citet{Straatman2016} crossmatched with \citet{vanderWel2014} ($\log(M_{\ast}/M_{\odot}) > 10.3$; small light blue open circles). Spectroscopically-confirmed quiescent galaxies are shown as filled symbols: \citet{Esdaile2021} ($11 < \log(M_{\ast}/M_{\odot}) < 11.3$), \citet{Lustig2021} ($10.8 < \log(M_{\ast}/M_{\odot}) < 11.3$), and SXDS-27434 \citep{Ito2024, Tanaka2019, Valentino2020} ($\log(M_{\ast}/M_{\odot}) = 11.06$). Our F277W measurements for GS-9209 \citep{Carnall2023Nature} ($\log(M_{\ast}/M_{\odot}) = 10.58$) and RUBIES-EGS-QG1 \citep{deGraaff2024} ($\log(M_{\ast}/M_{\odot}) = 10.9$) are also shown. Measurements for photometrically-selected massive galaxies ($\log(M_{\ast}/M_{\odot}) > 11.25$) from \citet{Marsan2019} are shown as open squares. In the \textbf{right panel}, median Sérsic indices are indicated with larger symbols matching the colors and styles of individual points. Crimson stars represent the median Sérsic indices for this work in the F277W (open) and F444W (filled) bands, with error bars derived using bootstrap resampling.}}
    \label{fig:sersic_redshift}
\end{figure*}

\section{Morphological Properties of Massive Quiescent Galaxies at $3<z<4$\label{sec:result_morphology}} 
\subsection{S\'{e}rsic index}

We present the S\'{e}rsic index of massive quiescent galaxies measured in F277W, corresponding to rest-frame morphology at $\sim0.5\mu$m at $z\sim3-4$ in Figure~\ref{fig:sersic_redshift}. The measured S\'{e}rsic index ranges from $n_{\mathrm{F277W}}=1.10-8.74$, with a median value \footnote{ZF-COS-20133 is excluded from the calculation due to its extremely small size ($R_{e}=0\farcs02$), well below the half of the FWHM of the F444W PSF, resulting in uncertain S\'{e}rsic parameter (see Section~\ref{sec:sersicfit}) for details} of $n_{\mathrm{F277W}}=4.54^{+1.31}_{-1.65}$. The uncertainties on these median values are computed using bootstrapping.  We also measure S\'{e}rsic index of our sample in F444W filter, which range from $n_{\mathrm{F444W}}=1.14-8.94$ \footnote{\edittwo{The highest values of $n_{\mathrm{F444W}}$ are close to the upper fitting limit of $n=9$, suggesting that they may be influenced by this boundary rather than purely reflecting the underlying light profile. However, these high indices still indicate a highly concentrated structure, consistent with expectations for quiescent galaxies.}}, with a median value of $n_{\mathrm{F444W}}=3.59^{+1.18}_{-0.54}$. Notably, we find that 14 out of 16 galaxies have $n>2.5$, consistent with bulge-dominated morphologies. This suggests that, even at $z\sim3.5$, the majority of massive quiescent galaxies ($\sim87\%$) are already bulge-dominated. 

\par In Figure~\ref{fig:sersic_redshift}, we compare our measurements with previous studies of massive quiescent galaxies at a similar redshift range  \citep{Straatman2015, Esdaile2021, Ito2024}  and at lower redshifts \citep{vanderWel2012,Marsan2019, Lustig2021}. \editone{We also present our measurements using F277W images for two spectroscopically confirmed massive quiescent galaxies for GS-9209 at $z=4.658$ \citep{Carnall2023Nature} and RUBIES-EGS-QG-1 at $z=4.886$ \citep{deGraaff2024}}. 
\par \editone{Despite significant scatter among individual measurements—likely arising from differences in rest-frame wavelengths probed, sample selection, or the diversity in morphologies of massive quiescent galaxies—the median S\'{e}rsic indices across studies show good agreement. Over the range $z\sim1.5-4$, the  median S\'{e}rsic index is consistent around $n\sim4$, suggesting that most massive quiescent galaxies predominantly exhibit bulge-dominated morphologies even at these early epochs. }

\par Interestingly, two galaxies of our sample, ZF-COS-19589 ($n=1.14\pm0.03$) and ZF-UDS-7542 ($n=1.70\pm0.03$) are consistent with being disk-dominated. \edittwo{This is further supported by their low bulge-to-total (B/T) ratios of 0.31 and 0.25, respectively, as discussed in Section~\ref{sec:result_bulgediskdecomp}.}  Similarly, SXDS-27434, a spectroscopically confirmed quiescent galaxy at $z=4.01$ \cite{Tanaka2019,Valentino2020}, has a low S\'{e}rsic index of $n=1.24^{+0.65}_{-0.91}$\citep{Ito2024}. These findings may reflect ongoing morphological transformations of massive galaxies from disk-dominated to bulge-dominated systems within this redshift range.
\par \editone{As noted in Section~\ref{sec:sersicfit}, ZF-COS-20133 has a S\'{e}rsic index of $n\sim0.5$, likely due to its extremely small size ($R_{e}\sim0\farcs02$), which is below the half of the FWHM of the F444W PSF. This unresolved nature introduces degeneracies in S\'{e}rsic fitting, leading to artificially low S\'{e}rsic indices. Thus, the structural parameters of this galaxy should be interpreted with caution.}


\begin{figure*}
    \centering
    \includegraphics[width=\textwidth]{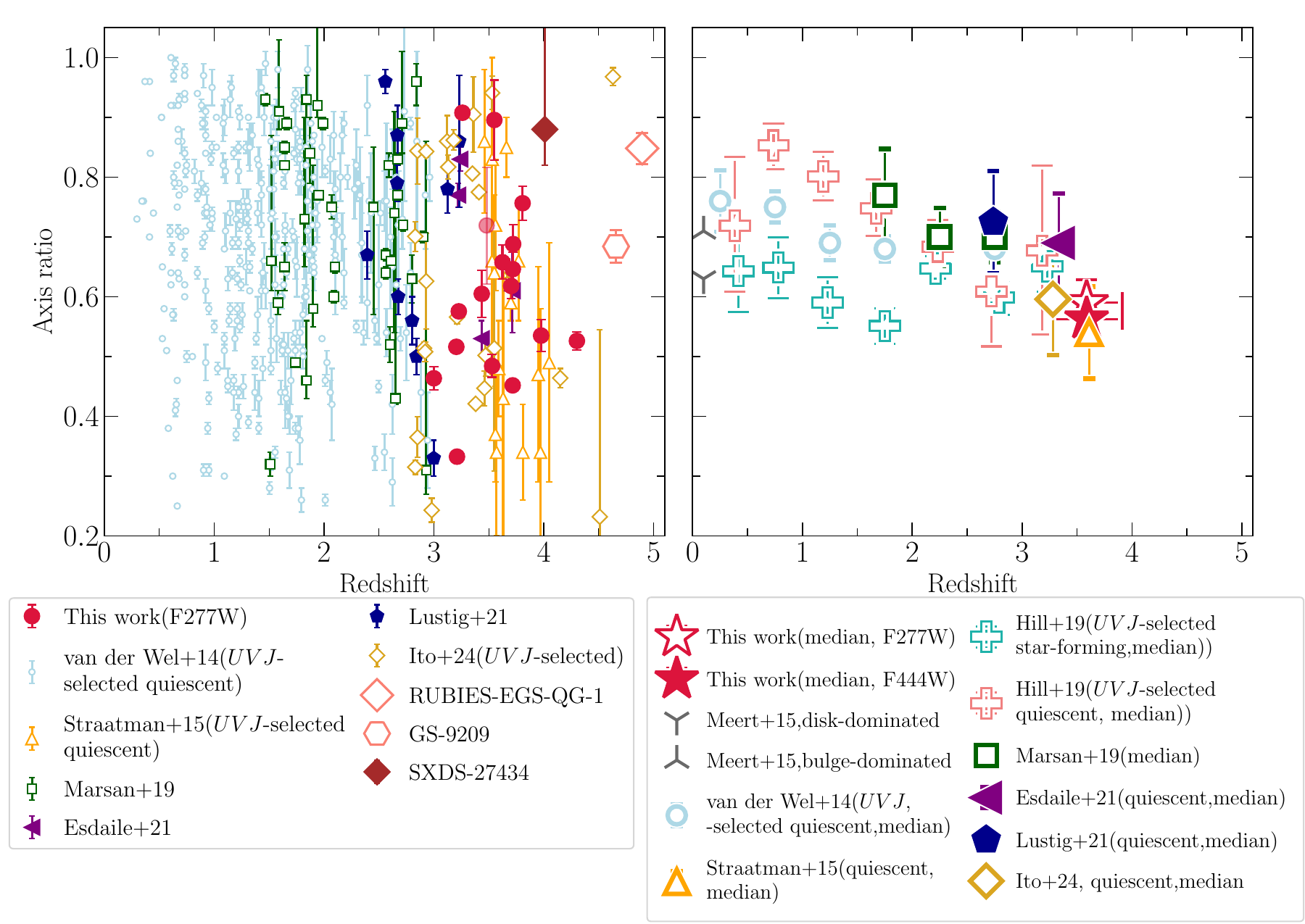}
\caption{\editone{The axis ratio ($q$) of massive galaxies as a function of redshift.
The \textbf{left panel} shows individual measurements from this work (crimson filled circles) and literature datasets, while the \textbf{right panel} highlights median $q$ values for each dataset. Literature data include $UVJ$-selected quiescent galaxies (open symbols): \citet{Straatman2015}, \citet{Hill2019}, \citet{Ito2024}, and \citet{Straatman2016} with morphologies from \citet{vanderWel2014}. Spectroscopically-confirmed quiescent galaxies are shown as filled symbols: \citet{Esdaile2021}, \citet{Lustig2021}, and SXDS-27434 \citep{Ito2024, Tanaka2019, Valentino2020}. Measurements for GS-9209 \citep{Carnall2023Nature} and RUBIES-EGS-QG1 \citep{deGraaff2024} are also included. Open squares represent photometrically-selected massive galaxies from \citet{Marsan2019}. Median values for disk-dominated ($B/T<0.5$) and bulge-dominated ($B/T>0.5$) nearby galaxies ($z\sim0.1$) are from \citet{Meert2015}. For comparison, medians for star-forming galaxies with $\log(M_{\ast}/M_{\odot})>11$ from \citet{Hill2019} are shown. Crimson stars indicate the median $q$ in F277W (open) and F444W (filled) for this work, with bootstrapped error bars.}}
    \label{fig:axisratio_redshift} 
\end{figure*}

\subsection{Axis ratio\label{sec:axisratio}}
\par \editone{We present the projected axis ratio ($q$) of massive quiescent galaxies measured in F277W in Figure~\ref{fig:axisratio_redshift}. We define this parameter as the ratio of the half-light semi-minor axis ($b$) to the half-light semi-major axis ($a$). The axis ratio ranges from $q_{\mathrm{F277W}}=0.33-0.91$, with a median value of $q_{\mathrm{F277W}}=0.59^{+0.04}_{-0.06}$. In F444W, the median value is $q_{\mathrm{F444W}}=0.56^{+0.02}_{-0.06}$. As we noted ealier, significant variation among individual measurements are likely arising from differences in rest-frame wavelengths probed, sample selection, or the diversity in morphologies of massive quiescent galaxies. Nevertheless, our results at $3<z<4$ are consistent with the median values reported at similar redshifts by \citet{Straatman2015}, \citet{Esdaile2021}, and \citet{Ito2024}, all within uncertainties.}
\par When compared to results at lower redshifts ($z\lesssim3$) from \citet{Lustig2021}, \citet{Marsan2019}, and \citet{vanderWel2014}, the median axis ratio shows a gradual increase with decreasing redshift, from $q\sim0.6$ at $z\sim3.5$ to $q\sim0.7-0.8$ by $z\sim0.5$. This evolution is consistent with findings from \citet{Hill2019} for $UVJ$-selected quiescent galaxies with $\log(M_{\ast}/M_{\odot})>11$. At $z\sim0.1$, the trend continues as we derive median axis ratios from structural parameter catalogs provided by \citet{Meert2015} for SDSS DR7, measured in $r$-band. At this low redshift, bulge-dominated galaxies ($B/T>0.5$) exhibit a median axis ratio of $q=0.71$, while disk-dominated galaxies ($B/T<0.5$) have a median of $q=0.63$. \edittwo{While this low-redshift sample is not $UVJ$-selected, selecting galaxies by $B/T$ provides a reasonable proxy for quiescent or early-type galaxies \citep[e.g.,][]{Lang2014, Brennan2015,Bluck2022,Dimauro2022}.}

\par \editthree{A similar trend in axis ratio and bulge-to-total ratio is already evident in our $z=3$-4 sample: bulge-dominated galaxies ($B/T > 0.5$) have a median of $q=0.62$, while disk-dominated galaxies ($B/T < 0.5$) are flatter, with a median $q=0.48$. This indicates that the correlation between projected axis ratio and $B/T$ is already in place at early cosmic times, even though overall axis ratios are lower than those observed at low redshift.}

\par Figure~\ref{fig:axisratio_redshift} also includes median $q$ values for $UVJ$-selected massive ($\log(M_{\ast}/M_{\odot})>11$) star-forming galaxies from \citet{Hill2019}, which remain roughly constant at $q\sim0.60-0.65$ across $0.4<z<3$. The comparable median axis ratios of massive star-forming galaxies and quiescent galaxies ($q=0.62$ in F277W) at $z\sim3$ suggest that high-redshift quiescent galaxies often retain the flattened, elongated shapes of their star-forming progenitors \citep[e.g.,][]{vanderWel2011,Wuyts2011,Bruce2012,Buitrago2013,Chang2013,Newman2015,Hill2017,Hill2019}. These galaxies have only recently quenched their star formation, leaving insufficient time for significant morphological evolution. The gradual increase in $q$ from $z\sim3.5$ to $z\sim0.1$ supports the idea that these galaxies become rounder and more bulge-dominated over time. This is consistent with a scenario where quiescent galaxies grow through minor mergers (mass ratios of 1:4 to 1:10) \citep{vanDokkum2010}, which progressively enhance their roundness. \edittwo{However, as shown by \citet{Karademir2019}, while minor mergers (1:10 to 1:5) significantly alter morphology and increase the S\'{e}rsic index, mini-mergers ($<$1:10) primarily enlarge or slightly puff up the disk rather than making the galaxy rounder, depending on the infall direction of the satellite relative to the host galaxy disk.}

\begin{figure*}
 \centering
    \includegraphics[width=0.95\textwidth]{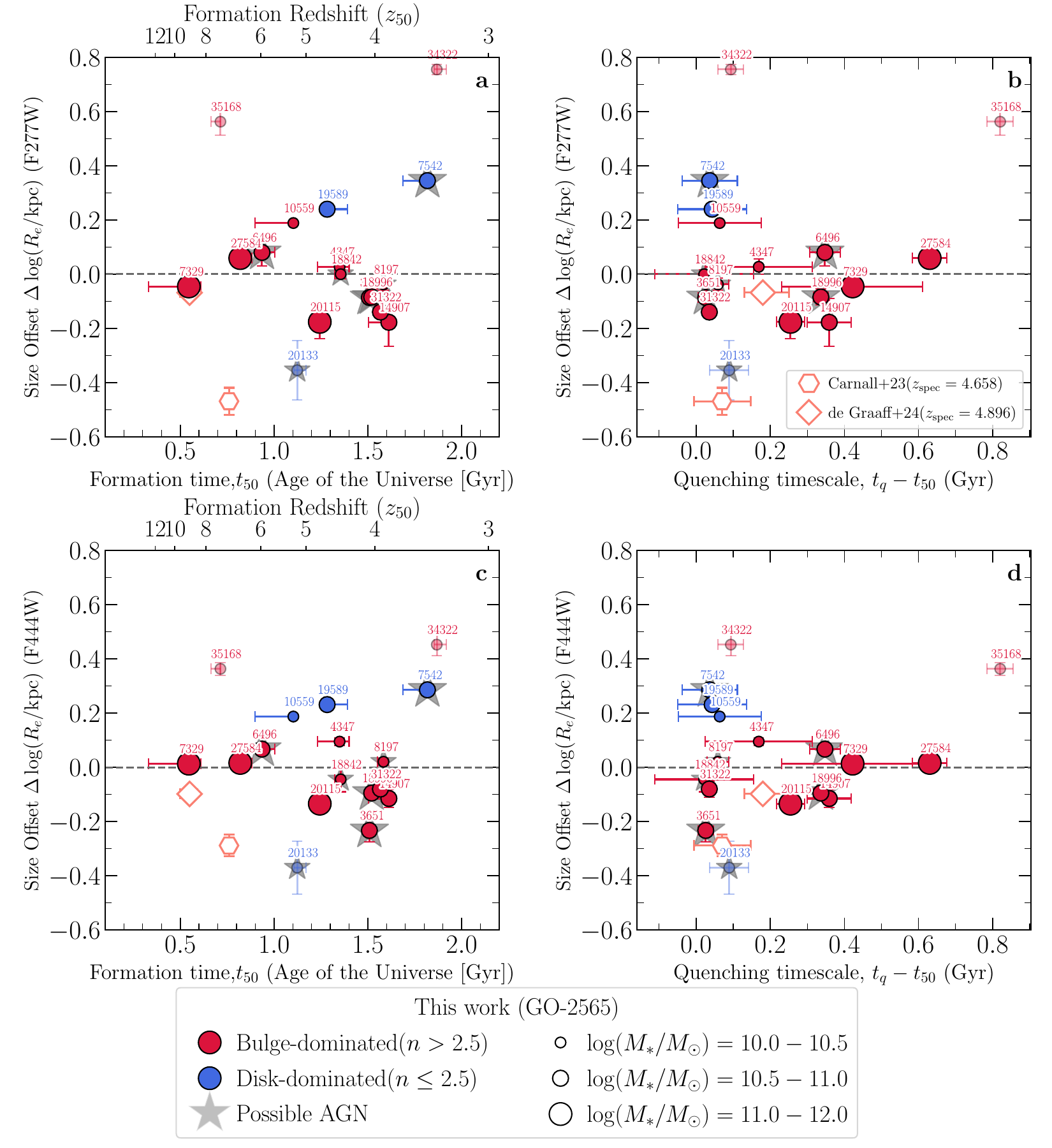}
    \caption{Relationship between the logarithmic size offsets ($\Delta \log(R_{e}/\mathrm{kpc})$) of massive quiescent galaxies at $3 < z < 4$ relative to the best-fit size–mass relation and their star formation histories. Panels (a) and (c) show size offsets measured in F277W and F444W, respectively, as a function of formation time ($t_{50}$), defined as the time when a galaxy forms 50\% of its stellar mass. Panels (b) and (d) show size offsets in F277W and F444W as a function of quenching timescale ($\Delta t_{q} = t_{q} - t_{50}$), where $t_{q}$ is the time at which the galaxy’s SFR drops below 10\% of its time-averaged value across the full SFH. In all panels, the horizontal dashed line marks $\Delta \log(R_{e}/\mathrm{kpc}) = 0$, corresponding to sizes consistent with the median size–mass relation. Galaxies are color-coded by morphology: disk-dominated systems ($n \leq 2.5$) are shown in blue, bulge-dominated systems ($n > 2.5$) in red, and candidate AGNs are indicated with star symbols. Galaxies with low stellar masses ($\log(M_{\ast}/M_{\odot}) < 10.3$), excluded from the size–mass relation fit, are shown with faint-colored symbols; their size offsets may be overestimated due to extrapolation beyond the fitted mass range, particularly if the relation flattens at low masses.Spectroscopically confirmed compact quiescent galaxies from the literature are also shown: GS-9209 ($\log(M_{\ast}/M_{\odot}) = 10.58$, $z = 4.658$; orange hexagon, \citealt{Carnall2023Nature}) and RUBIES-EGS-QG-1 ($\log(M_{\ast}/M_{\odot}) = 10.9$, $z = 4.896$; orange diamond, \citealt{deGraaff2024}), with F277W and F444W size measurements derived from our analysis. While galaxies with earlier formation times ($z_{50} \gtrsim 5$) generally have sizes consistent with the size–mass relation, \editfour{recently formed bulge-dominated galaxies tend to be more compact by up to $\sim$0.2–0.3 dex. No statistically significant correlation is found between size offset and quenching timescale in either bandpass, highlighting the large scatter and diversity of evolutionary pathways at fixed stellar mass.}
}

       \label{fig:logsizeoffset_tform_tq}
\end{figure*}

\section{Size-Mass Relations of Massive Quiescent Galaxies at $3<z<4$ and Their Star Formation Histories\label{sec:result_smr_sfh}}

\par \editone{In this section, we explore how galaxy size depends on other properties at fixed stellar mass. To do so, we measure the size offsets $\Delta \log(R_{e}/\mathrm{kpc})$ for each galaxy relative to the best-fit size-mass relation. A size offset of $\Delta \log(R_{e}/\mathrm{kpc})=0$ indicates a galaxy consistent with the expected size for its stellar mass, while positive value ($\Delta \log(R_{e}/\mathrm{kpc})>0$) indicate larger, more extended galaxies. This approach isolates the effects of stellar mass, enabling us to investigate how other parameters, such as star formation histories, influence galaxy size.}

\par \editone{Figure~\ref{fig:logsizeoffset_tform_tq}a illustrates the relationship between galaxy size offsets in F277W and formation times $t_{50}$, defined as the time when a galaxy formed $50\%$ of its observed stellar mass since the Big Bang. We adopt the best-fit $t_{50}$ parameters for our sample from \cite{Nanayakkara2025}, which were inferred using \textsc{FAST++}. Details of the method used to constrain $t_{50}$ can be found in \citeauthor{Nanayakkara2025}}


\editfour{We find that galaxies with earlier formation times ($z_{50} \gtrsim 5$) generally have sizes consistent with the expected size–mass relation. In contrast, several late-forming systems ($z_{50} \lesssim 5$) appear more compact—particularly those with bulge-dominated morphologies, such as ZF-UDS-8197, 3D-EGS-18996, 3D-EGS-31322, ZF-COS-14907, and ZF-UDS-3651—which exhibit size deficits of up to $\sim$0.2–0.3 dex. Meanwhile, late-forming \textit{disk-dominated} galaxies, including ZF-COS-19589 and ZF-UDS-7542, are significantly larger than expected. A Spearman rank correlation test between formation time ($t_{50}$) and F277W size offset confirms a statistically significant negative correlation for the bulge-dominated subsample ($\rho = -0.59$, $p = 0.045$), but no significant trend in the full sample ($\rho = -0.24$, $p = 0.42$) or when using F444W sizes. This discrepancy likely reflects the differing sensitivity of the two filters: F277W traces rest-frame optical light and is more responsive to recent star formation, while F444W probes older stellar populations and more closely reflects the mass-weighted structure. As a result, structural imprints tied to formation time are more evident in the F277W sizes. These findings are further supported by our Random Forest regression analysis (Appendix~\ref{appendix:RF_size}), which identifies $t_{50}$ as the most important predictor of size after stellar mass in the F277W filter. Taken together, the results suggest that the observed compactness among late-forming galaxies is largely confined to bulge-dominated systems—potentially reflecting distinct evolutionary mechanisms such as dissipative collapse or centrally concentrated starbursts (see Section~\ref{sec:implications} for further discussion).}

\par \editthree{While a morphology-dependent trend is observed, several galaxies still deviate significantly from the size–formation time pattern, highlighting the structural diversity within the quiescent population. Notable outliers include ZF-UDS-7542, ZF-COS-19589, 3D-EGS-34322, and ZF-COS-20133, all of which appear $0.2$–$0.8$ dex larger than predicted by the best-fit $R_{e}$–$M_{\ast}$ relation. ZF-UDS-7542 and ZF-COS-19589 are disk-dominated systems, with S\'{e}rsic indices of $n_{\mathrm{F277W}} = 1.1$ and 1.7 and bulge-to-total ($B/T$) flux ratios of 0.3 and 0.4, respectively—morphologies that naturally result in more extended light profiles than their spheroidal counterparts. ZF-COS-20133, on the other hand, is unresolved or only marginally resolved in both F277W and F444W imaging, making its size measurement particularly uncertain.}
\par 3D-EGS-34322 presents a more complex case. Although it possesses a substantial bulge component ($n_{\mathrm{F277W}} = 2.87 \pm 0.19$, $B/T = 0.73 \pm 0.02$), it formed half of its stellar mass only $\lesssim$100 Myr before the observed epoch ($z = 3.226$) and quenched just $\sim$20 Myr earlier \citep{Nanayakkara2025}. It is also the dustiest galaxy in our sample ($\tau_{\text{dust}} \sim 1.7$), occupying the dusty star-forming region of the $U-V$ vs. $V-J$ diagram \citep[][see their Figure 18]{Nanayakkara2025}, and may still host residual star formation—raising the possibility that its observed size does not yet reflect that of a fully quenched system.

\par \editthree{Finally, both 3D-UDS-35168 and 3D-EGS-34322 have stellar masses below our adopted threshold ($\log(M_{\ast}/M_{\odot}) = 10.3$) and were therefore excluded from the derivation of the best-fit size–mass relation. Their inferred size offsets are thus extrapolated beyond the calibrated mass range and may be overestimated—especially if the size–mass relation flattens at the low-mass end, as reported in previous studies \citep[e.g.,][]{Mowla2019a,Kawinwanichakij2021,Nedkova2021,Cutler2022}. These caveats should be considered when interpreting their positions relative to the broader size–formation time trend.}

\par \editthree{We assess whether the apparent trend between galaxy compactness and formation time could be driven by differences in observed redshift across the sample. This concern arises because $t_{50}$, by definition, cannot exceed the age of the Universe at the galaxy’s observed redshift. If formation time were tightly coupled to redshift, and size also evolves with redshift, any trend with $t_{50}$ might simply reflect redshift evolution. However, we find that galaxies with similarly recent formation times (e.g., $t_{50} \gtrsim 1$ Gyr) span a wide range of observed redshifts, indicating that $t_{50}$ and redshift are not strictly degenerate. A Spearman rank correlation test between $t_{50}$ and redshift yields $\rho = -0.40$, indicating only a moderate anti-correlation. This suggests that while there is some redshift dependence, it is not strong enough to fully account for the observed trend between size offset and formation time.}

\par In addition to formation timescales, quenching timescales provide valuable insights into the physical processes driving galaxy evolution. Following \cite{Carnall2018} and \cite{Nanayakkara2025}, we define the quenching time, $t_{q}$, as the time when the galaxy’s SFR falls below $10\%$ of the time-averaged SFR across its full star formation history (SFH). The quenching timescale, $\Delta t_{q} \equiv t_{q} - t_{50}$, is then the difference between the quenching time ($t_{q}$) and the formation time ($t_{50}$). For our massive quiescent galaxies at $3 < z < 4$, we find quenching timescales ranging from $\Delta t_{q} = 0.02\text{–}0.82$ Gyr. This rapid quenching aligns with the downsizing trend, where more massive galaxies quench earlier and do so over shorter timescales \citep[e.g.,][]{Cowie1996, Bell2004, Faber2007, Carnall2018,Tacchella2022} 

\par \editfour{We present the relationship between $\Delta \log(R_{e}/\mathrm{kpc})$ and quenching timescale ($\Delta t_{q} = t_q - t_{50}$) in Figure~\ref{fig:logsizeoffset_tform_tq}b. While some quiescent galaxies, such as ZF-UDS-3651, 3D-EGS-31322, ZF-COS-20115, ZF-UDS-8197, with short quenching timescales ($\Delta t_q \lesssim 0.4$ Gyr) appear slightly more compact than the median size–mass relation\footnote{We note that two galaxies in our sample, 3D-EGS-34322 and 3D-UDS-35168, fall outside the stellar mass range used to fit the size–mass relation; their size offsets may therefore be overestimated due to extrapolation. Consequently, they are excluded from the interpretation of trends discussed here.}, this is not a general trend. In fact, several late-forming, disk-dominated galaxies—such as ZF-UDS-7542, ZF-COS-19589, and ZF-COS-10559—exhibit large size excesses ($\sim$0.3–0.4 dex) despite rapid quenching, highlighting significant scatter. A Spearman rank correlation test confirms the lack of a statistically significant correlation between $\Delta t_q$ and size offset in either bandpass, with correlation coefficient of $|\rho| < 0.2$ and $p$-values $\gg 0.05$ for both the full and bulge-dominated subsamples. The same trends are shown in Figure~\ref{fig:logsizeoffset_tform_tq}d using sizes measured in F444W, with similar scatter and no clear correlation. This lack of a direct trend is not unexpected: while our Random Forest regression (Appendix~\ref{appendix:RF_size}) ranks $\Delta t_q$ as a secondary predictor of size, its importance arises from joint interactions with other physical parameters—such as stellar mass and formation time—that are captured in the multivariate Random Forest framework but not in one-dimensional correlation tests like the Spearman correlation.}


\par \editone{Interestingly, some individual quiescent galaxies from the literature exhibit both very compact morphologies and rapid quenching, in line with our expectations for short $\Delta t_q$. For example, the massive galaxy GS-9209 ($\log(M_{\ast}/M_{\odot}) = 10.58$, $z = 4.658$) has a remarkably small size of $R_{e} = 0.25 \pm 0.03$ kpc (F277W), which is $\sim$0.5 dex below our size–mass relation, and quenched within $\sim$70 Myr after forming the bulk of its stellar mass \citep{Carnall2023Nature}.}

\par \editone{Similarly, RUBIES-EGS-QG-1 at $z = 4.896$ \citep{deGraaff2024}, with $\log(M_{\ast}/M_{\odot}) = 10.9$, has a compact size of $R_{e} = 0.83 \pm 0.04$ kpc and quenched over a timescale of $\sim$180 Myr. While both galaxies are consistent with the idea that rapid quenching may be linked to compact sizes, they lie at higher redshifts ($z > 4.5$) and represent extreme cases, limiting direct comparison to our sample.}

\par Taken together, the observed trends highlight the complex interplay between star formation history and structural evolution. At $3 < z < 4$, galaxies with earlier formation times ($z_{50} \gtrsim 5$) tend to have sizes consistent with the expected relation, \editfour{while late-forming, bulge-dominated galaxies are often more compact (by $\lesssim$0.3 dex).} However, the large scatter and the presence of both compact and extended systems with rapid quenching timescales—both in our sample and in the literature—suggest that compactness is not solely determined by either formation time or quenching duration. Other physical processes, such as the mode of gas accretion, merger-driven central starbursts, or environmental effects, likely play a role.

\par \editthree{To quantitatively assess which physical properties most strongly predict galaxy size, we perform a Random Forest regression analysis (Appendix~\ref{appendix:RF_size}), using the same method later applied to bulge-to-total ratios. The results confirm that stellar mass is the dominant predictor, while formation time and quenching timescale also contribute modestly—consistent with the empirical trends discussed above. This method also reveals that local environmental processes significantly contribute to the observed diversity in galaxy morphology, as quantified by the bulge-to-total ($B/T$) light ratio (see Section~\ref{sec:randomforest_analysis}).}

\begin{figure}
  \centering
 \includegraphics[width=0.48\textwidth]{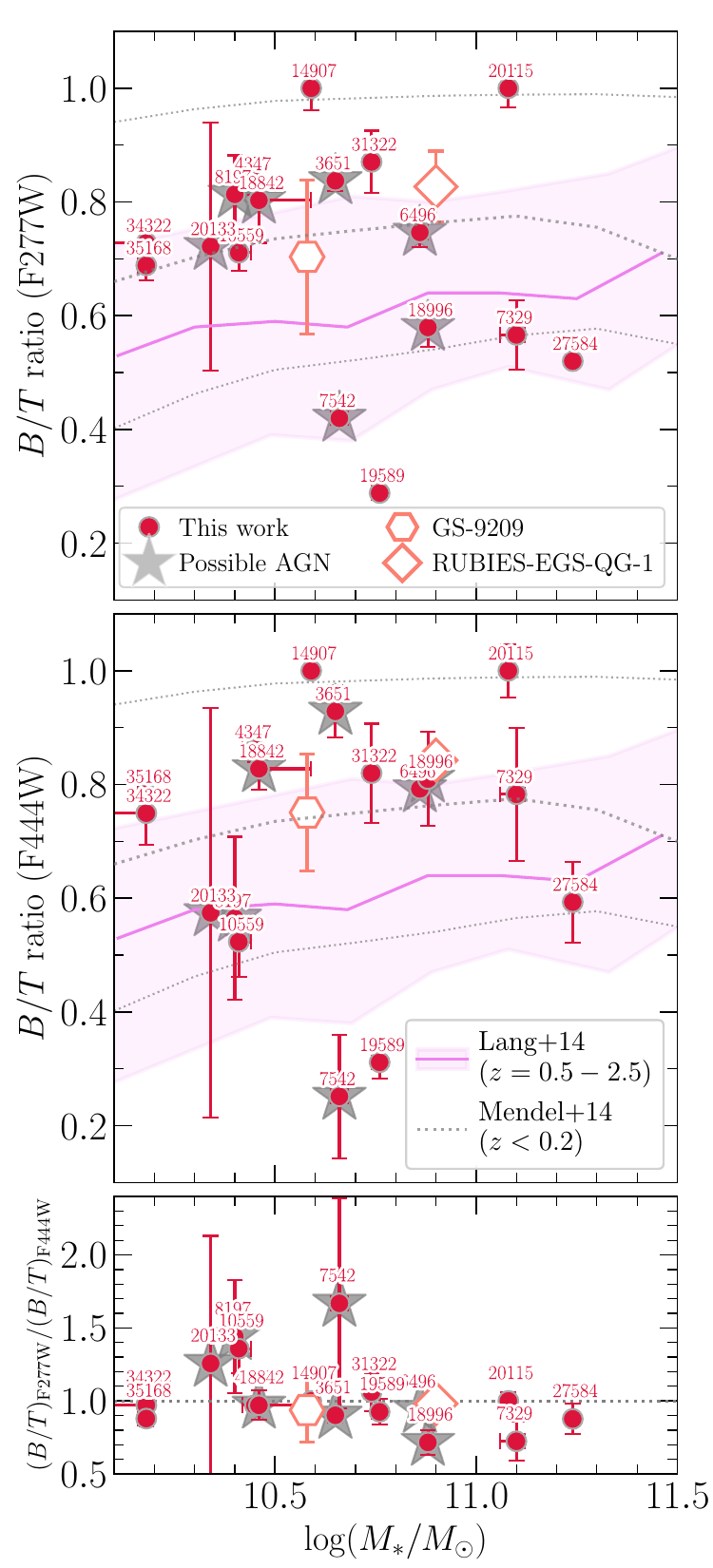}
\caption{\editone{\textbf{Top:} Bulge-to-total ($B/T$) ratios in F277W vs. total stellar mass ($M_{\ast}$) for massive quiescent galaxies at $3 < z < 4$ (red circles). \textbf{Middle:} Same as the top, but for F444W.  \textbf{Bottom:} Ratio of $B/T$ in F277W to F444W vs. $M_{\ast}$. Our measurements for spectroscopically confirmed quiescent galaxies from the literature are included: GS-9209 ($z = 4.658$; \citealt{Carnall2023Nature}) and RUBIES-EGS-QG1 ($z = 4.886$; \citealt{deGraaff2024}).The solid line and shaded region in the middle panel represent the median $B/T$ and 50th-percentile scatter for quiescent galaxies at $0.5 < z < 2.5$ (CANDELS/3DHST; \citealt{Lang2014}). Dotted lines show the median $B/T$ (and its scatter) for quiescent galaxies at $z < 0.2$ using the sample of \cite{Mendel2014} for SDSS. }} 
 \label{fig:btratio_totalmass}
\end{figure}

\begin{figure*}
  \centering
 \includegraphics[width=\textwidth]{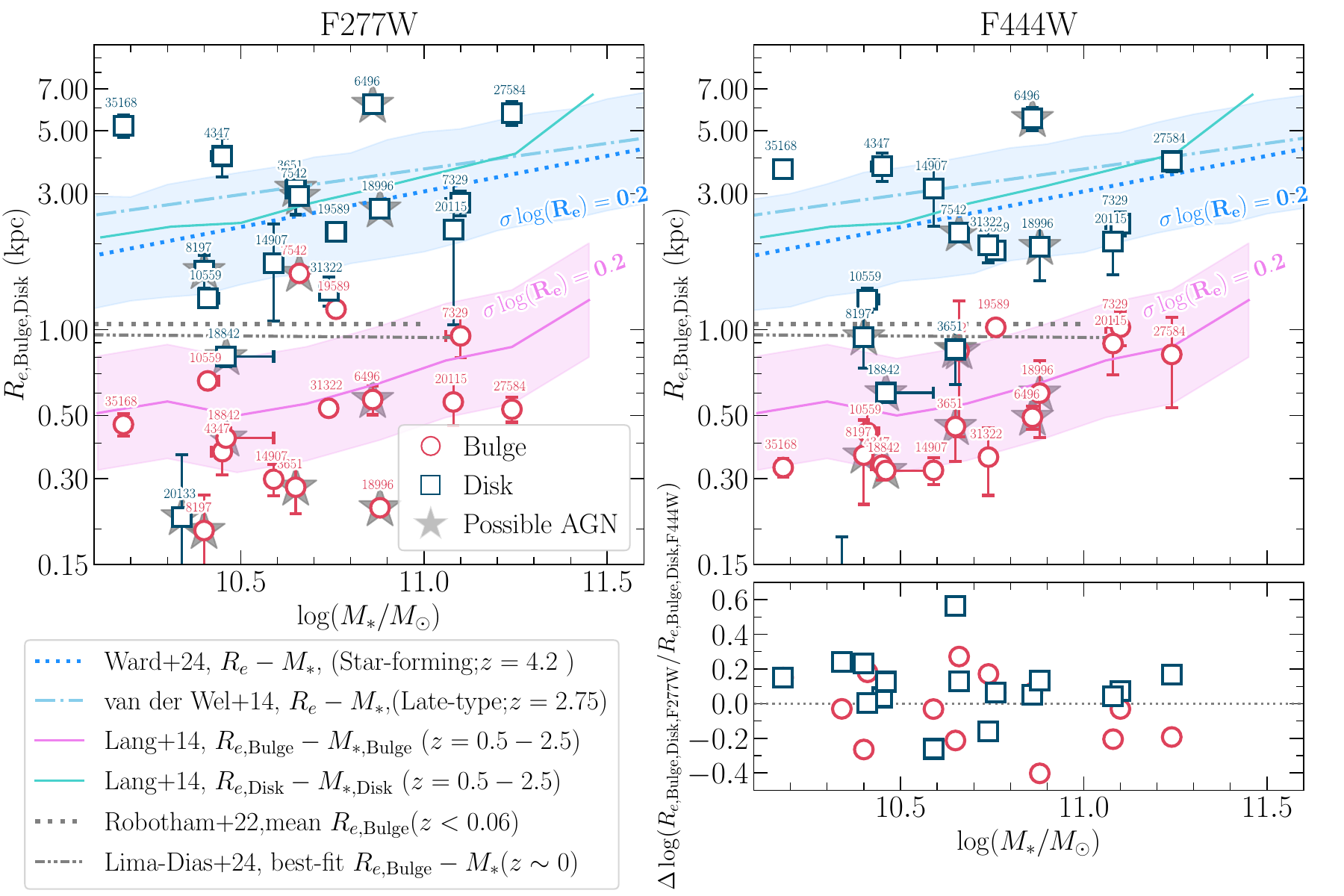}
\caption{\editone{
\textbf{Left:} Effective radii of the disk ($R_{e,\mathrm{Disk}}$, blue squares) and bulge ($R_{e,\mathrm{Bulge}}$, red circles) components as a function of total stellar mass ($M_{\ast}$) for massive quiescent galaxies at $3 < z < 4$, measured in F277W (rest-frame $\sim0.6 \, \mu\mathrm{m}$). 
\textbf{Right:} Same as the left panel, but for F444W (rest-frame $\sim1 \, \mu\mathrm{m}$). Stars indicate that galaxies are
potentially AGN based on significant detections of prominent rest-frame optical emission lines \citep{Nanayakkara2025}. Light blue dotted and dot-dashed lines show size-mass relations ($R_e - M_{\ast}$) for star-forming galaxies at $z \sim 4.2$ \citep{Ward2024} and late-type galaxies at $z = 2.75$ \citep{vanderWel2014}, with intrinsic scatter ($\sigma_{\log R_e} \sim 0.2$ dex) indicated by the shaded region. Pink and blue solid lines show median $R_{e,\mathrm{Bulge}}$ and $R_{e,\mathrm{Disk}}$ as a function of bulge ($M_{\ast,\mathrm{Bulge}}$) and disk ($M_{\ast,\mathrm{Disk}}$) stellar masses, respectively, for quiescent galaxies at $0.5 < z < 2.5$ \citep{Lang2014}, with scatter shown by the pink shaded region. Results for early-type galaxies at $z \sim 0$ from \citet{Robotham2022} and \citet{Lima-Dias2024} are also shown. Disk sizes for our sample align with size-mass relations of star-forming and late-type galaxies, as well as the $R_{e,\mathrm{Disk}} - M_{\ast}$ relation at $0.5 < z < 2.5$, suggesting disk components primarily evolve through star formation. \textbf{Bottom right:} Logarithmic size differences between disk and bulge components in F277W and F444W as a function of stellar mass.}} 
 \label{fig:bulgedisk_sizemass}

\end{figure*}
\section{Bulge and Disk components sizes and stellar mass relation}
\label{sec:result_bulgediskdecomp}
In this section, we explore the $B/T$ ratio of massive quiescent galaxies at $z=3-4$ as a function of galaxy mass (Figure~\ref{fig:btratio_totalmass}). For the $B/T$ ratio measured in F277W, our massive quiescent galaxies have $(B/T)_{\mathrm{F277W}}=0.29-1.00$ with the median value of $(B/T)_{\mathrm{F277W}}=0.73\pm0.05$. For the $B/T$ ratio measured in F444W, the galaxies have $(B/T)_{\mathrm{F444W}}=0.25-1.00$ with the median value of $(B/T)_{\mathrm{F444W}}=0.78\pm0.06$.  For both bandpasses, nearly $90\%$ of our galaxy sample (14 out of 16) have $B/T$ exceeding 0.5. ZF-UDS-7542 and ZF-COS-19589 have low $B/T$ ratios, corresponding to $(B/T)_{\mathrm{F444W}}=0.25$ and $0.31$, respectively, consistent with being disk-dominated galaxies. On average, the $B/T$ ratios determined in F277W and F444W are nearly consistent with each other, and the median value of the ratio of the $B/T$ ratios for these two bandpasses is  $(B/T)_{\mathrm{F277W}}/(B/T)_{\mathrm{F444W}}=0.97\pm0.04$.  

\par \editone{In Figure~\ref{fig:btratio_totalmass}, we compare our results with the median $B/T$ ratio (and 50th-percentile scatter) as a function of stellar mass for quiescent galaxies at $0.5 < z < 2.5$, based on CANDELS/3DHST measurements from \cite{Lang2014}. Their $B/T$ ratios are derived from bulge-disk decompositions of stellar mass maps reconstructed through resolved stellar population modeling. The comparison shows that approximately $80\%$ (11 out of 16) of our massive quiescent galaxies at $3 < z < 4$ exhibit significantly higher $B/T$ ratios than quiescent galaxies at $0.5 < z < 2.5$ of similar stellar mass. Notable exceptions include ZF-UDS-7542 and ZF-COS-19589, which have $B/T$ ratios lower than expected from the $B/T-M_{\ast}$ distribution at $0.5 < z < 2.5$.}

\par \editone{We also compare our results to galaxies in the local universe using bulge, disk, and total stellar mass measurements from SDSS DR7 by \cite{Mendel2014}, combined with star formation rate (SFR) estimates from \cite{Brinchmann2004}. Following the best-fit relation for the star-forming main sequence (MS) and its evolution across cosmic time from \cite{Popesso2023} (their Equation 14), we classify galaxies with SFRs at least one dex below the MS as quiescent. Restricting the sample to quiescent galaxies at $z_{\mathrm{spec}} < 0.2$ (median $z = 0.1$), we compute their median $B/T$ ratios as a function of stellar mass. Figure~\ref{fig:btratio_totalmass} shows the median $B/T$ ratios and 50th-percentile scatter for these galaxies. Massive quiescent galaxies at $z \sim 0.1$ with $\log(M_{\ast}/M_{\odot}) > 10$ exhibit high $B/T$ ratios, with a median $B/T \sim 0.7$, consistent with our findings for massive quiescent galaxies at $z = 3{-}4$ of similar stellar mass. This result suggests that the majority of our sample at $z=3{-}4$ have significantly formed their bulges, similar to their counterparts at $z \sim 0.1$. This is consistent with our comparison between the axis ratios of our sample with those of $z\sim0.1$ galaxies (Section~\ref{sec:axisratio})}.

\par \editone{ We present the sizes of disk and bulge components for our galaxies as a function of total stellar mass in Figure~\ref{fig:bulgedisk_sizemass}. The bulge sizes measured in F277W (F444W) range from $R_{e,\mathrm{Bulge}} = 0.20$–$1.58$ ($0.32$–$1.02$) kpc, with a median value of $R_{e,\mathrm{Bulge}} = 0.53_{-0.23}^{+0.04}$ ($0.46_{-0.10}^{+0.39}$) kpc. Similarly, the disk sizes measured in F277W (F444W) range from $R_{e,\mathrm{Disk}} = 0.80$–$6.18$ ($0.60$–$5.69$) kpc, with a median value of $R_{e,\mathrm{Disk}} = 2.65_{-1.04}^{+0.50}$ ($2.04_{-0.76}^{+1.10}$) kpc. Uncertainties in the median values are estimated using bootstrapping. We infer a median size ratio $R_{e,\mathrm{Bulge}}/R_{e,\mathrm{Disk}}$ of $0.17_{-0.08}^{+0.21}$ ($0.34_{-0.24}^{+0.09}$) for F277W (F444W). Despite significant scatter, these median size ratios are consistent with measurements of galaxies at $0.5 < z < 2.5$ from \cite{Lang2014} ($R_{e,\mathrm{Bulge}}/R_{e,\mathrm{Disk}} \sim 0.2$). }

\par \editone{Furthermore, we find that $R_{e,\mathrm{Disk}}$  measured in F277W (rest-frame $\sim 0.6\mu\mathrm{m}$) are up to $\sim 0.15$ dex larger than those measured in F444W (rest-frame $\sim 1 \mu\mathrm{m}$), as shown in the bottom panel of Figure~\ref{fig:bulgedisk_sizemass}. This suggests a negative color gradient in the disk components of these massive quiescent galaxies. Notable exceptions are ZF-UDS-8197 and ZF-UDS-3651, where the disk sizes in F444W are smaller by $0.2$ and $0.6$ dex, respectively, compared to those in F277W.}

\par \editone{In Figure~\ref{fig:bulgedisk_sizemass}, we also compare our results with the best-fit total size-mass relations ($R_e - M_{\ast}$) of star-forming galaxies at $3 < z < 5.5$ \citep{Ward2024} and late-type galaxies at $z = 2.75$ \citep{vanderWel2014}. These relations are based on size measurements corrected to the rest-frame 5000\AA. Disk sizes measured in F277W for four galaxies have lower normalization than the total size-mass relations of star-forming galaxies but remain within the intrinsic scatter of $\sigma_{\log R_e} \sim 0.2$ dex \citep[e.g.,][]{vanderWel2014, Ward2024}. This consistency suggests that the disks of these quiescent galaxies primarily evolved via star formation, in qualitative agreement with results at $z = 0.2$–$1.5$ from \cite{Nedkova2024}. Notable outliers include 3D-UDS-35168 and ZF-UDS-6496, whose disk sizes exceed the expected size-mass relation by $\sim 0.3$ dex, and ZF-COS-18842, which lies $\sim 0.4$ dex below the relation.} 

\par We also compare our measurements with the median $R_{e,\mathrm{Disk}} - M_{\ast}$ relation for quiescent galaxies at $0.5 < z < 2.5$ from \cite{Lang2014}, derived using bulge-disk decompositions from stellar mass maps. The $R_{e,\mathrm{Disk}}$ of our quiescent galaxies at $3 < z < 4$ are generally consistent with this relation within its intrinsic scatter, except for 3D-UDS-35168 and ZF-UDS-6496. \edittwo{Additionally, our F444W-based $R_{e,\mathrm{Bulge}} - M_{\ast}$  and $R_{e,\mathrm{Disk}} - M_{\ast}$ relations are consistent with the median relations for quiescent galaxies at $0.5 < z < 2.5$ \citep{Lang2014}, within the intrinsic scatter.}

\par Examining the $R_{e,\mathrm{Bulge}} - M_{\ast}$ relation, we find significant scatter in F277W but less in F444W, consistent with the expectation that F444W better reflects mass-weighted morphology at these redshifts. The relation flattens at $\log(M_{\ast}/M_{\odot}) \sim 10.5$, below which it shows little stellar mass dependence. This agrees with findings at $z = 0.2$–$1.5$ \citep{Nedkova2024}, $z \sim 0$ \citep{Robotham2022,Lima-Dias2024}. \edittwo{A similar flattening in the quiescent size–mass relation at $\log(M_{\ast}/M_{\odot}) \sim 10.3$ for galaxies at $1<z<3$ was recently reported by \citet{Cutler2024}, who identified two distinct populations : lower-mass quiescent galaxies with disky structures and younger ages, and more massive quiescent galaxies with spheroidal morphologies and older stellar populations. This transition mass aligns well with our observed flattening in the bulge size–mass relation, reinforcing the idea that different evolutionary mechanisms (e.g., environmental/feedback-driven quenching at lower masses vs. merger-driven growth at higher masses) shape the morphologies of quiescent galaxies \citep[see also][who reach similar conclusion for quiescent galaxies at $z<1.0$]{Morishita2017, Kawinwanichakij2021}.}
 
\section{Ranking of Parameters -- Machine Learning Analysis\label{sec:randomforest_analysis}}

\par The properties of galaxies are shaped by a complex interplay of factors, such as stellar mass, structural properties, star formation history, and the environment in which a galaxy resides. Many of these parameters are highly inter-correlated, making it challenging to disentangle their individual effects. The small size of our sample (17 galaxies) further limits our ability to robustly analyze correlations while controlling for other variables (e.g., within specific parameter bins). To address these challenges, we employ a Random Forest (RF) regressor \citep{Breiman2001,Geurts2006}, a machine learning technique well-suited for analyzing complex, nonlinear relationships, and quantifying the relative importance of various parameters \citep[e.g.,][]{Bluck2020,Bluck2022,Piotrowska2022,Baker2024,Li2025}. In this study, we use the RF regressor to focus on identifying the key factors that influence galaxy structure, as characterized by the $B/T$ measurement.

\par Building on this approach, in this section, we aim to identify which parameters are the most predictive of the morphological properties of massive quiescent galaxies at $z=3-4$. Specifically, we utilize a Random Forest (RF) regressor to predict the morphological properties of galaxies, focusing on the $B/T$ ratio. \edittwo{Similar techniques have been applied at lower redshifts; for example, \citet{Bluck2022} used a Random Forest classifier to study quenching and found bulge mass to be the strongest predictor. Although their focus was on quenching rather than morphology, their findings reinforce the significance of bulge-dominated structures in galaxy evolution, motivating our high-redshift analysis.} 
\par In this study, our primary goal is not to predict unknown morphological properties of galaxies (though the technique could be applied for that purpose) but rather to investigate how various parameters influence morphological transformation. To achieve this, we adopt the \textsc{RandomForestRegressor} from the \textsc{Scikit-Learn} Python package\footnote{\url{https://scikit-learn.org/stable/}}\citep{scikit-learn}. We use 100 estimators ($N_{\mathrm{est}}$), corresponding to independent decision trees, in this implementation. For the maximum depth of the trees, we use the default setting of \texttt{max\_depth$=$None}, which allows the trees to grow as deep as needed to fully capture patterns in the data. The relative importance (commonly referred to as ``feature importance") is calculated based on how much each feature reduces prediction errors (measured using the mean squared error for regression) at each split in the trees. These values are then averaged across the $N_{\mathrm{est}}$ independent decision trees to produce the final feature importance. 

\par \editone{To assess the relative importance of various parameters, we include the following set of parameters (features) in the RF regressor: total stellar mass ($M_{\ast}$);  the overdensity contrast of a galaxy ($\log(1+\delta^{\prime}_{3})$), derived using the Bayesian-motivated density metric based on the distance to the three nearest neighbors (Section~\ref{sec:measureoverdensity});  formation time ($t_{50}$), quenching timescale ($\Delta t_{q} \equiv t_{q} - t_{50}$); and AGN flag, which assigned a value of 1 for galaxies with significant detections of H$\alpha$ and/or $\mathrm{[OIII]}\lambda5007$ emission (indicating potential AGN activity) and 0 otherwise. \editthree{We do not include redshift as an input parameter in the RF regressor because our sample spans a relatively narrow redshift range ($z=3-4.3$). Within this interval, we do not observe a strong trend between redshift and B/T ratio, suggesting that redshift does not drive significant morphological evolution in our sample. Including it in the model would therefore add little predictive power while potentially introducing noise or redundancy.}}

\par To account for the potential influence of random noise and ensure the robustness of our RF regression results, we conducted 10000 independent RF regressions with random initialization (\texttt{random\_state$=$None}). Each RF regressor was trained using \texttt{bootstrap$=$True}, ensuring that a unique bootstrap sample of the input dataset was used for each decision tree in every run. This introduced variability in the sampled data for each tree and across regressions, enhancing the independence of each RF model even when the input dataset remained the same. For each parameter (``feature"), we calculated the median feature importance across these runs to mitigate variability introduced by small sample size. Parameters were then ranked based on their median importance values. This approach evaluates whether the observed predictivity is consistent and not an artifact of random noise, providing a more reliable assessment of feature relevance.  Additionally, in each run, we introduce a set of random numbers (denoted as ``Rand") as an additional parameter. \edittwo{We calculate the probability, $p$, as the fraction of RF runs in which the relative importance of a given physical parameter exceeds that of the random parameter (``Rand"). This approach provides a quantitative measure of the robustness of feature importance by leveraging multiple independent RF regressions and comparisons against a randomized baseline.}

\editthree{The final RF regression is applied to a subset of 15 galaxies with reliable measurements for all input parameters. We exclude ZF-COS-20115, ZF-COS-19589, and ZF-COS-20133 due to missing overdensity estimates or unresolved morphology. Additionally, RUBIES-EGS-QG1 and GS-9209 are not included in the analysis, as they do not belong to our core sample and are shown only for comparison with previous work.}


\begin{figure*}
  \centering
\includegraphics[width=\textwidth]{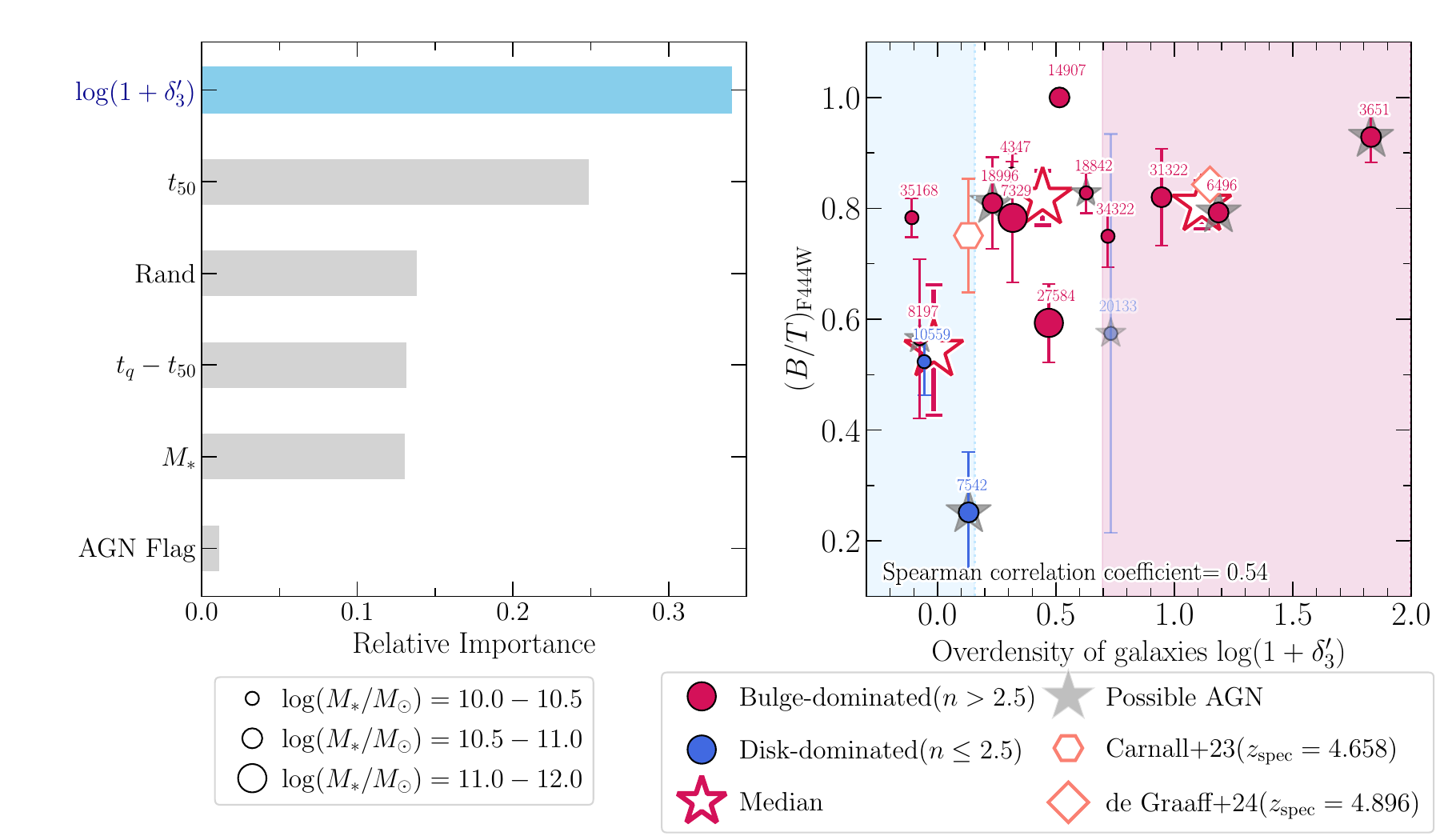}
   \caption{\editone{Left: Results of the RF regression analysis predicting $B/T$ ratios in F444W ($(B/T)_{\mathrm{F444W}}$) for massive quiescent galaxies. The $x$-axis shows the relative importance of each parameter, ranked by predictive power (listed on the $y$-axis). The overdensity of a galaxy, $\log(1+\delta^{\prime}_{3})$, derived from the projected distance all the three nearest neighbors (see Section~\ref{sec:measureoverdensity}), is identified as the most influential predictor of $(B/T)_{\mathrm{F444W}}$.  Right: $(B/T)_{\mathrm{F444W}}$ as a function of  $\log(1+\delta^{\prime}_{3})$. Open circles represent individual galaxies, with shaded regions indicating the lower and upper 25th percentiles of $\log(1+\delta^{\prime}_{3})$ distribution of our massive quiescent galaxies, defining low- and high-density environments, respectively. Our measurements for two spectroscopically confirmed quiescent galaxies from the literature are also included: GS-9209 ($\log(M_{\ast}/M_{\odot}) = 10.58, z = 4.658$; open orange hexagon, \citealt{Carnall2023Nature}) and RUBIES-EGS-QG1 ($\log(M_{\ast}/M_{\odot}) = 10.9, z = 4.886$; open orange diamond, \citealt{deGraaff2024}). ZF-COS-20133, with a size well below half of the FWHM of the F444W PSF (Figure~\ref{fig:size_mass_relation_f277w_f444w}), has uncertain structural measurements and is excluded from the RF analysis but shown here with reduced opacity. Large filled stars show the median $(B/T)_{\mathrm{F444W}}$ for galaxies for three bins: the lower 25th percentile, the interquartile range, and the upper 25th percentile of $\log(1+\delta^{\prime}_{3})$. The figure reveals a clear trend of increasing $(B/T)_{\mathrm{F444W}}$ with higher galaxy overdensities. }} 
 \label{fig:RF_BTratio}
\end{figure*}

\begin{figure}
  \centering
\includegraphics[width=0.48\textwidth]{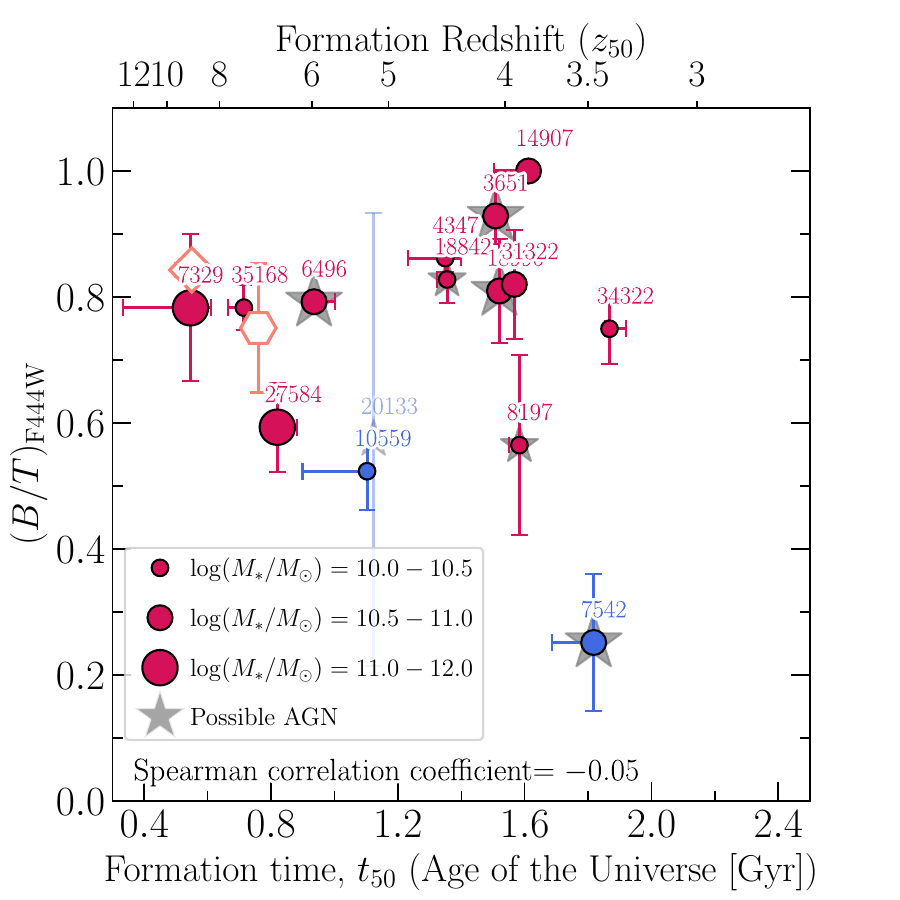}
   \caption{\edittwo{The relationship between th bulge-to-total flux ratio in F444W ($(B/T)_{\mathrm{F444W}}$) and galaxy formation time $t_{50}$, defined as the time since the Big Bang when a galaxy formed 50\% of its observed stellar mass. Marker sizes represent galaxies in three stellar mass bins, as indicated in the legend. Red and blue symbols represent bulge- and disk-dominated galaxies, respectively. Overall, earlier-forming galaxies tend to have high $(B/T)_{\mathrm{F444W}}$, suggesting that rapid early mass assembly is linked to bulge-dominated structures. Notably, a subset (5 out of 9) of recently formed galaxies ($z\lesssim5$) also exhibit high $(B/T)_{\mathrm{F444W}}$ despite residing in intermediate-density environments, implying that additional non-environmental processes may also contribute to bulge growth in these systems (see Section~\ref{sec:discussion}). }} 
 \label{fig:btratio_tform}
\end{figure}

\subsection{Which parameter is the most important in predicting the $B/T$ ratio of massive quiescent galaxies at $z=3-4$?}
\par \editone{In Figure~\ref{fig:RF_BTratio}, we present the results of our RF regression analysis to predict the $B/T$ ratio of massive quiescent galaxies at $3<z<4$ in the F444W filter, $(B/T)_{\mathrm{F444W}}$. The analysis identifies the overdensity of galaxies, $\log(1+\delta^{\prime}_{3})$, derived using the projected distance to the three nearest neighbors (see Section~\ref{sec:measureoverdensity}), as the most important parameter, with a relative importance of $0.26$. This result is highly robust, as demonstrated by 10,000 independent RF regression runs with random initialization, which show a near-unity probability that $\log(1+\delta^{\prime}_{3})$ outperforms a randomly generated variable (``Rand") in predictive power.}

\par \edittwo{Following $\log(1+\delta^{\prime}_{3})$, the next most important predictor of $(B/T)_{\mathrm{F444W}}$ is the galaxy formation time $t_{50}$ -- the time since the Big Bang at which a galaxy formed 50\% of its stellar mass. As shown in Figure~\ref{fig:btratio_tform}, galaxies with lower $t_{50}$ values -- i.e., earlier formation -- tend to have high $(B/T)_{\mathrm{F444W}}$ ratios, suggesting that rapid early mass assembly contributes to significant bulge growth. Interestingly, we also identify a subset (5 out of 9) of recently formed galaxies (around $z\sim4$) that already exhibit high $B/T$ and primarily reside in intermediate-density environments. We further explore the physical mechanisms behind this trend in Section~\ref{sec:discussion}.}

\par \editone{When all $\log(1+\delta^{\prime}_{N})$ metrics ($N=3,5,7,10$) were included alongside other parameters (e.g., $M_{\ast}$, $t_{50}$, $\Delta t_{q}$), $\log(1+\delta^{\prime})_3$ consistently emerged as the most predictive. The contributions of $\log(1+\delta^{\prime}_{N})$ for $N > 3$ were significantly diminished, with relative importances that sometimes ranked below random noise. However, when individual $\log(1+\delta^{\prime}_{N})$ metrics were tested in isolation (excluding others), those at $N=5$ and $10$ ranked as the second most important parameter after galaxy formation time ($t_{50}$). Nevertheless, $\log(1+\delta^{\prime})_3$ remained the strongest predictor, emphasizing that while large-scale environments contribute to some extent, the local environment at very small scales ($N=3$) plays the dominant role in shaping galaxy structure.   }
 
\par The relationship between local environment and $B/T$ is further evident in the observed enhancement of $B/T$ with increasing overdensity. As shown in the right panel of Figure~\ref{fig:RF_BTratio}, the median $(B/T)_{\mathrm{F444W}}$ of massive quiescent galaxies increases from $0.54\pm0.12$ in the lowest-density quartile ($\log(1+\delta^{\prime}_{3})<0.16$) to $0.81\pm0.04$ in the highest-density quartile ($\log(1+\delta^{\prime}_{3})>0.70$). \editthree{For our sample of 15 massive quiescent galaxies (RUBIES-EGS-QG1 and GS-9209 are not included)}, the Spearman rank correlation coefficient between $\log(1+\delta^{\prime}_{3})$ and $(B/T)_{\mathrm{F444W}}$ is 0.54, \editthree{with a p-value of 0.04}. This is stronger than the correlations found using alternative density definitions ($N=5$, 7, and 10), which yield lower coefficients ($\rho \lesssim 0.5$) and \editthree{higher p-values ($\sim$0.1–0.2)}. These results support the conclusion that $\log(1+\delta^{\prime}_{3})$ is the most predictive parameter for $(B/T)_{\mathrm{F444W}}$, outperforming other environmental metrics and intrinsic galaxy properties explored in this work. In Section~\ref{sec:discussion}, we extend our RF regression analysis to the IllustrisTNG simulation using the same methodology, enabling direct comparison between observations and theoretical predictions for the link between environment and morphology.

\section{Discussion}
\label{sec:discussion}
\par In this section,  \editone{we place the size-mass relation of massive quiescent galaxies at $z\sim3-4$ in the context of previous studies, discuss the implications of our findings for quenching mechanisms, and use predictions from hydrodynamical simulations to explore the impact of environments on the morphological properties of massive quiescent galaxies at this redshift.}

\begin{figure}
    \centering
    \includegraphics[width=0.49\textwidth]{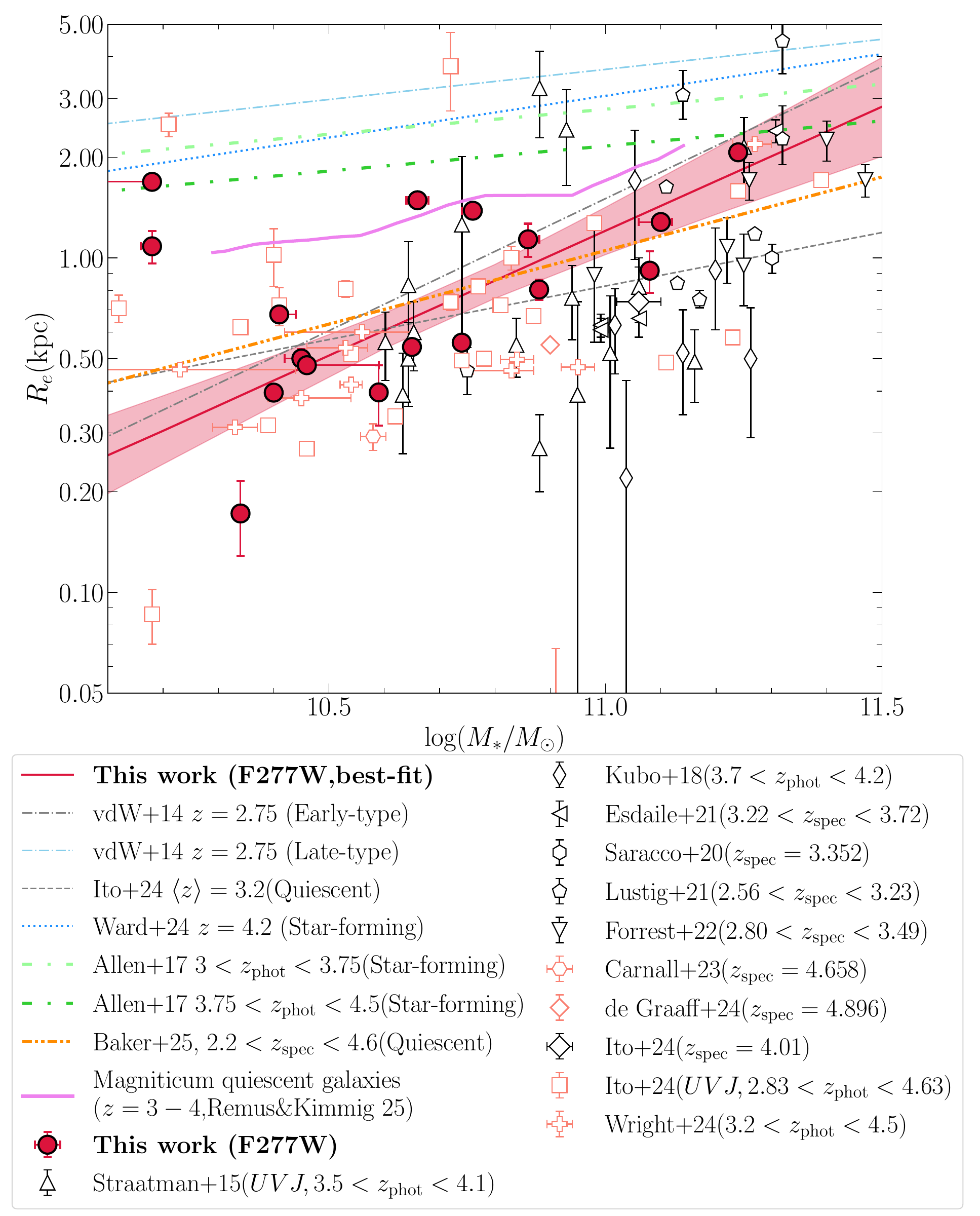}
 \caption{Compilation of the $R_{e}$–$M_{\ast}$ relation for quiescent galaxies up to $z\sim4$ (labeled). Our best-fit $R_{e}$–$M_{\ast}$ relation, based on $2.7\mu$m size measurements, is shown as a solid line with a shaded region indicating the uncertainty, reproduced from Figure~\ref{fig:size_mass_relation_f277w_f444w}. Individual $2.7\mu$m size measurements are represented as filled red circles. Open markers show size measurements of individual quiescent galaxies from the literature at $z\ge2.5$ \citep{Straatman2015,Kubo2018,Esdaile2021, Saracco2020,Lustig2021, Forrest2022,Carnall2023Nature,deGraaff2024,Ito2024,Wright2024}, with open light red markers indicating JWST-based measurements and black open markers representing ground-based or HST measurements. The relations for star-forming galaxies at the similar redshift range are taken from \cite{vanderWel2014}, \cite{Allen2017}, and \cite{Ward2024}, respectively. The relations for quiescent galaxies are taken from  \cite{vanderWel2014}, \cite{Ito2024}, \cite{Baker2025}. While the observed wavelengths differ among the samples, many sizes are not corrected to the rest-frame 5000\, \text{\AA}, except in \cite{Lustig2021} and \cite{Forrest2022}. Overall, the compiled results are consistent with our findings. \edittwo{The median $R_{e}$–$M_{\ast}$ relation for quiescent galaxies at $3<z<4$ from cosmological simulation suite Magneticum Pathfinder \citep{Remus2025} are also shown}. }
    \label{fig:sizemassrelation_otherworks}
\end{figure}

\subsection{Comparison of the Size-Mass Relation of Massive Quiescent Galaxies at $3<z<4$ with Previous Studies}
\label{sec:compare_otherworks}

\par We compare our size-mass relations and best-fit parameters in F277W and F444W with previous studies in Figure~\ref{fig:sizemassrelation_otherworks}, which compiles the $R_{e}$–$M_{\ast}$ relations for quiescent galaxies from various studies spanning $z\sim2.5$ to $z\sim4.5$ \citep{Straatman2015,Kubo2018,Esdaile2021,Saracco2020,Lustig2021,Forrest2022,Carnall2023Nature,Ito2024,Wright2024,Baker2025}, based on observations with JWST, HST, and ground-based telescopes. The figure also includes the size-mass relations for star-forming galaxies at $z=2.75$ and $z=4.2$ from \cite{vanderWel2014} and \cite{Ward2024} for comparison. Despite variations in the rest-frame wavelengths of size measurements among these studies—only some of which are corrected to the rest-frame 5000\, \text{\AA} \citep[e.g.,][]{Lustig2021,Forrest2022,Ito2024}—the compiled $R_{e}$–$M_{\ast}$ relations are broadly consistent with our findings.

\par \editone{While the overall trend agree, discrepancies arise due to sample selection, redshift coverage, and measurement methods.} For example, massive quiescent galaxies with $\log(M_{\ast}/M_{\odot})\gtrsim11$ in \cite{Lustig2021} have larger sizes ($R_{e}\sim2$–$4$ kpc) than predicted by our relation ($\sim1.3$ kpc at $M_{\ast}=10^{11.3}~M_{\odot}$). These galaxies are at lower redshifts ($z=2.7$–$3.1$) than our sample’s median redshift ($z\sim3.5$). 

\par Moreoever, variations in instrumentation contribute to differences between studies using JWST and HST data. \editone{At similar observed wavelengths, the smaller PSF sizes of JWST/NIRCam filter and higher sensitivity enable the detection of compact light profile and/or faint, extended structures at larger radii, resulting in larger effective radii and higher S\'{e}rsic indices.} In contrast, the broader PSF and lower sensitivity of HST/WFC3 may smooth out the compact centers of galaxies and/or underestimate their sizes by missing faint extended components (see Appendix~\ref{sec:compare_size_nircam_wfc3} for the test and discussion).

\par A more direct comparison can be made with \cite{Ito2024}, as both studies use JWST/NIRCam, minimizing instrumental differences. Additionally, the overlapping redshift range ($3.0<z<4.3$ for our sample and $2.8<z<4.6$ for \citeauthor{Ito2024}) allows for detailed examination of similarities and differences. \citeauthor{Ito2024} analyzed the size-mass relation of 26 quiescent galaxies, \edittwo{selected using photometric redshifts and rest-frame $UVJ$ colors, with $9.8<\log(M_{\ast}/M_{\odot})<11.4$,} and corrected their sizes to a rest-frame wavelength of $0.5\mu$m. Due to our smaller sample size, we cannot constrain the size gradient ($\Delta \log(R_{e})/\Delta \log (\lambda)$) as done by \citeauthor{Ito2024} and others \citep[e.g.,][]{vanderWel2014,Kawinwanichakij2021,Nedkova2021}. However, our narrower redshift ($3.0<z<4.3$) and stellar mass range ($10.2<\log(M_{\ast}/M_{\odot})<11.2$) ensure that we are measuring the size and structural properties of our galaxies at relatively narrow rest-frame wavelength: $\lambda_{\mathrm{rest}}\sim0.5$–$0.7\mu$m for F277W and $\lambda_{\mathrm{rest}}\sim0.8$–$1.1\mu$m for F444W.

\par Our F277W-based size–mass relation has a normalization of $\log(A/\mathrm{kpc}) = -0.14 \pm 0.05$ at $M_{\ast} = 5 \times 10^{10},M_{\odot}$, which is $0.04$ dex higher than \citeauthor{Ito2024}’s value of $-0.18 \pm 0.06$, though consistent within uncertainties. Our best-fit slope of $\alpha = 0.74 \pm 0.20$ is notably steeper than the \citeauthor{Ito2024}’s slope of $\alpha = 0.32 \pm 0.20$. \editone{While both studies adopt the same parameterization from \citet{vanderWel2014}, differences in normalization and slope may arise from sample selection effects, such as contamination by dusty star-forming galaxies in $UVJ$-selected samples \citep[e.g.,][]{Martis2019,PerezGonzales2023,Long2024}. Additionally, \citeauthor{Ito2024}’s broader redshift range may include more compact, higher-redshift galaxies, lowering the overall normalization.}

\par Our slope is in better agreement with that reported by \citet{vanderWel2014} for quiescent galaxies at $z\sim3$, who find $\alpha = 0.79 \pm 0.07$. Notably, all three studies—ours, \citet{Ito2024}, and \citet{vanderWel2014}—adopt a consistent stellar mass threshold of $\log(M_{\ast}/M_{\odot}) > 10.3$. However, when we include all galaxies in our sample with $\log(M_{\ast}/M_{\odot}) > 10.18$, the best-fit slope flattens to $\alpha = 0.36 \pm 0.20$, more closely matching that of \citet{Ito2024}. This suggests that the shallower slope found by \citeauthor{Ito2024} may reflect contamination from intrinsically lower-mass galaxies with larger sizes that scatter into the high-mass regime due to stellar mass uncertainties. If the size–mass relation indeed flattens below a pivot mass of $\log(M_{\ast}/M_{\odot}) \sim 10.3$, as observed at lower redshifts, such contamination would bias the slope toward shallower values.

\par Despite these differences, our results and those of \citeauthor{Ito2024} are consistent with the redshift evolution of quiescent galaxy sizes extrapolated from \cite{vanderWel2014} and \cite{Straatman2015}, as well as measurements at low redshifts \citep[e.g.,][]{Nedkova2021,Kawinwanichakij2021,Mowla2019}. \editone {This evolution suggests a nearly sevenfold size increase} at fixed stellar mass from $z\sim4$ to the present day (Figure~\ref{fig:size_evol}).

\editfour{In addition to comparisons with observational studies, we find a qualitative agreement between our observed size--mass relation and predictions from the Magneticum Pathfinder simulations \citep{Remus2025}. Specifically, both show that more quenched galaxies at high redshift tend to be more compact, consistent with the trend illustrated in Fig.~3 of \citet{Remus2025}. While Magneticum’s quiescent galaxies are systematically larger than observed---likely due to limited mass and spatial resolution that prevent the most compact systems from being accurately resolved with enough star particles---the simulation nonetheless captures the overall evolutionary trend. We highlight this comparison because Magneticum is among the few cosmological simulations that provide structural predictions for quiescent galaxies at $z>3$, and it produces a relatively large population of massive quiescent systems at these redshifts. This is a consequence of its earlier quenching timescales compared to other simulations, resulting in a higher number density of massive quiescent sources at $z \sim 4$ \citep{Kimmig2025}. The discrepancy in size normalization underscores the need for higher-resolution simulations to enable more precise comparisons.}


\par Finally, our best-fit intrinsic scatter in the size–mass relation at $3<z<4$ in F277W is $\sigma_{\log R_{e}} = 0.18^{+0.05}_{-0.03}$. Within the uncertainties, this value is consistent with the scatter reported for early-type galaxies at $2.5 < z < 3.0$ by \citet{vanderWel2014} ($\sigma_{\log R_{e}} = 0.14 \pm 0.03$), as well as with the broader scatter measured by \citet{Ito2024} ($\sigma_{\log R_{e}} = 0.28^{+0.13}_{-0.11}$). We also note that our measured intrinsic scatter is broadly consistent with the $0.22$–$0.25$ dex scatter in the halo spin parameter \citep[$\lambda$;][]{Bullock2001},\footnote{Galaxy sizes are known to correlate with their parent halo properties, particularly the virial radius and angular momentum acquired through tidal torques during cosmological collapse. Since the halo spin parameter ($\lambda$) reflects variations in angular momentum, it is natural to expect that its scatter contributes to the observed dispersion in galaxy sizes \citep[e.g.,][]{Kravtsov2013}.} suggesting that angular momentum diversity may contribute to the observed size scatter.

\par The inferred intrinsic scatter also reflects a broader diversity in the structural properties of massive quiescent galaxies at high redshift, implying that size is influenced by additional parameters beyond stellar mass. This diversity may reflect a range of evolutionary pathways shaped by progenitor populations \citep[e.g.,][]{Forrest2020a,Forrest2020b,Valentino2020,Carnall2023Nature}, merger histories \citep[e.g.,][]{Matharu2019,Ghosh2024}, and local environments \citep[e.g.,][]{McConachie2022,Ito2023,Espinoza2024,Kakimoto2024}.

\begin{figure}
    \centering
    \includegraphics[width=0.48\textwidth]{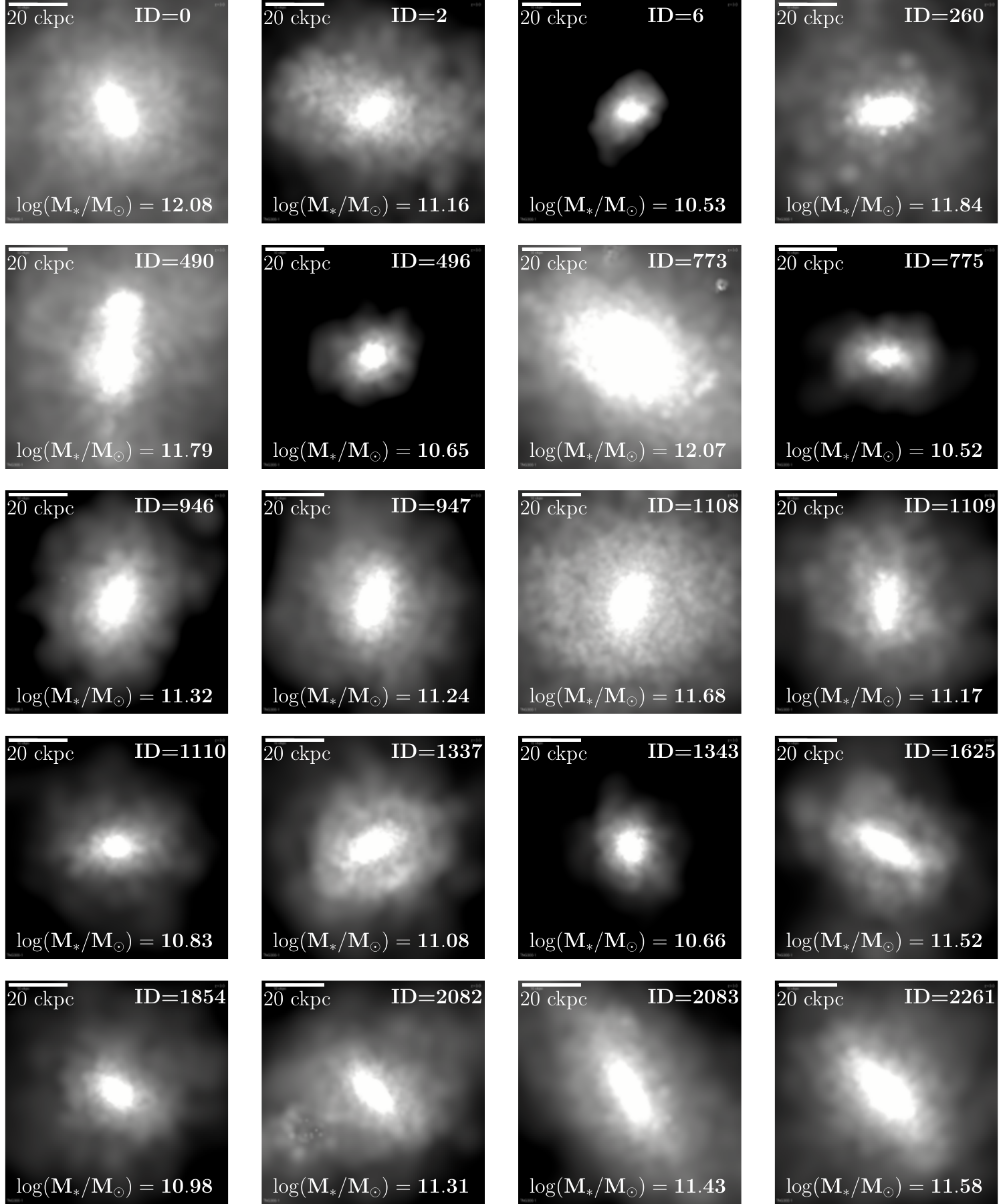}
 \caption{\edittwo{Synthetic JWST/NIRCam F277W images of 20 massive quiescent galaxies selected from the TNG300-1 simulation at $z = 3$. Each galaxy has a stellar mass of $\log(M_{\ast}/M_{\odot})>10.5$ (labeled). All images are $2\farcs5\times2\farcs5$, centered on each galaxy, and are displayed on the same color scale, with F277W surface brightness in units of mag/arcsec$^{2}$. The galaxies are ordered by subhalo ID.}}
    \label{fig:TNGmontage}
\end{figure}

\subsection{Insights from Simulations: Ex-Situ Growth and Environmental Density as Drivers of $B/T$ of Massive Galaxies\label{sec:tng_comparison}}
\par We aim to further understand the observed correlation between $B/T$ and environmental density ($\log(1+\delta^{\prime}_{3})$) of massive quiescent galaxies at $3<z<4$ (Figure~\ref{fig:RF_BTratio}).  To achieve this, we use the IllustrisTNG 300-1 simulation \citep{Springel2018} to select massive galaxies with $\log(M_{\ast}/M_{\odot})>10.5$ at $z=3$. We identify quiescent galaxies in the simulation by requiring  $\mathrm{sSFR} < 1.5 \times 10^{-10}\mathrm{yr}^{-1}$ \citep{Schreiber2018}, resulting in a sample of 668 simulated quiescent galaxies at $z=3$. Among these, 580 galaxies are classified as centrals, meaning they do not reside within the virial radius of a larger halo, and 88 galaxies are classified as satellites, which are located within the virial radius of a central galaxy. \edittwo{A subset of 20 of these massive quiescent galaxies is shown in Figure~\ref{fig:TNGmontage}, where we present synthetic JWST/NIRCam F277W images of the selected sample. These images illustrate the structural diversity of the quiescent population in TNG300-1, highlighting variations in morphology and central concentration. }

\par For each simulated galaxy, we quantify the $B/T$ ratio using the supplementary catalog provided by \citet{Genel2015}. Following \citet{Muller2023} \citep[see also][]{Tacchella2019}, the $B/T$ ratio is calculated as twice the fractional mass of stars with a circularity parameter $\epsilon < 0$, referred to as ``CircTwiceBelow0" \footnote{\url{https://www.tng-project.org/data/docs/specifications/\#sec5c}}. \editthree{This kinematic definition differs from the photometric bulge–disk decomposition adopted for the observed galaxies. We discuss the implications of this methodological discrepancy in more detail in Section~\ref{sec:caveats}.}

\par \editone{Additionally, we quantify the impact of merging history on $B/T$ ratios of massive quiescent galaxies by using the \emph{ex-situ} stellar mass fraction ($f_{\ast,\mathrm{ex-situ}}$). This quantity represents the fraction of a galaxy's stellar mass that originated from stars that formed in other galaxies and subsequently accreated. \cite{Rodriguez-Gomez2016} demonstrated that in the Illustris simulation,  $f_{\ast,\mathrm{ex-situ}}$ is strongly correlated with stellar mass and other galaxy properties, making it a robust and well-measured quantity. In particular, for central galaxies at $z=0$, $f_{\ast,\mathrm{ex-situ}}$  is a good proxy for massive, recent, and gas-poor (dry) merging history, which become ubiquitous at $M_{\ast}>10^{11}M_{\odot}$ (see their Figure 6). For our RF regressor analysis, we use the supplementary catalog provided by \citet{Rodriguez-Gomez2015, Rodriguez-Gomez2016, Rodriguez-Gomez2017} for the measurement of $f_{\ast,\mathrm{ex-situ}}$. It is important to note, however, that $f_{\ast,\mathrm{ex-situ}}$ does not directly capture the entire merging history but rather hightlights the relative role of mergers compared to dissipative processes such as in-situ star formation \citep[e.g.,][]{Oser2010}}. 

\par \editone{To directly compare our result from RF regressor analysis of observed sample with the prediction from the simulation, we use the catalog of nearest neighbors of galaxies provided by \citet{Flores-Freitas2024}, constructed using Euclidean 3D distances between galaxies in the simulation box with stellar masses above $\log(M_{\ast}/M_{\odot}) = 9.1$, consistent with the resolution of the TNG300-1 simulation. Environmental densities are esimated using a Bayesian-motivated approach (Appendix~\ref{appendix:measure_overdensity}; Equation~\ref{eq:bayesian_density_cowan}), similar to our sample, but computed volumetrically using 3D distances, instead of 2D projections. The volumetric density is calculated as: } 
\begin{equation}
\rho^{\prime}_{N} \propto \frac{1}{\sum_{i=1}^{N} d_{i,\mathrm{3D}}^{3}},
\end{equation}
\noindent where $d_{i,\mathrm{3D}}$ is the 3D distance to the $i$th nearest neighbor, as listed in the \citet{Flores-Freitas2024} catalog.
\par The overdensity is defined as $(1+\delta^{\prime\mathrm{3D}}_{N})=  \rho^{\prime}_{N}/\left<\rho^{\prime}_{N} \right>_{\mathrm{uniform}}$, where $ \left<\rho^{\prime}_{N} \right>_{\mathrm{uniform}}$ represents the median volumetric density derived from uniform density maps. We construct these maps using galaxies uniformly distrubuted within the simulation box volume, converted to physical units at $z=3$, ensuring consistency with the methodology used for the observed sample.

\par We perform a RF regression analysis to identify the most important parameters for predicting the $B/T$ ratio in the TNG sample, following the method described in Section~\ref{sec:randomforest_analysis}. To explore the relative importance of various parameters, we consider the following set of features: black hole mass ($M_{\mathrm{BH}}$), gas mass fraction ($f_{\mathrm{gas}}$), subhalo mass ($M_{\mathrm{H}}$), group halo mass ($M_{\mathrm{group,200c}}$), local overdensity derived using the distance to the three nearest neighbors ($\log(1+\delta^{\prime\mathrm{3D}}_{3})$), and the fraction of stellar mass formed ex-situ ($f_{\ast,\mathrm{ex-situ}}$).

\par \editthree{We also include formation time ($t_{50}$) and quenching timescale ($t_{\mathrm{q}} - t_{50}$) in the simulation-based Random Forest (RF) analysis. This enables a direct comparison between the physical drivers of bulge growth in simulated and observed galaxies. We find that the distributions of $t_{50}$ and $t_q - t_{50}$ in TNG300-1 are broadly consistent with those inferred for massive quiescent galaxies at $z \sim 3$ \citep[e.g.,][]{Nanayakkara2025}, supporting the use of these quantities in the RF analysis. The main exception is ZF-UDS-7329, the oldest system in our observed sample, which exhibits an unusually early formation epoch that is not well matched by any simulated analogs in TNG300-1 \citep[see also][]{Glazebrook2024Nature}. While this limits direct comparisons on a galaxy-by-galaxy basis for such extreme cases, the overall population-level agreement justifies the inclusion of $t_{50}$ and $t_q - t_{50}$ in our predictive modeling framework.
}
\par We also exclude total stellar mass ($M_{\ast}$) from the initial set of input features because it is explicitly defined in the simulation as the sum of in-situ and ex-situ stellar mass. Since $f_{\ast,\mathrm{ex-situ}}$ is already included, adding $M_{\ast}$ could introduce redundancy and potentially obscure the importance of other parameters in driving $B/T$. However, to assess the impact of this choice, we later perform an additional RF test including $M_{\ast}$ (see Appendix~\ref{appendix:Mstar_RF_test}).

\par \editthree{Finally, we test whether the predictive power of overdensity is independent of group halo mass, given their known correlation \citep[e.g.,][]{Chiang2013}. To do so, we perform an additional RF analysis excluding $M_{\mathrm{group,200c}}$ (see Appendix~\ref{appendix:RF_tests_TNG} and discussion below).}



\begin{figure*}
  \centering
 \includegraphics[width=\textwidth]{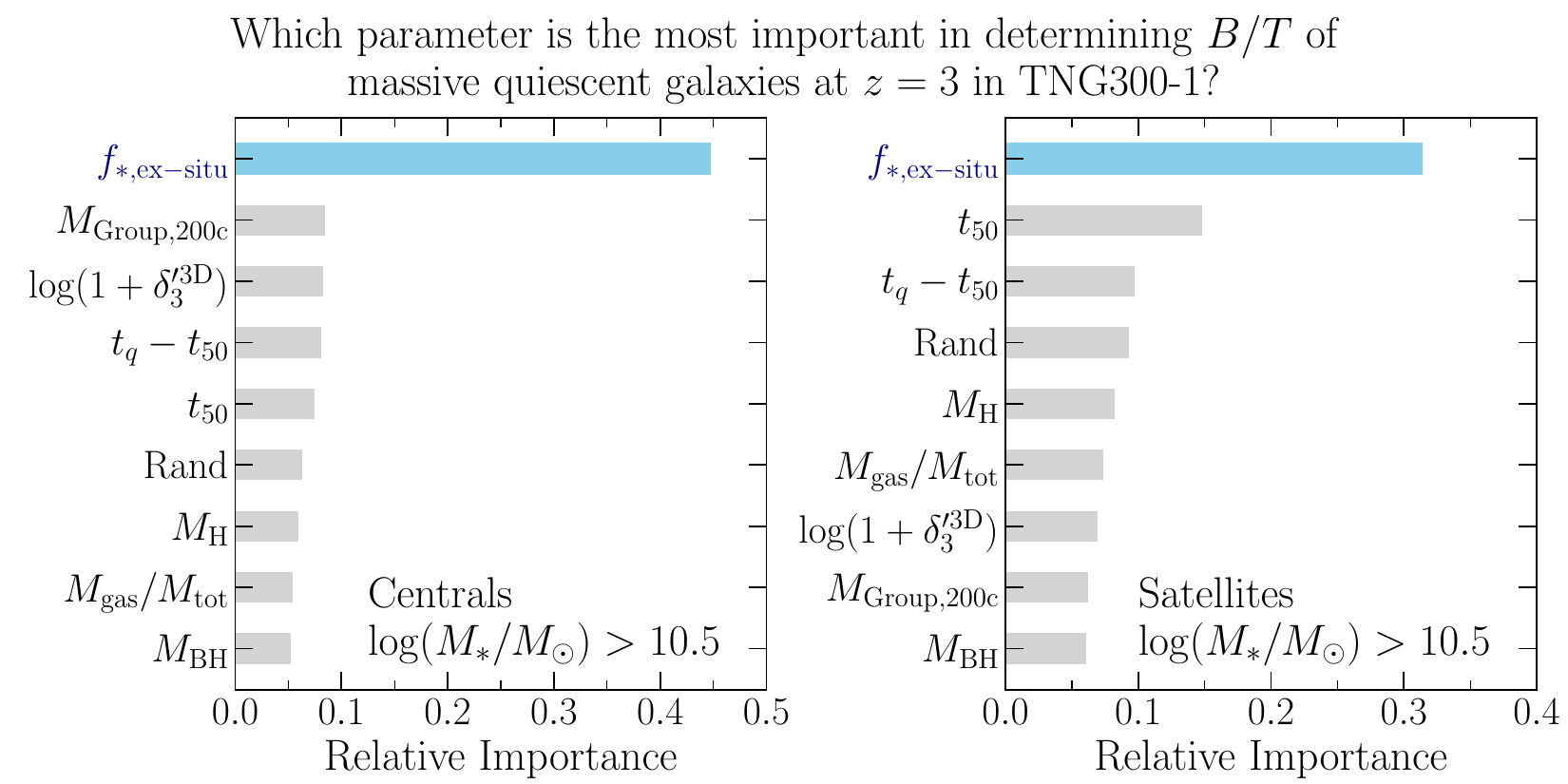}
\caption{Results from a Random Forest (RF) regression analysis to predict the bulge-to-total ratio ($B/T$) of massive quiescent galaxies ($\log(M_{\ast}/M_{\odot}) > 10.5$, $z=3$, $\mathrm{sSFR} < 1.5 \times 10^{-10} \, \mathrm{yr}^{-1}$) selected from the TNG300-1 simulation. The $x$-axis shows the relative importance of each parameter (listed on the $y$-axis) for predicting $B/T$. The fraction of stellar mass formed ex-situ ($f_{\ast,\mathrm{ex-situ}}$) is the most important parameter for both centrals (left) and satellites (right), highlighting the dominant role of mergers and accretion in driving bulge growth. \editthree{For centrals, group halo mass ($M_{\mathrm{Group,200c}}$) and environmental density based on the distance to the three nearest neighbors ($\log(1+\delta^{\prime\mathrm{3D}}_3)$) are the next most important features, suggesting that bulge buildup is also influenced by the local environment—likely by modulating merger rates and dynamical interactions within group-scale halos. In contrast, for satellites, intrinsic formation history parameters—specifically the formation time ($t_{50}$) and the quenching timescale ($t_q - t_{50}$)—rank immediately after $f_{\ast,\mathrm{ex-situ}}$. These features trace internal processes such as early star formation, compact gas collapse, and rapid quenching, which collectively shape bulge structure prior to infall. The low importance of group halo mass and local overdensity for satellites indicates that their environments at $z=3$ plays a limited role, and that their structural properties are primarily imprinted by early-time, internal evolutionary pathways.}}

\label{fig:tng_RF_btratio}
\end{figure*}

\begin{figure*}
  \centering
\includegraphics[width=\textwidth]{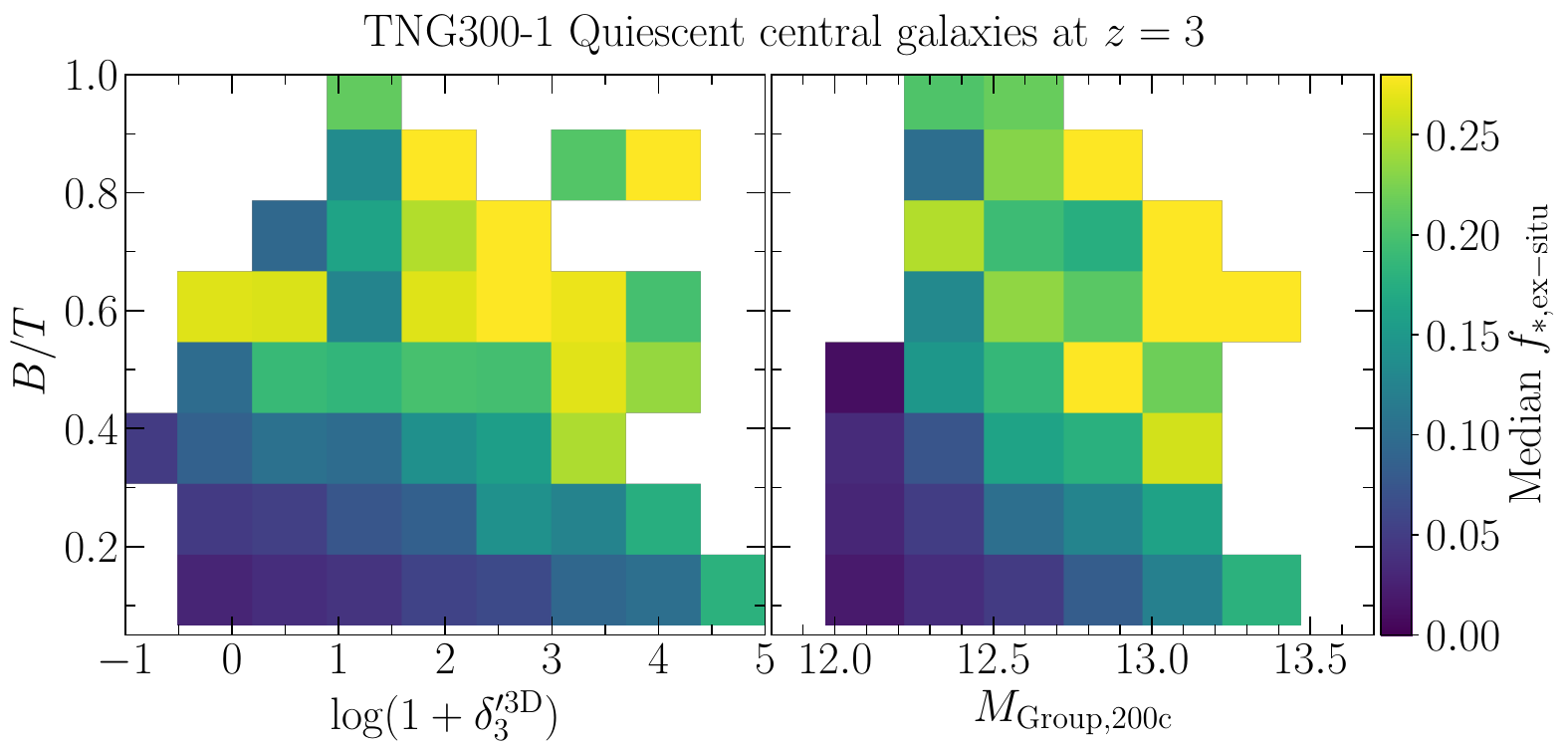}
\caption{\edittwo{Left: Median fraction of stellar mass formed ex-situ ($f_{\ast,\mathrm{ex-situ}}$) as a function of galaxy overdensity ($\log(1+\delta^{\prime\mathrm{3D}}_{3})$) and bulge-to-total ratio ($B/T$) for massive central quiescent galaxies from the IllustrisTNG300-1 simulations \citep{Springel2018} at $z=3$ with $\log(M_{\ast}/M_{\odot}) > 10.5$. The median $f_{\ast,\mathrm{ex-situ}}$ is calculated in bins of $\Delta \log(1+\delta^{\prime\mathrm{3D}}_{3})=0.7$ and $\Delta B/T=0.12$. Right: Median fraction of $f_{\ast,\mathrm{ex-situ}}$ as a function of group halo mass $M_{\mathrm{group,200c}}$ and $B/T$. The median $f_{\ast,\mathrm{ex-situ}}$ is calculated in bins of $\Delta \log M_{\mathrm{group,200c}}=0.25$ and $\Delta B/T=0.12$. Galaxies with high $B/T$ ratios and residing in denser environments or more massive group halo mass have higher median $f_{\ast,\mathrm{ex-situ}}$, suggesting that mergers are more frequent in denser environment and play significant role in driving bulge formation in massive galaxies at high redshift.}} 
\label{fig:btratio_1pdelta_medianfexsitu}
\end{figure*}

\par Figure~\ref{fig:tng_RF_btratio} shows the RF regression results for TNG300-1 massive quiescent galaxies, separated into centrals and satellites. In both subsamples, the fraction of stellar mass formed ex-situ, $f_{\ast,\mathrm{ex-situ}}$, is identified as the most important parameter for predicting $B/T$, with relative importances of 0.45 for centrals and 0.31 for satellites. Other parameters, such as black hole mass ($M_{\mathrm{BH}}$) and gas mass fraction ($f_{\mathrm{gas}}$), show significantly lower relative importances ($\lesssim0.1$). \edittwo{The strong dominance of $f_{\ast,\mathrm{ex-situ}}$ reinforces the connection between mergers and bulge growth, highlighting the critical role of accreted stellar mass in shaping the structure of massive quiescent galaxies.}

\par \edittwo{Although their relative importance is much lower than that of $f_{\ast,\mathrm{ex-situ}}$, group halo mass ($M_{\mathrm{group,200c}}$) and local overdensity ($\log(1+\delta^{\prime\mathrm{3D}}_3)$) rank second and third, respectively, for centrals, each with a relative importance of $\sim$0.08. This suggests that environmental effects—both on group scales and at smaller scales captured by overdensity—may play a modest but non-negligible role in bulge formation. The group halo virial radius ($R_{\mathrm{Group},200c}$)\footnote{$R_{\mathrm{Group},200c}$ is defined as the radius of a sphere centered on the group halo within which the mean density is 200 times the critical density of the Universe.} typically spans $0.3{-}1$ cMpc, while the 3D distance to the third nearest neighbor ranges from $0.03{-}10$ cMpc. The comparable importance of both $M_{\mathrm{group,200c}}$ and $\log(1+\delta^{\prime\mathrm{3D}}_3)$ implies that environmental influence on bulge structure may arise from a combination of group-scale dynamics and local galaxy interactions.}


\par \editthree{For satellites, the most important predictors of $B/T$ after the ex-situ stellar mass fraction are the formation time ($t_{50}$) and the quenching timescale ($t_q - t_{50}$), both of which strongly outperform environmental metrics. In contrast, local overdensity ($\log(1+\delta^{\prime\mathrm{3D}}_{3})$) and group halo mass ($M_{\mathrm{group,200c}}$) have low relative importance—comparable to or even below that of a randomly generated feature. These findings indicate that the bulge structure of massive quiescent satellites is primarily shaped by internal processes tied to early star formation and quenching, rather than by their environments at $z=3$. Specifically, early gas collapse, merger-driven assembly, and rapid self-regulated quenching likely dominate their structural evolution. This supports a scenario in which many satellites undergo morphological preprocessing before infall into more massive halos \citep[e.g.,][]{Wilman2009,Roberts2017,Brambila2023,Oxland2024,Sampaio2024}, such that their $B/T$ is largely set prior to their incorporation into larger environments. While post-infall environmental effects may suppress star formation, they appear to play a secondary role in determining bulge prominence.}

\par To assess the robustness of our RF results, we perform two additional tests. First, we examine whether including total stellar mass ($M_{\ast}$) alters the ranking of other features. For centrals, $M_{\ast}$ emerges as the second most important parameter, with a relative importance of 0.09 (see Appendix~\ref{appendix:Mstar_RF_test}). This is not surprising given its known correlation with $f_{\ast,\mathrm{ex-situ}}$, the dominant predictor of $B/T$. However, the inclusion of $M_{\ast}$ slightly reduces the relative importance of $M_{\mathrm{group,200c}}$ and $\log(1+\delta^{\prime\mathrm{3D}}_{3})$, potentially masking the contribution of environmental effects. 

\par \editthree{For satellites, the inclusion of $M_{\ast}$ has minimal impact on the feature ranking: $f_{\ast,\mathrm{ex-situ}}$, $t_{50}$, and $t_q - t_{50}$ remain the top predictors of $B/T$, while environmental parameters continue to exhibit low importance. This suggests that the structural properties of massive quiescent satellites are primarily governed by intrinsic star formation histories, rather than by stellar mass correlations or the local environment at $z=3$.}

\par \editthree{Second, motivated by the known correlation between overdensity and group halo mass \citep[e.g.,][]{Chiang2013}, we rerun the RF analysis excluding $M_{\mathrm{group,200c}}$ to test whether the predictive power of $\log(1+\delta^{\prime\mathrm{3D}}_{3})$ is independent of group halo mass. We find that overdensity remains among the top predictors of $B/T$ in central galaxies—even when $M_{\mathrm{group,200c}}$ is excluded—although it still ranks below $f_{\ast,\mathrm{ex-situ}}$, which consistently emerges as the most important driver of bulge growth. Full results are presented in Appendix~\ref{appendix:RF_tests_TNG}.}

\par In addition, we investigate the relationship between $f_{\ast,\mathrm{ex-situ}}$, $B/T$, and local environmental density ($\log(1+\delta^{\prime\mathrm{3D}}_3)$) for TNG300-1 massive quiescent central galaxies in the left panel of Figure~\ref{fig:btratio_1pdelta_medianfexsitu}. The median $f_{\ast,\mathrm{ex-situ}}$ is computed in bins of $\Delta \log(1+\delta^{\prime\mathrm{3D}}_3)=0.7$ and $\Delta B/T=0.12$. Due to the low fraction of quiescent satellite galaxies (88 galaxies, $\sim0.13$ of the sample), we focus exclusively on central galaxies. We find that massive quiescent central galaxies with high $B/T$ ratios exhibit significantly higher $f_{\ast,\mathrm{ex-situ}}$, ($\sim0.2-0.3$), particularly in dense environments ($\log(1+\delta^{\prime\mathrm{3D}}_{3}) \gtrsim 1.5$). The trend is strongest for the most massive systems with $M_{\ast}\gtrsim10^{11}M_{\odot}$, \edittwo{suggesting that mergers play a role in bulge growth in overdense regions, particularly for the most massive quiescent galaxies.}

\par \edittwo{In the right panel of Figure~\ref{fig:btratio_1pdelta_medianfexsitu}, we examine the relationship between $f_{\ast,\mathrm{ex-situ}}$, $B/T$, and group halo mass ($M_{\mathrm{group,200c}}$). The median $f_{\ast,\mathrm{ex-situ}}$ is computed in bins of $\Delta \log (M_{\mathrm{group,200c}}/M_{\ast})=0.25$ and $\Delta B/T=0.12$. Similar to the trend with $\log(1+\delta^{\prime\mathrm{3D}}_3)$, high B/T galaxies tend to have higher $f_{\ast,\mathrm{ex-situ}}$, particularly in massive halos of $\log (M_{\mathrm{group,200c}}/M_{\odot}) \gtrsim13$. Together with the local density trend, this supports a consistent picture in which both local overdensities and group halo mass contribute to enhanced merger activity and bulge growth in massive quiescent centrals.}

\par \edittwo{However, we also identify a population of high $B/T$ galaxies residing in lower-density environments ($\log(1+\delta^{\prime\mathrm{3D}}_{3}) \lesssim1$) and in lower-mass halos ($\log (M_{\mathrm{group,200c}}/M_{\odot}) \lesssim12.5$), where the median $f_{\ast,\mathrm{ex-situ}}$  is notably lower.  This trend is more pronouced for galaxies with $M_{\ast}\lesssim10^{11}M_{\odot}$. The presence of bulge-dominated galaxies in low- to intermediate-density environments -- consistent with the trend observed in Figure~\ref{fig:RF_BTratio} -- suggests that the multiple pathways contribute to bulge growth, possibly depending on stellar mass. We explore these possibilities in the following subsections.}

\par \edittwo{Additionally, we identify a distinct population of quiescent galaxies in dense environments and high-mass halos ($\log (M_{\mathrm{group,200c}}/M_{\odot}) \gtrsim12.5$) but with relatively low $B/T$ ratios. These galaxies have systematically lower $f_{\ast,\mathrm{ex-situ}}$, indicating that not all massive quiescent galaxies in dense environments have undergone significant bulge growth via ex-situ mass accretion. One possibility is that these overdensities represent dynamically young structures, such as proto-groups or proto-clusters, where galaxies have not yet had sufficient time to build up their bulges. In this scenario, morphological transformation lag behind quenching, meaning that these galaxies could still retain lower $B/T$ ratios. The presence of this population highlights the diversity in structural evolution within dense regions, suggesting that environmental effects on bulge growth vary across different quiescent galaxies.}

\subsection{Unveiling the Connection: Mergers, Environmental Density, and Bulge Growth}
\par \edittwo{For the observed sample of massive quiescent galaxies at $z=3-4$, where direct measurements of $f_{\ast,\mathrm{ex-situ}}$ \edittwo{and $M_{\mathrm{group,200c}}$} are unavailable, we quantify the local environment using the density estimator, $\log(1+\delta^{\prime}_{3})$, measured from the distances to the three nearest neighbors. Our RF regression analysis (Section~\ref{sec:randomforest_analysis}) identifies $\log(1+\delta^{\prime}_{3})$ as the strongest predictor of $B/T$. Even when all $\log(1+\delta^{\prime}_{N})$ metrics ($N=$ 3,5,7,10) were included, $\log(1+\delta^{\prime}_{3})$ remained dominant, reinforcing the key role of local environment in shaping galaxy morphology.}

\par \edittwo{One particularly compelling case in our sample is 3D-EGS-31322, a massive quiescent galaxy residing in a dense environment with $\log(1+\delta^{\prime}_{3})=0.95$. This galaxy exhibits a high bulge-to-total ratio of $B/T = 0.82 \pm 0.09$ and displays a tidal tail, indicative of a recent or ongoing interaction (Figure~\ref{fig:BDdecomp_f277w_galightfitting})\citep[see also][]{Jin2024}. \citet{Ito2025} confirmed that it is part of a merging pair within the core of the Cosmic Vine structure, with a projected separation of 4.5 kpc and a velocity offset of 680 km s$^{-1}$. Additionally, its SED fitting reveals a bursty star formation episode lasting $\sim30$ Myr \citep[][see their Figure 4]{Nanayakkara2025}, consistent with a merger-induced starburst scenario in which gas-rich interactions trigger rapid, compact star formation \citep[e.g.,][]{Armus1987,Kennicutt1987,Sanders1996,Ellison2008,Knapen2015,Moreno2021,Xu2021}.  Similar merger-induced starbursts have been observed and simulated at high redshift \citep[e.g.,][]{Fensch2017,Horstman2021,Renaud2022,Faisst2025,McClymont2025}, where such episodes are both frequent and intense. Given that the galaxy has been quenched for $\sim$300 Myr by the time of observation, this suggests the starburst was both brief and followed swiftly by quenching. While the link between starbursts and quenching remains under investigation \citep[e.g.,][]{RodriguezMontero2019,Petersson2023}, the observed properties of 3D-EGS-31322—its compact starburst, tidal features, and high $B/T$—suggest that merger-induced starburst may play a key role in bulge growth and the onset of quenching in dense environments.}

\par \edittwo{This interpretation aligns with $N$-body simulations showing that tidal fields in group- to cluster-scale environments can trigger starbursts \citep[e.g.,][]{Mihos1994,Cibinel2019,Patton2020}, particularly at high redshift ($z\gtrsim1$) when such structures are still dynamically assembling \citep{Martig2008}. Moreover, results from \citet{Husko2023}, based on N-body simulations, predict that merger-induced bursts contribute an increasing fraction of bulge mass at earlier epochs—rising from $\sim5\%$ at $z=4$ to $\sim35\%$ at $z=10$ (see their Figure 8). This trend suggests that merger-driven starbursts may play an increasingly important role in bulge formation in the early universe.} 

 \par \edittwo{We further examined environments of massive quiescent galaxies in the TNG300-1 simulation to compare the model predictions with our observations. The virial radii of simulated galaxy groups hosting such galaxies at $z=3$ span $0.3-1$ cMpc (median of $0.5$ cMpc), closely matching the 3rd nearest neigbors distances in our observed sample. These galaxies predominantly reside in halos of $10^{12}-10^{13.6}~M_{\odot}$ (median $10^{12.5}~M_{\odot}$), indicating that our density estimator is sensitive to environments ranging from within individual halos (at the low-mass end) to (proto-)group scales at the high-mass end \citep[e.g.,][]{Muldrew2012}.  Having established this, we now explore its implications for structural formation of massive galaxies.}
 
\par \edittwo{The clear correlation between $B/T$, $f_{\ast,\mathrm{ex-situ}}$, and environmental density in TNG300-1 
supports the interpretation that mergers play a significant role in bulge formation, particularly in overdense regions. Galaxise with the highest $B/T$ in dense environments have already acquired $\sim20\%-30\%$ of their stellar mass through ex-situ accretion. This trend is consistent with the numerical model of \cite{Puskas2025}, which predict similar ex-situ mass fractions for galaxies with $10.5<\log(M_{\ast}/M_{\odot})<11$ at $z=4$ (see their Figure 14). }

\par \edittwo{Given the observed connection between environment and morphology, it is important to recognize that different studies define environmental density in different ways, which affects interpretations of the role of mergers and overdensity-driven quenching. Here, we define density contrast, $\log(1+\delta^{\prime}_{3})$, as the excess number density relative to a uniform galaxy field. In contrast, \citet{Kimmig2025} used the Magneticum Pathfinder simulation \citep{Remus2025} to measure deviations relative to mass-matched star-forming galaxies, reporting it as $\delta_{\sigma,\mathrm{rel}}$ (see their Equation 4). In TNG300-1, massive quiescent galaxies reside in denser environments relative to a uniform galaxy field, with a median $\log(1+\delta^{\prime\mathrm{3D}}_{3})=1.45$. Meanwhile, \citet{Kimmig2025} found that, within 0.5 cMpc -- a scale comparable to the median 3rd-nearest neighbor distance in our TNG300-1 sample --massive quiescent galaxies in Magneticum are $-1.7\sigma$ outliers, residing in less dense environments than mass-matched star-forming galaxies. However, this does not imply that quiescent galaxies in Magneticum inhabit underdense regions relative to the full simulation volume, but rather that their local densities differ from mass-matched star-forming counterparts. These variations highlight the importance of considering different environmental density definitions when comparing studies.}

\par \edittwo{Broader observational evidence supports the link between mergers and overdense environments at $z\sim3$. \citet{Jin2024} estimated the halo mass of the densest region of the Cosmic Vine structure, where 3D-EGS-31322 resides, and derived a halo mass of $10^{12.7}~M_{\odot}$, consistent with our TNG300-1-based inference of typical (proto-)group halo masses. Recently, \cite{Shibuya2024} further supports the connection between galaxy environments and merger activity out to even higher redshifts. The authors measured the relative galaxy merger fraction ($f^{\mathrm{rel,col}}_{\mathrm{merger}}$) as a function of galaxy overdensity ($\delta$) using a sample of $\sim32,000$ galaxies at $z\sim2-5$ from the HSC Strategic Survey Program and the CFHT Large Area U-band survey. Their studies, \edittwo{based on morphologically identified mergers}, demonstrated that $f^{\mathrm{rel,col}}_{\mathrm{merger}}$ increases with increasing overdensity, consistent with earlier results \citep{Hine2016, Liu2023}.  Crucially, \citet{Shibuya2024} extended this trend to  $z\sim4-5$, providing robust observational evidence that mergers occur more frequently in dense environment even at early cosmic times. }
\par Our finding align with observational studies at lower redshifts, which have extensively examined the relationship between environmental effects and galaxy structure -- quantified by metrics such as $B/T$, central stellar mass density, S\'{e}rsic index. These studues span both low redshifts ($z \sim 0$) \citep[e.g.,][]{Weinmann2009,Woo2017,Bluck2019,Paulino-Afonso2019, Wang2020} and intermediate redshifts \cite[$z\sim0.5-3$; e.g.,][]{Papovich2012, Bassett2013, Kawinwanichakij2017, Gu2020, Ge2020,Sazanova2020,Song2023}. For example, \citet{Bluck2019} analyzed galaxies at $z < 0.2$ and found a significant correlation between the $B/T$ and local environmental densities (evaluted at the 5th nearest neighbors) for both central and satellite galaxies. Their results show that $B/T$ gradually increases with moderate overdensities, but rises steeply in highly overdense regions (see their Figure 3).  Extending these trends to higher redshifts ($3<z<4$), our study demonstrates that local environmental density remains a key driver of massive galaxy morphology even during these early epochs, enabled by JWST/NIRCam’s unprecedented spatial resolution and depth.
\par \edittwo{Together with results from \citet{Bluck2019} and \citet{Shibuya2024}, our findings underscore a consistent link between local environment and bulge-dominated morphologies across cosmic time. However, the nature of mergers likely evolves: while low-redshift mergers are typically gas-poor and lead to gradual morphological transformation, high-redshift mergers are more gas-rich and often trigger starbursts. Our results at $z=3$–4 suggest that such merger-induced starbursts are a key pathway for bulge growth in dense environments at early epochs.}

\par \edittwo{While our results support merger-induced starbursts as a key driver of bulge growth in dense environments, we also find evidence for additional pathways in lower-density environments. We observed a population of high-$B/T$ quiescent central galaxies residing in such environments and exhibiting low ex-situ mass fractions, particularly those with $M_{\ast}<10^{11}M_{\odot}$ (Figure~\ref{fig:btratio_1pdelta_medianfexsitu}) -- suggesting that mergers are not the dominant mechanism in these cases. We also identify a population of quiescent galaxies in dense environments with low $B/T$ and ex-situ fractions, possibly located in dynamically young proto-groups or proto-clusters where quenching has occurred, but morphological transformation via mergers may not yet have taken place. These trends point to multiple, environment- and possibly mass-dependent pathways for bulge growth, which we explore in the following section.}

\subsection{Beyond Mergers: Alternative Pathways to Bulge Growth}
\par \edittwo{In our sample, 5 out of 14 massive quiescent galaxies reside in low- to intermediate-density environments ($\log(1+\delta^{\prime}_{3}) \lesssim0.6$) yet already exhibit high bulge-to-total ratios ($B/T\gtrsim0.8$). A similar trend appears in the TNG300-1 simulation, where quiescent central galaxies in comparable environments ($\log(1+\delta^{\prime\mathrm{3D}}_{3}) \lesssim1$) show high $B/T$ ratios and low ex-situ mass fractions (Figure~\ref{fig:btratio_1pdelta_medianfexsitu}). This pattern is most pronounced for galaxies with $M_{\ast}<10^{11}~M_{\odot}$, suggesting that mergers are not the primary driver of bulge formation these systems.}

\par \edittwo{One possible mechanism for bulge growth in these systems is violent gravitational disk instabilities (VDIs). At high redshift, galaxy disks are highly turbulent and gravitationally unstable, characterized by high velocity dispersion and massive perturbations \citep[e.g.,][]{Dekel2009,Cacciato2012,Romeo2016}. In this scenario, gas-rich, thick, turbulent disks fragment into massive clumps due to Toomre instability \citep[][]{Toomre1964}. These clumps migrate toward the galaxy center due to dynamical friction, where they coalesce to form bulges, leading to morphological transformation over time \citep[e.g.,][]{vandenBergh1996,Elmegreen2005,Bournaud2008,Genzel2008}. } 

\par \edittwo{The dominant role of disk instabilities-induced starbursts in spheroid growth has been demonstrated in N-body simulations by \citet{Husko2023}, who find that such starbursts account for the majority of spheroid mass at all redshifts -- peaking at 70\% at $z = 4$ and remaining significant (60\%) at $z = 10$. At $z = 4$, these starbursts dominate spheroid growth in all but the most massive galaxies ($M_{\ast}>10^{11}~M_{\odot}$). While their study focuses on the total spheroid component, these results are consistent with our observational and simulation findings: quiescent galaxies with $M_{\ast} < 10^{11}~M_{\odot}$ in low- to intermediate-density environments already exhibit prominent bulge components. This supports the interpretation that violent disk instabilities—rather than mergers—are a key pathway for bulge growth in lower-mass galaxies and in less dense environments at high redshift.}
\par \edittwo{The structural and star formation properties of our massive quiescent galaxies in low- to intermediate environments are consistent with this VDI-driven scenario. These galaxies exhibit compact sizes, high $B/T$ ratios, high specific star formation rate (sSFR $\gtrsim2~\mathrm{Gyr}^{-1}$) during their peak formation epoch \citep{Nanayakkara2025}. This observation matched the expectation that disk instabilities trigger rapid star formation \citep{Lehnert1996,Romeo2016,Tadaki2018,Fujimoto2024,Faisst2025} and drive structural evolution, even in the absence of external interactions. }
\par \edittwo{Recently, \citet{Faisst2025} found an enhanced fraction of star-forming galaxies with clumpy morphologies (``clumpy disks") in the starburst population of isolated disk galaxies at $z<4$. Additionally, the fraction of clumpy disks increases significantly with star formation efficiency, out to at least  $z=4$. Taken together, these findings provide direct observational evidence that disk instabilities fuel rapid star formation and driving structural evolution of massive galaxies at high redshifts, particularly in environments where mergers are less frequent.}

\par \edittwo{A key caveat to this scenario is that we do not observe clear star-forming clumps in our quiescent galaxy sample. If VDIs played a role in their evolution, these clumps may have already migrated to the center, forming a bulge, or dissipated as star formation ceased. Observations of star-forming galaxies at similar redshifts reveal clumpy morphologies indicative of ongoing disk instabilities \citep[e.g.,][]{Guo2015,Shibuya2016,Tanaka2024,Faisst2025}, suggesting that clumps are a transient phase in bulge formation.}
\par \edittwo{Recent JWST observations by \citet{Claeyssens2025} of 18 lensed galaxies at $z=1-8.5$ in the SMACS0723 field show that clumps in thick, gas-rich disks have short survival timescales, rarely exceeding 1 Gyr. Most clumps dissolve within hundreds of Myr due to strong shear forces, and older ($>100$ Myr) clumps are typically gravitationally bound. If similar processes occurred in our galaxies, any clumpy features may have already dispersed before quenching, further explaining their absence.  Likewise, \citet{Faisst2025} suggest that compact early-universe galaxies may still harbor small-scale structures at $\sim100$ pc resolution, detectable only through gravitational lensing or future diffraction-limited observations.}

\subsection{Implications for the Formation of Massive Quiescent Galaxies\label{sec:implications}} 

\par \editone{A key finding from this work is that old massive quiescent galaxies with formation redshifts of $z_{50} \gtrsim 5$ ($t_{50} \lesssim 1$ Gyr after the Big Bang) tend to exhibit extended morphologies with median sizes of $R_{e} \sim 1$ kpc—consistent with the predicted $R_{e}$–$M_{\ast}$ relation (Figures~\ref{fig:size_mass_relation_f277w_f444w} and~\ref{fig:logsizeoffset_tform_tq}a)—and relatively low bulge-to-total light ratios ($B/T \lesssim 0.7$; Figure~\ref{fig:btratio_totalmass}). These galaxies also span a broad range of quenching timescales ($\Delta t_{q} \sim 0.4\text{--}0.8$ Gyr).}

\par \editfour{In contrast, nearly half of the sample consists of younger quiescent galaxies with formation redshifts $z_{50} \lesssim 5$ and short quenching timescales ($\Delta t_{q} \lesssim 0.4$ Gyr). While this late-forming subsample shows a wider range of structural properties, we find that \textit{bulge-dominated} systems in this group tend to be more compact—by up to $\sim0.2$–$0.3$ dex below the median size–mass relation—whereas \textit{disk-dominated} systems are significantly more extended. This trend is more apparent in the F277W band, which traces rest-frame optical light and is more sensitive to recent star formation. In contrast, the correlation between formation time and size is weaker and not statistically significant in the F444W band, which reflects older, mass-weighted structure. These findings suggest a nuanced interplay between formation time, quenching history, and morphology in shaping galaxy structure during early quiescence. Together, our results point to multiple evolutionary pathways—ranging from early-forming systems with extended morphologies to late-forming, compact bulge-dominated galaxies and more extended disk-dominated ones—shaped by distinct formation histories, quenching mechanisms, and morphological transformations. In the following sections, we explore these implications in the context of different formation scenarios.}

\subsubsection{On the Origin of Massive Quiescent Galaxies with Compact Morphology }
\par Our sample of late-forming ($z_{50}<5$) massive quiescent galaxies with compact structures, e.g., ZF-UDS-8197, ZF-UDS-3651, 3D-EGS-31322, and ZF-COS-18842, aligns with the ``gas-rich compaction" scenario predicted by cosmological simulations for the buildup of central regions in high-redshift galaxies. This scenario involves highly dissipative gas accretion toward the galaxy center, which enhances central star formation activity. This is consistent with the high specific star formation rates (sSFRs $\gtrsim10$ Gyr$^{-1}$) during their peak activity of these late-forming compact galaxies \citep{Nanayakkara2025}. The subsequent depletion of central gas leads to the quenching of star formation in the core \citep[e.g.,][]{Dekel2009Apj,Dekel2014,Zolotov2015,Tacchella2016a,Lapiner2023,Cenci2024,McClymont2025arXiv}, leaving behind a younger central stellar population from the recent starbursts fueled by dissipative gas inflow. 
\par However, compact galaxies are not exclusively late-forming. The cases of ZF-UDS-7329, GS-9209, and RUBIES-EGS-QG1 illustrate that some massive compact galaxies at $z\sim3-5$, particularly those that formed very early ($z_{50}>5$), likely originated from gravitational collapse at high redshifts \citep[e.g.,][]{Wellons2015,Lilly2016}. These early-formed galaxies began with high central stellar mass densities and quenched rapidly on short timescales ($\lesssim200$ Myr), thereby maintaining their compact morphologies.

\par \edittwo{The rapid quenching and compact morphologies observed in both late- and early-forming massive quiescent galaxies suggest a common underlying mechanism. This is supported by results from the Magneticum Pathfinder cosmological simulations \citep{Remus2025}, where \citet{Kimmig2025} find that massive quiescent galaxies at $z\sim3.4$ experience rapid bursts of star formation, followed by AGN feedback triggered by the isotropic collapse of surrounding gas on short timescales ($\sim200$ Myr). This unified quenching pathway --characterized by rapid gas accretion, intense central starbursts, and AGN feedback -- applies to both late- and early-forming compact galaxies. In particular, the simulations reveal that some galaxies form as early structures by $z=9.43$, undergoing isotropic gas inflow and early quenching. This scenario naturally explains the formation of early-quenched systems such as ZF-7329, GS-9209 and RUBIES-EGS-QG1. Furthermore, \citet{Kimmig2025} report that the stellar mass profiles of quiescent galaxies exhibit enhanced central density, suggesting that isotropic collapse funnels gas more efficiently into the center, rather than building extended disks. This supports a direct link between the predicted quenching mechanism and the compact morphologies of these galaxies.}

\par \edittwo{Similar conclusions are drawn from the THESAN-ZOOM simulations by \citet{McClymont2025}, which show that star formation at high redshift is more bursty than at later times. They identify two modes of starbursts—externally driven, fueled by rapid gas inflows, and internally driven, triggered by interstellar medium recycling. The externally driven mode, dominant at $z>3$, mirrors the isotropic inflow scenario in Magneticum, reinforcing the idea that intense central starbursts triggered by large-scale accretion are a key driver of early quenching. These results together point to a common high-redshift quenching pathway across different simulation frameworks, connecting bursty star formation, rapid gas inflow, and compact morphologies in massive galaxies.}

\par \edittwo{There are multiple pathways to quench a galaxy, and not all quiescent systems undergo a compaction phase that builds a dense core. At high redshifts, quenching of massive galaxies can proceed through cosmological starvation—a reduction in gas accretion once a galaxy becomes a satellite within a larger halo \citep{Wetzel2013, Feldmann2015}—followed by an overconsumption phase, during which the remaining gas is rapidly depleted by star formation and outflows \citep{McGee2014, Balogh2016}. The overconsumption model predicts shorter quenching timescales at higher redshifts and in more massive galaxies, as these systems exhaust their gas more quickly after cosmological inflow is suppressed.}

\par \edittwo{While molecular gas depletion timescales (i.e., the gas mass divided by the SFR; $\tau_{\mathrm{depl}}$) show only a mild decrease with increasing stellar mass in observations \citep[e.g.,][]{Saintonge2016, Tacconi2018, Tacconi2020}, simulations often predict a steeper decline in $\tau_{\mathrm{depl}}$ for more massive galaxies \citep[e.g.,][]{Dave2011, Lagos2015, Tacchella2016a, Kudritzki2021}. In both cases, however, molecular gas fractions drop significantly with increasing stellar mass. This implies that, in massive galaxies, quenching is primarily driven by reduced gas supply rather than enhanced star formation efficiency (i.e., shorter $\tau_{\mathrm{depl}}$), consistent with the overconsumption scenario. Observational studies of dense environments provide further support for this picture; for example, \citet{Gururajan2025} find that galaxies in the densest regions of the Hyperion proto-supercluster at $z\sim2.5$ exhibit declining molecular gas fractions and shortened depletion timescales, accompanied by elevated SFRs. These results indicate that accelerated gas consumption and limited replenishment drive the quenching process in dense environments.}

\par \edittwo{In addition, the deep gravitational potentials of massive galaxies, along with AGN feedback, can heat infalling gas and suppress fresh accretion \citep[e.g.,][]{Dekel2006, RodriguezMontero2019, Donnari2021}, further limiting the gas supply and accelerating quenching. These physical processes reinforce the overconsumption framework. Together with the observed SFR–stellar mass correlation \citep[e.g.,][]{Noeske2007, Tomczak2016}, they imply quenching timescales of a few hundred Myr for massive galaxies. Environmental quenching also becomes more efficient with increasing stellar mass at $z\sim1$–4.5 \citep[e.g.,][]{Kawinwanichakij2017, Chartab2020, Shahidi2020, McConachie2022}, consistent with an environmentally driven overconsumption scenario. The evidence from \citet{Gururajan2025} further suggests that processes such as gas stripping, cessation of cold flows, and enhanced star formation efficiency (SFE) in dense regions may act in concert to accelerate quenching in massive galaxies at high redshift.}

\par Moreover, the overconsumption model, in its simplest form, implies that quenching may not necessarily involve significant morphological transformation: galaxies simply exhaust their gas supply while retaining their original structures, aside from disk fading.

\par Examining the environments of massive quiescent galaxies at $3<z<4$, we find that 11 out of 14 galaxies reside in overdense regions ($\log(1+\delta^{\prime})_{3}\gtrsim0$; see Figure~\ref{fig:RF_BTratio}). Additionally, we identify two galaxies in our sample (ZF-UDS-7542, and ZF-COS-19589) with large size ($R_{e}\sim1.5$ kpc, $\gtrsim0.2$ dex above the best-fit $R_{e}-M_{\ast}$ relation). Recently, \cite{Espinoza2024} studied member galaxies of overdensities (with dark matter halo masses of about $10^{13}M_{\odot}$) identified at $3<z<4$. The authors found two overdensities at $z\simeq3.55$ and $z\simeq3.43$ containing massive ($\log(M_{\ast}/M_{\odot})\sim 11.3$) and quiescent members. These galaxies also have sizes of $R_{e}\sim1-1.5$ kpc and reside in higher local evironmental densities than quiescent galaxies in the field. 
\par \edittwo{A similar trend is found in the TNG300-1 simulation, where we identify a population of quiescent galaxies residing in dense environments and massive halos ($\log (M_{\mathrm{group,200c}}/M_{\odot}) \gtrsim12.5$) yet exhibiting relatively low $B/T$ ratios and systematically lower ex-situ mass fraction ($f_{\ast,\mathrm{ex-situ}}$). These galaxies have not undergone significant bulge growth via ex-situ accretion, suggesting that structural transformation lags behind quenching. This is consistent with the overconsumption model, in which galaxies rapidly deplete their gas through star formation and outflows, quenching without substantial morphological change. The existence of such galaxies in TNG300-1 indicates that quenching can precede bulge growth, even in dense environments. Observationally, we also find quiescent galaxies with low $B/T$ and large sizes in average-density environments ($\log(1+\delta^{\prime}_{3}) \sim 0$), suggesting that the overconsumption need not be restricted to the densest regions at this high redshift.}

\subsection{The Contributions from Active Galactic Nuclei}

\par Among the galaxies in our sample, five systems—ZF-UDS-6496, ZF-UDS-3651, ZF-COS-18842, ZF-UDS-8197, and 3D-EGS-18996—exhibit significant detections of H$\alpha$ and/or $\mathrm{[OIII]}\lambda5007$ emission lines, with $S/N>5$ and observed equivalent widths (EW) greater than $10\AA$. \edittwo{Given such prominent emission features, these galaxies may host active galactic nuclei (AGN) \citep{Nanayakkara2025}. In particular, \citet{Nanayakkara2024Nature} reported that ZF-UDS-8197 has a $\mathrm{[OIII]}\lambda5007$/H$\beta$ flux ratio of 1.4, consistent with AGN-powered line emission according to the classification scheme of \citet{Juneau2014}.}

\par These five galaxies also exhibit notably short quenching timescales, with a median value of $\sim$100 Myr, and are systematically $\sim$0.1–0.2 dex smaller than the sizes predicted from the stellar mass–size relation. Taken together, these observations suggest that AGN activity may play a dual role in both quenching star formation and driving structural compaction. A plausible scenario involves merger-induced starbursts funneling gas into the central regions, triggering AGN activity and subsequently quenching star formation on short dynamical timescales of $t_{\mathrm{dyn}} \sim 100$ Myr \citep[][]{Hopkins2008,Alexander2012,Toft2014}.

\par We also consider whether AGN continuum emission might bias our structural measurements by contributing unresolved central light, potentially leading to artificially small sizes. To test this, we compare the effective radii derived from single Sérsic fits with those from point source $+$ Sérsic models for the AGN candidates in our sample with high-resolution imaging. We find that the sizes increase by a factor of $\sim$1.3–1.6 when a central point source is included. \editfour{This confirms that central AGN light can indeed bias the measured sizes toward smaller values if not modeled separately, supporting the interpretation that some compact morphologies may be partially influenced by unresolved AGN emission. However, even after accounting for a point source component, these galaxies remain up to $\sim$0.1–0.2 dex smaller than the expected size–mass relation, indicating that unresolved AGN emission may contribute but does not fully account for their compactness.}

\par Our findings are consistent with previous studies linking rapid quenching to AGN activity. \cite{Carnall2018} performed a detailed analysis of star formation histories (SFHs) of quiescent galaxies from UltraVISTA with $\log(M_{\ast}/M_{\odot})>10$ and redshifts $0.25<z<3.75$ and found that at $z\gtrsim1$ galaxies with rapid quenching time scales of $<1$ Gyr are more common, and the quenching timescale is more rapid at higher stellar masses. The authors tentatively identified that these galaxies could be affected by quasar-mode AGN feedback,  potentially triggered by mergers. In \cite{Carnall2023Nature}, they further showed that the observed spectrum of GS-9209 exhibits the high line ratio of $\log_{10}([\mathrm{NII}]/\mathrm{H}\alpha)=-0.01\pm0.04$, significantly higher than would be expected as a result of ongoing star formation, and is consistent with excitation due to and AGN or shock resulting from galactic outflows.

\subsection{Caveats}
\label{sec:caveats}
\par \edittwo{In this section, we highlight several caveats that may impact our analysis and interpretation. First, we consider the systematic uncertainties introduced by using photometric redshifts in our environmental density measurements. Our photometric redshift uncertainties, typically ranging from $\sigma_{z}/(1+z) = 0.01-0.03$, influence the selection of neighboring galaxies when computing the excess surface number density, $\log(1+\delta_{N}$). Given our selection criteria of $\left | \Delta z \right | = 2\times \sigma_z (1+z)\sim0.1-0.3$, this corresponds to a comoving distance of approximatedly 80 to 240 cMpc at  $z=3.5$. As a result, our metric may lack sensivity to identify structures on scales smaller than 80 cMpc in overdense regions.} 
\par \edittwo{ Additionally, photometric redshift uncertainties could introduce the contamination effects in two competing ways. First, increasing  $\left | \Delta z \right | $ smooths the density field, potentially diluting the significance of true protocluster overdensities. This occurs because low-density outskirts of proto-cluster and physically unassociated interlopers can be projected into the same redshift slice, blending with the actual overdensity \citep[][see their Figure 13]{Chiang2013}.  Conversely, overlapping structures along the line of sight can artificially enhance the observed overdensity by stacking multiple physically separate structures. As a result, the dense environments identified using photometric redshifts are subject to systematic effects that may lead to an underestimation of local surface densities. Consequently, our environmental density values should be interpreted as relative measurements rather than absolute densities.}

\par \edittwo{While our study leverages TNG300-1 to explore the evolution of quenched galaxies, we acknowledge its limitations in reproducing the observed quenching histories and number densities of quiescent galaxies at high redshifts ($z\gtrsim3$). TNG300-1 underpredicts number densities of massive quiescent galaxies ($M_{\ast}\gtrsim10^{10}~M_{\odot}$) at these epochs before experiencing a sharp rise at $z \approx 3$ \citep[e.g.,][]{Valentino2023,DeLucia2024,Remus2025,Weller2025}. This discrepancy arises because quenching in TNG is primarily governed by a threshold black hole mass, which determines when the kinetic AGN feedback mode activates \citep[e.g.,][]{Weinberger2017,Terrazas2020,Zinger2020,Hartley2023,Piotrowska2022,Kurinchi-Vendhan2024}.}
\par  \edittwo{In IllustrisTNG, quenching occurs once the kinetic AGN mode becomes efficient, which is triggered when the central black hole mass surpasses a critical threshold \citep{Hartley2023,Kurinchi-Vendhan2024}.  Since black hole growth correlates with stellar mass, galaxies above this threshold inevitably quench, resulting in a sudden transition rather than a gradual buildup of quiescent galaxies \citep[e.g.,][]{Kimmig2025,Remus2025,Weller2025}. Observational studies reveal discrepancies between TNG’s predictions and observed quiescent galaxy number densities at $z\sim4$, suggesting that TNG delays quenching compared to observations \citep[e.g.,][]{deGraaff2024,Glazebrook2024Nature,Nanayakkara2025}. This delay likely results from its strong dependence on a mass-based black hole feedback threshold. In TNG, quenching occurs only after the kinetic AGN mode dominates, which requires kinetic feedback energy to exceed 1\% of the thermal mode energy \citep[][]{Weinberger2018}. These discrepancies suggest that AGN feedback alone is insufficient to fully explain quenching, and additional mechanisms—such as environmental effects—may play a significant role \citep[e.g.,][]{Kurinchi-Vendhan2024,Kimmig2025,Weller2025}.}
\par \edittwo{ More flexible feedback models, such as those implemented by the Magneticum
Pathfinder hydrodynamical cosmological simulations \citep{Remus2025}, incorporate additional quenching mechanisms beyond AGN feedback. These models account for processes like star formation feedback and gas expulsion in lower-density environments, enabling earlier and more gradual quenching \citep[][]{Kimmig2025,Remus2025}. Unlike TNG, these models allow for continuous gas accretion even as AGN-driven outflows develop, rather than requiring a strict black hole mass threshold for quenching \citep[][]{Steinborn2015}. As a result, they produce stronger and more sustained feedback, leading to an earlier decline in star formation and a higher number of quenched galaxies at $z>3$ compared to simulations that primarily rely on AGN-driven quenching.}

\edittwo{Finally, a caveat in our analysis is the use of kinematic bulge-to-total ($B/T$) ratios in TNG300-1, given the known discrepancies with photometric decomposition methods in simulations \citep[e.g.,][]{Scannapieco2010,Jang2023}. Photometric disk-to-total ($D/T$) ratios, and consequently $B/T$ ratios, tend to be systematically higher than their kinematic counterparts. \citet{Scannapieco2010} attributed this to photometric decompositions assuming an exponential disk extending to the galaxy center, while kinematic decompositions often exclude stars in the central $\sim2$ kpc. \citet{Jang2023} further demonstrated that photometric $D/T$ ratios exhibit substantial scatter compared to kinematic estimates in the NEWHORIZON simulation, with a correlation coefficient of only 0.35. This discrepancy is most pronounced in low-mass galaxies ($\log(M_{\ast}/M_{\odot})<9.5$), where photometric methods frequently classify galaxies as pure disks despite kinematic evidence of significant bulge components.}
\par \edittwo{While these differences highlight the need for caution when directly comparing simulated and observed $B/T$ ratios, it is important to note that the key environmental trends we identify in TNG300-1 remain physically meaningful and are consistent with observational results. Both the simulations and observations show that galaxies in denser environments tend to have higher $B/T$ values, with environmental density ($\log(1+\delta^{\prime}_{3})$) in observations) and ex-situ stellar mass fraction ($f_{\ast,\mathrm{ex-situ}}$ in simulations) emerging as key predictors of bulge prominence. The strong correlation between $B/T$, merger history, and local density in TNG300-1 supports the interpretation that environmental processes -- including tidal interactions and mergers—play a significant role in shaping galaxy morphology. Thus, while kinematic and photometric decompositions may differ in absolute values of $B/T$, the qualitative agreement in the environmental dependence of bulge growth reinforces the robustness of our conclusions.} 

\section{Conclusion} \label{sec:conclusion}
We have analyzed the structural properties of 17 spectroscopically confirmed, massive quiescent galaxies at $3.0<z<4.3$ ($\bar{z}\sim3.5$) from \cite{Nanayakkara2025}, observed using JWST/NIRSpec. Using S\'{e}rsic profile fitting F277W  (rest-frame $\sim0.6\mu$m) and F444W  (rest-frame $\sim1\mu$m), we measured sizes ($R_{e}$), S\'{e}rsic indices ($n$), axis ratios ($q$), bulge-to-total ratio ($B/T$).  We derive the size-mass ($R_{e}-M_{\ast}$) relation for these galaxies, assumming a power-law form over the stellar mass range $\log(M_{\ast}/M_{\odot})>10.3$. The main conclusions from the analysis of the $R_{e}-M_{\ast}$ relation and morphological properties are summarized as follows:
\begin{itemize}
\item  The power-law slope of the size-mass relation corresponds to $\alpha=0.7$, consistent with previous works at similar redshifts.  The size at $M_{\ast} =5\times10^{10}M_{\odot}$ is $R_{e}\sim0.7$ kpc (0.6 kpc) based on F277W (F444W) image, reflecting a size evolution (at fixed stellar mass) of nearly a factor of 7 from $z \sim 4$ to the present day \citep[e.g.,][]{vanderWel2014, Straatman2015}.

\item Nearly $90\%$ of the sample exhibits bulge-dominated morphologies ($n > 2.5$), with median S\'{e}rsic indices measured as $n = 4.54^{+1.31}_{-1.65}$ in F277W and $n = 3.59^{+1.18}_{-0.54}$ in F444W. A smaller fraction ($\sim10\%$) displays disk-dominated morphologies ($n < 2.5$), potentially undergoing morphological transformations.
\item The median axis ratios of these galaxies, measured in F277W and F444W images, are consistent with $q \sim 0.6$, comparable to massive star-forming galaxies at similar redshifts \citep{Hill2019}. The gradual increase in axis ratios from $q \sim 0.6$ at $z \sim 3.5$ to $q = 0.71$ at $z \sim 0.1$ suggests that these galaxies become progressively rounder over time, consistent with morphological evolution driven by minor mergers \citep{vanDokkum2010}.
\end{itemize}

We explored the relationship between galaxy size and star formation history, quantified by formation times ($t_{50}$) and quenching timescales ($\Delta t_q$). These parameters, derived using the \textsc{FAST++} SED fitting code \citep{Nanayakkara2025}, yielded the following insights:
\begin{itemize}
\item About $\sim30\%$ of our sample formed early as $t_{50}\sim0.6-1.1$ Gyr after the Big Bang ($z_{50}\sim8.7-5.2$) and quenched over relatively long timescales ($\Delta t_{q}\sim0.4-0.8$ Gyr). These galaxies typically have sizes ($R_{e} \sim 1$ kpc) in agreement with the predicted $R_{e}$–$M_{\ast}$ relation at this redshift.
\item Nearly half of our sample consists of late-forming galaxies with formation time of $t_{50}\sim1.2-1.6$ Gyr ($z_{50}\sim4.8-3.8$) and quenched rapidly ($\Delta t_{q}\lesssim0.4$ Gyr). \editfour{While this subsample shows a wider scatter in structural properties, we find that late-forming, bulge-dominated galaxies tend to be more compact—up to $\sim0.2-0.3$ dex below the size–mass relation—whereas late-forming, disk-dominated systems are larger than expected. This highlights a possible interplay between morphology and formation history in shaping galaxy structure at early times.}
\end{itemize}
\par We use a Random Forest regressor (RF) to identify the primary drivers of galaxy morphology, quantified by $B/T$ in F444W. Our analysis reveals that the local environment, quantified by the Bayesian-motivated overdensity metric $\log(1+\delta^{\prime}_{3})$ (based on the distances to the three nearest neighbors), is the most important predictor of $B/T$, outperforming other environmental metrics and intrinsic galaxy properties explored in this work. This is evidenced by the observed increase in $B/T$, from $\sim0.5$ in low-density environments ($\log(1+\delta^{\prime}_{3})<0.16$) to $\sim0.8$ in high-density environments ($\log(1+\delta^{\prime}_{3})>0.70$).  To further investigate the observed correlation between $B/T$ and environmental densities, we extend our RF regression analysis to the IllustrisTNG simulation. Our main conclusions are as follows: 
\begin{itemize}
\item The fraction of stellar mass formed ex-situ, $f_{\ast,\mathrm{ex-situ}}$, is identified as the most important predictor of $B/T$ for massive quiescent galaxies at $z=3$ in the simulation. This highlights the critical role of mergers in shaping galaxy structure, surpassing the influence of other environmental metrics and intrinsic galaxy properties.

\item \edittwo{The simulations predicts that massive quiescent galaxies with high $B/T$ in dense local environments shave acquired $\sim20\%-30\%$ of their stellar mass through mergers. Combined with observational evidence for increasing merger fractions with increasing local environmental densitiy \citep[e.g.,][] {Shibuya2024}, this reinforces the observed correlation between $B/T$ and $\log(1+\delta^{\prime}_{3})$ and underscores the role of local environmental processes -- partcularly mergers -- in shaping morphology of massive galaxies even at early cosmic epochs.}

\item \edittwo{Observationally, galaxies like 3D-EGS-31322 exhibit high $B/T$, reside in dense environments, and display a tidal tail, indicative of a recent or ongoing interaction. Its bursty star formation history, lasting $\sim30$ Myr \citep{Nanayakkara2025}, supports a merger-induced starburst scenario, where gas-rich mergers at high redshift trigger rapid starbursts that build bulges and initiate quenching. This scenario aligns with predictions from N-body simulations, which show that merger-induced starbursts contribute significantly to bulge growth at earlier cosmic times, increasing from $\sim5\%$ at $z=4$ to $\sim35\%$ at $z=10$ \citep{Husko2023}.}
\end{itemize}

\par \edittwo{While our observations in dense environments support a merger-induced starburst scenario, we also identify a complementary pathway—violent disk gravitational instabilities (VDIs)—in which gas-rich, turbulent disks fragment into clumps that migrate inward and build bulges. This mechanism may dominate in high-$B/T$ quiescent galaxies in low- to intermediate-density environments, where internal processes likely shape their structure. N-body simulations predict that VDI-driven starbursts contribute up to 70\% of spheroid mass at $z=4$, remaining significant at $z=10$ \citep{Husko2023}. Observed increases in clumpy disk fractions in isolated starbursts out to $z\sim4$ further support this scenario in merger-poor environments \citep{Faisst2025}. The lack of visible clumps in our sample may reflect their short lifetimes or post-quenching evolution \citep{Claeyssens2025}.}

\par \edittwo{Bringing these findings together, our sample of massive quiescent galaxies at $3<z<4$ reveals diverse yet physically connected evolutionary pathways.  Late-forming systems ($z_{50}\lesssim5$) are consistent with gas-rich compaction \citep[e.g.,][]{Tacchella2016a}, where inflows trigger central starbursts, rapid quenching ($\lesssim0.4$ Gyr), and compact morphologies. Others galaxies, with extended structures, align with overconsumption models involving gas starvation and rapid depletion, often as satellites \citep[e.g.,][]{McGee2014}. These processes likely act alongside environmental mechanisms, such as mergers, to explain the observed trend of increasing compactness with local density. A unifying mechanism —rapid gas accretion, central starbursts, and AGN feedback triggered by the isotropic gas collapse—may underlie both pathways, particularly for massive galaxies at high redshift. This is supported by the predictions from recent cosmological simulations \citep[e.g.,][]{Remus2025,Kimmig2025,McClymont2025}, which reproduce compact, fast-quenched massive system with both early and late formation histories.}

\par In closing, our study underscores the capability of JWST/NIRCam to unveil the rest-frame near-IR morphological properties of high-redshift galaxies at $z > 3$. However, spatially resolved stellar population studies of larger samples are crucial for robustly constraining the physical mechanisms driving the observed relationships between galaxy morphology, star formation history, and environment.

\begin{acknowledgments}
The authors would like to thank the anonymous referee for a
comprehensive and constructive report that allowed us to
improve the overall quality of the manuscript. This work is based on observations made with the
NASA/ESA/CSA James Webb Space Telescope. The data
were obtained from the Mikulski Archive for Space Telescopes
at the Space Telescope Science Institute, which is operated
by the Association of Universities for Research in Astronomy,
Inc., under NASA contract NAS 5-03127 for JWST.
These observations are associated with program JWST-GO-
2565. All of the data presented in this article were obtained from the Mikulski Archive for Space Telescopes (MAST) at the Space Telescope Science Institute. The specific observations analyzed can be accessed via \dataset[doi:10.17909/rkd4-gr92]{https://doi.org/10.17909/rkd4-gr92}. L.K., K.G., T.N., and C.J. acknowledge support
from Australian Research Council Laureate Fellowship
FL180100060. This project made use of \textsc{Astropy} \citep{astropy:2013, astropy:2018, astropy:2022}, \textsc{Matplotlib}  
\citep{Hunter2007}. The data products presented herein were retrieved from the Dawn JWST
Archive (DJA). DJA is an initiative of the Cosmic Dawn Center
(DAWN), which is funded by the Danish National Research
Foundation under grant DNRF140.
\end{acknowledgments}

%

\vspace{5mm}
\facilities{JWST (NIRSpec and NIRCam)}


\software{Astropy \citep{astropy:2022}}



\appendix
\section{Summary of Morphological Parameters Measurements of Massive Quiescent Galaxies at $z=3-4$\label{app:morphotables}}
We provide the best-fit morphological parameters for a single S\'{e}rsic fitting in Table~\ref{tab:morphomeas} in Appendix~\ref{app:morphotables}. 

\begin{rotatetable}
\begin{deluxetable}{ccccccccccc}
\tabletypesize{\scriptsize}
\tablecaption{Summary of Morphological Parameters Measurements for a single S\'{e}rsic fitting \label{tab:morphomeas}}
\tablehead{
\colhead{ID} & \colhead{$\log(M_{\ast}/M_{\odot})$}& \colhead{$z_{\mathrm{spec}}$} & \colhead{$R_{e,\mathrm{F277W}}(\mathrm{kpc})$} & \colhead{$n_{\mathrm{F277W}}$} & \colhead{$q_{\mathrm{F277W}}$}  & \colhead{$\chi^{2}_{\mathrm{red,F277W}}$}  & \colhead{$R_{e,\mathrm{F444W}}(\mathrm{kpc})$}  & \colhead{$n_{\mathrm{F444W}}$} & \colhead{$q_{\mathrm{F444W}}$} & \colhead{$\chi^{2}_{\mathrm{red,F444W}}$}
} 
\colnumbers
\startdata 
ZF-UDS-3651 & $10.65_{-0.01}^{+0.01}$ & $3.81$ & $0.54\pm0.02$ & $5.98\pm1.98$ & $0.76\pm0.03$ & $1.54$ & $0.32\pm0.03$ & $6.95\pm1.20$ & $0.76\pm0.01$ & $0.63$ \\
ZF-UDS-4347 & $10.45_{-0.03}^{+0.00}$ & $3.70$ & $0.50\pm0.03$ & $7.64\pm0.57$ & $0.62\pm0.02$ & $2.23$ & $0.50\pm0.02$ & $6.65\pm0.66$ & $0.73\pm0.01$ & $0.71$ \\
ZF-UDS-6496 & $10.86_{-0.00}^{+0.02}$ & $3.98$ & $1.14\pm0.13$ & $5.97\pm1.43$ & $0.54\pm0.03$ & $1.02$ & $0.90\pm0.04$ & $5.94\pm0.41$ & $0.58\pm0.01$ & $0.56$ \\
ZF-UDS-7329 & $11.10_{-0.04}^{+0.02}$ & $3.21$ & $1.28\pm0.03$ & $2.58\pm0.20$ & $0.33\pm0.00$ & $3.43$ & $1.16\pm0.02$ & $2.60\pm0.09$ & $0.33\pm0.00$ & $1.87$ \\
ZF-UDS-7542 & $10.66_{-0.02}^{+0.02}$ & $3.20$ & $1.49\pm0.01$ & $1.70\pm0.08$ & $0.52\pm0.01$ & $0.72$ & $1.08\pm0.00$ & $1.70\pm0.03$ & $0.46\pm0.01$ & $0.71$ \\
ZF-UDS-8197 & $10.40_{-0.01}^{+0.01}$ & $3.55$ & $0.40\pm0.02$ & $3.75\pm1.11$ & $0.90\pm0.07$ & $0.86$ & $0.39\pm0.01$ & $2.51\pm0.18$ & $0.85\pm0.01$ & $0.50$ \\
ZF-COS-10559 & $10.41_{-0.00}^{+0.03}$ & $4.30$ & $0.68\pm0.01$ & $2.90\pm0.25$ & $0.53\pm0.02$ & $0.50$ & $0.58\pm0.01$ & $2.19\pm0.09$ & $0.55\pm0.03$ & $0.38$ \\
ZF-COS-14907 & $10.59_{-0.01}^{+0.01}$ & $3.00$ & $0.40\pm0.08$ & $8.74\pm0.32$ & $0.46\pm0.02$ & $1.63$ & $0.39\pm0.03$ & $5.78\pm0.70$ & $0.41\pm0.01$ & $1.03$ \\
ZF-COS-18842 & $10.46_{-0.00}^{+0.13}$ & $3.72$ & $0.48\pm0.02$ & $2.71\pm0.41$ & $0.69\pm0.03$ & $0.96$ & $0.37\pm0.04$ & $3.13\pm0.66$ & $0.77\pm0.01$ & $0.58$ \\
ZF-COS-20133 & $10.34_{-0.00}^{+0.01}$ & $3.48$ & $0.17\pm0.04$ & $0.47\pm0.22$ & $0.72\pm0.10$ & $1.10$ & $0.14\pm0.03$ & $0.57\pm1.08$ & $0.54\pm0.12$ & $0.50$ \\
ZF-COS-19589 & $10.76_{-0.02}^{+0.00}$ & $3.72$ & $1.39\pm0.01$ & $1.10\pm0.02$ & $0.45\pm0.01$ & $0.45$ & $1.12\pm0.01$ & $1.14\pm0.03$ & $0.42\pm0.01$ & $0.37$ \\
ZF-COS-20115 & $11.08_{-0.01}^{+0.00}$ & $3.72$ & $0.92\pm0.13$ & $5.73\pm2.13$ & $0.65\pm0.03$ & $1.32$ & $0.80\pm0.03$ & $4.17\pm0.27$ & $0.68\pm0.02$ & $0.78$ \\
3D-EGS-18996 & $10.88_{-0.01}^{+0.00}$ & $3.26$ & $0.81\pm0.06$ & $6.16\pm1.32$ & $0.91\pm0.01$ & $18.50$ & $0.64\pm0.01$ & $5.36\pm0.38$ & $0.84\pm0.01$ & $4.44$ \\
3D-EGS-27584 & $11.24_{-0.01}^{+0.01}$ & $3.62$ & $2.07\pm0.04$ & $5.33\pm0.41$ & $0.66\pm0.03$ & $8.75$ & $1.45\pm0.06$ & $4.05\pm0.29$ & $0.57\pm0.03$ & $2.33$ \\
3D-EGS-31322 & $10.74_{-0.00}^{+0.01}$ & $3.44$ & $0.56\pm0.02$ & $2.88\pm0.57$ & $0.60\pm0.04$ & $5.12$ & $0.53\pm0.04$ & $3.09\pm0.28$ & $0.55\pm0.05$ & $3.97$ \\
3D-EGS-34322 & $10.18_{-0.10}^{+0.01}$ & $3.23$ & $1.69\pm0.07$ & $2.87\pm0.19$ & $0.58\pm0.00$ & $0.71$ & $0.75\pm0.07$ & $3.01\pm0.08$ & $0.54\pm0.04$ & $0.47$ \\
3D-UDS-35168 & $10.18_{-0.02}^{+0.00}$ & $3.53$ & $1.08\pm0.12$ & $8.30\pm0.97$ & $0.48\pm0.02$ & $0.52$ & $0.61\pm0.03$ & $8.94\pm0.05$ & $0.48\pm0.01$ & $0.40$ \\
\enddata
\tablecomments{...}
\end{deluxetable}
\end{rotatetable}

\begin{rotatetable}
\begin{deluxetable}{ccccccccc}
\tabletypesize{\scriptsize}
\tablecaption{Summary of Morphological Parameters Measurements for Two-Component Fitting (Bulge-Disk Decomposition)\label{tab:morphomeas_BD}}
\tablehead{
\colhead{ID} & \colhead{$R_{e,\mathrm{Disk,F277W}}$} & \colhead{$R_{e,\mathrm{Bulge,F277W}}$} & \colhead{$(B/T)_{\mathrm{F277W}}$}  & \colhead{$\chi^{2}_{\mathrm{red,F277W}}$} & \colhead{$R_{e,\mathrm{Disk,F444W}}$} & \colhead{$R_{e,\mathrm{Bulge,F444W}}$} & \colhead{$(B/T)_{\mathrm{F444W}}$}  & \colhead{$\chi^{2}_{\mathrm{red,F444W}}$}  }
 \colnumbers
\startdata 
ZF-UDS-3651 & $1.87\pm0.36$ & $0.28\pm0.05$ & $0.84\pm0.02$ & $1.50$ & $0.51\pm0.13$ & $0.46\pm0.11$ & $0.93\pm0.05$ & $0.83$ \\
ZF-UDS-4347 & $2.43\pm0.38$ & $0.37\pm0.06$ & $0.84\pm0.02$ & $1.43$ & $2.23\pm0.26$ & $0.34\pm0.04$ & $0.86\pm0.02$ & $0.73$ \\
ZF-UDS-6496 & $3.69\pm0.16$ & $0.57\pm0.07$ & $0.75\pm0.03$ & $0.89$ & $3.29\pm0.30$ & $0.49\pm0.05$ & $0.79\pm0.01$ & $0.51$ \\
ZF-UDS-7329 & $1.66\pm0.17$ & $0.95\pm0.15$ & $0.57\pm0.06$ & $2.18$ & $1.41\pm0.14$ & $1.02\pm0.14$ & $0.78\pm0.12$ & $1.02$ \\
ZF-UDS-7542 & $1.75\pm0.02$ & $1.58\pm0.02$ & $0.42\pm0.01$ & $0.68$ & $1.30\pm0.01$ & $0.84\pm0.42$ & $0.25\pm0.11$ & $0.63$ \\
ZF-UDS-8197 & $0.96\pm0.13$ & $0.20\pm0.06$ & $0.81\pm0.07$ & $0.73$ & $0.56\pm0.13$ & $0.36\pm0.12$ & $0.56\pm0.14$ & $0.61$ \\
ZF-COS-10559 & $0.77\pm0.02$ & $0.66\pm0.02$ & $0.71\pm0.03$ & $0.48$ & $0.76\pm0.07$ & $0.44\pm0.08$ & $0.52\pm0.06$ & $0.37$ \\
ZF-COS-14907 & $1.02\pm0.38$ & $0.30\pm0.04$ & $1.00\pm0.04$ & $1.70$ & $1.86\pm0.49$ & $0.32\pm0.03$ & $1.00\pm0.01$ & $1.04$ \\
ZF-COS-18842 & $0.48\pm0.04$ & $0.42\pm0.04$ & $0.80\pm0.07$ & $0.84$ & $0.36\pm0.01$ & $0.32\pm0.01$ & $0.83\pm0.04$ & $0.52$ \\
ZF-COS-20133 & $0.13\pm0.09$ & $0.03\pm0.01$ & $0.72\pm0.22$ & $1.58$ & $0.08\pm0.04$ & $0.03\pm0.02$ & $0.57\pm0.36$ & $0.65$ \\
ZF-COS-19589 & $1.31\pm0.01$ & $1.18\pm0.01$ & $0.29\pm0.01$ & $0.44$ & $1.13\pm0.05$ & $1.02\pm0.05$ & $0.31\pm0.03$ & $0.36$ \\
ZF-COS-20115 & $1.34\pm0.72$ & $0.56\pm0.10$ & $1.00\pm0.03$ & $1.19$ & $1.21\pm0.28$ & $0.90\pm0.20$ & $1.00\pm0.05$ & $0.71$ \\
3D-EGS-18996 & $1.58\pm0.06$ & $0.24\pm0.01$ & $0.58\pm0.03$ & $15.27$ & $1.17\pm0.28$ & $0.60\pm0.18$ & $0.81\pm0.08$ & $3.71$ \\
3D-EGS-27584 & $3.44\pm0.32$ & $0.53\pm0.05$ & $0.52\pm0.01$ & $8.36$ & $2.33\pm0.19$ & $0.82\pm0.29$ & $0.59\pm0.07$ & $1.94$ \\
3D-EGS-31322 & $0.82\pm0.10$ & $0.53\pm0.03$ & $0.87\pm0.05$ & $3.99$ & $1.18\pm0.16$ & $0.36\pm0.10$ & $0.82\pm0.09$ & $3.18$ \\
3D-EGS-34322 & $2.67\pm0.56$ & $1.29\pm0.09$ & $0.73\pm0.02$ & $0.58$ & $1.00\pm0.45$ & $0.76\pm0.07$ & $0.75\pm0.06$ & $0.42$ \\
3D-UDS-35168 & $3.10\pm0.28$ & $0.47\pm0.04$ & $0.69\pm0.03$ & $0.50$ & $2.19\pm0.17$ & $0.33\pm0.03$ & $0.78\pm0.04$ & $0.39$ \\
\enddata
\tablecomments{The effective radius of bulge, $R_{e,\mathrm{Bulge}}$, and disk components, $R_{e,\mathrm{Disk}}$, are given in the unit of kpc.}
\end{deluxetable}
\end{rotatetable}

\begin{deluxetable}{lcccc}
\tabletypesize{\footnotesize}
\tablecaption{A Comparison Between Percentage Residual for Single S\'{e}rsic and Two-Component (Bulge-Disk Decomposition) Fits in the F277W Bandpass\label{tab:residual_comparison_f277w}}
\tablehead{
\colhead{Galaxy} & \multicolumn{2}{c}{Residual (\%)} & \colhead{Absolute Difference} & \colhead{Relative Difference (\%)} \\
\colhead{}       & \colhead{$R_{\%, \text{single S\'{e}rsic}}$} & \colhead{$R_{\%, \text{two-component}}$} & \colhead{$\Delta R_{\%}$} & \colhead{$100 \times \frac{\Delta R_{\%}}{R_{\%, \text{single~S\'{e}rsic}}} $}
}
\startdata
\hline
\multicolumn{5}{c}{\textbf{Significant Improvement with Two-Component Fit $\left(\frac{\Delta R_{\%}}{R_{\%, \text{single~S\'{e}rsic}}} > 10\%\right)$}} \\
\hline
3D-EGS-27584 & 5.8 & 2.3 & 3.6 & 61.4 \\
ZF-UDS-7329 & 2.5 & 1.8 & 0.6 & 26.5 \\
ZF-UDS-7542 & 3.2 & 2.5 & 0.8 & 23.8 \\
3D-EGS-34322 & 4.5 & 3.6 & 1.0 & 21.4 \\
3D-EGS-31322 & 2.1 & 1.7 & 0.4 & 19.6 \\
ZF-COS-20133 & 1.8 & 1.5 & 0.3 & 14.5 \\
ZF-COS-19589 & 7.7 & 6.8 & 0.9 & 12.0 \\
\hline
\multicolumn{5}{c}{\textbf{No Significant Improvement $\left( \left |\frac{\Delta R_{\%}}{R_{\%, \text{single~S\'{e}rsic}}} \right | \leq 10\%\right)$}} \\
\hline
ZF-UDS-8197 & 3.1 & 2.8 & 0.3 & 8.5 \\
3D-UDS-35168 & 6.6 & 6.1 & 0.4 & 6.5 \\
ZF-UDS-6496 & 4.0 & 4.0 & 0.0 & 0.6 \\
ZF-COS-18842 & 4.4 & 4.6 & -0.3 & -6.0 \\
\hline
\multicolumn{5}{c}{\textbf{Better Fit with Single S\'{e}rsic $\left(\frac{\Delta R_{\%}}{R_{\%, \text{single~S\'{e}rsic}}} < -10\%\right ) $}} \\
\hline
ZF-COS-14907 & 5.0 & 8.5 & -3.5 & -71.3 \\
ZF-UDS-4347 & 3.5 & 5.6 & -2.1 & -59.3 \\
ZF-UDS-3651 & 3.7 & 5.4 & -1.7 & -46.8 \\
3D-EGS-18996 & -2.1 & 2.7 & -0.6 & -30.8 \\
ZF-COS-10559 & 2.6 & 3.2 & -0.6 & -22.8 \\
ZF-COS-20115 & 2.2 & 2.5 & -0.3 & -16.1 \\
\hline
\enddata
\tablecomments{The relative improvement of the two-component model over the single Sérsic model is quantified using the residual difference $\Delta R_{\%} = | R_{\%, \mathrm{single~S\acute{e}rsic}} | - | R_{\%, \mathrm{two-component}}|$ and the relative percentage improvement $\Delta R_{\%} / R_{\%, \text{single~S\'{e}rsic}} \times 100$.}
\end{deluxetable}

\begin{deluxetable}{lcccc}
\tabletypesize{\footnotesize}
\tablecaption{A Comparison Between Percentage Residual for Single S\'{e}rsic and Two-Component (Bulge-Disk Decomposition) Fits in the F444w Bandpass\label{tab:residual_comparison_f444w}}
\tablehead{
\colhead{Galaxy} & \multicolumn{2}{c}{Residual (\%)} & \colhead{Absolute Difference} & \colhead{Relative Difference (\%)} \\
\colhead{}       & \colhead{$R_{\%, \text{single S\'{e}rsic}}$} & \colhead{$R_{\%, \text{two-component}}$} & \colhead{$\Delta R_{\%}$} & \colhead{$100 \times \frac{\Delta R_{\%}}{R_{\%, \text{Single~S\'{e}rsic}}} $}
}
\startdata
\multicolumn{5}{c}{\textbf{Significant Improvement with Two-Component Fit $\left(\frac{\Delta R_{\%}}{R_{\%, \text{single~S\'{e}rsic}}} > 10\%\right)$}} \\
\hline
3D-EGS-27584 & 0.6 & 0.4 & 0.2 & 34.8 \\
ZF-COS-20133 & 0.9 & 0.7 & 0.2 & 25.2 \\
ZF-UDS-6496 & 2.9 & 2.3 & 0.6 & 22.3 \\
ZF-COS-19589 & 5.4 & 4.4 & 1.0 & 18.7 \\
ZF-UDS-7329 & 2.9 & 2.4 & 0.5 & 18.6 \\
\hline
\multicolumn{5}{c}{\textbf{No Significant Improvement $\left( \left |\frac{\Delta R_{\%}}{R_{\%, \text{single~S\'{e}rsic}}} \right | \leq 10\%\right)$}} \\
\hline
ZF-UDS-8197 & 2.7 & 2.4 & 0.2 & 8.9 \\
3D-UDS-35168 & 4.2 & 4.1 & 0.1 & 2.9 \\
ZF-UDS-4347 & 2.7 & 2.6 & 0.1 & 2.7 \\
ZF-COS-18842 & 4.6 & 5.0 & -0.4 & -9.4 \\
\hline
\multicolumn{5}{c}{\textbf{Better Fit with Single S\'{e}rsic $\left(\frac{\Delta R_{\%}}{R_{\%, \text{single~S\'{e}rsic}}} < -10\%\right ) $}} \\
\hline
3D-EGS-18996 & -0.8 & 1.3 & -0.5 & -67.5 \\
ZF-COS-20115 & 0.7 & 1.0 & -0.3 & -45.0 \\
ZF-COS-14907 & 3.3 & 4.7 & -1.4 & -42.7 \\
3D-EGS-31322 & 1.2 & 1.7 & -0.5 & -37.6 \\
ZF-UDS-3651 & 2.2 & 3.0 & -0.7 & -32.3 \\
ZF-UDS-7542 & 1.6 & 2.2 & -0.5 & -31.1 \\
ZF-COS-10559 & 2.3 & 2.8 & -0.5 & -21.4 \\
3D-EGS-34322 & 3.3 & 3.8 & -0.5 & -15.0 \\
\hline
\enddata
\tablecomments{The relative improvement of the two-component model over the single Sérsic model is quantified using the residual difference $\Delta R_{\%} = | R_{\%, \mathrm{single~S\acute{e}rsic}} | - | R_{\%, \mathrm{two-component}}|$ and the relative percentage improvement $\Delta R_{\%} / R_{\%, \text{single~S\'{e}rsic}} \times 100$.}
\end{deluxetable}

\begin{figure}
\centering
\includegraphics[width=0.48\textwidth]{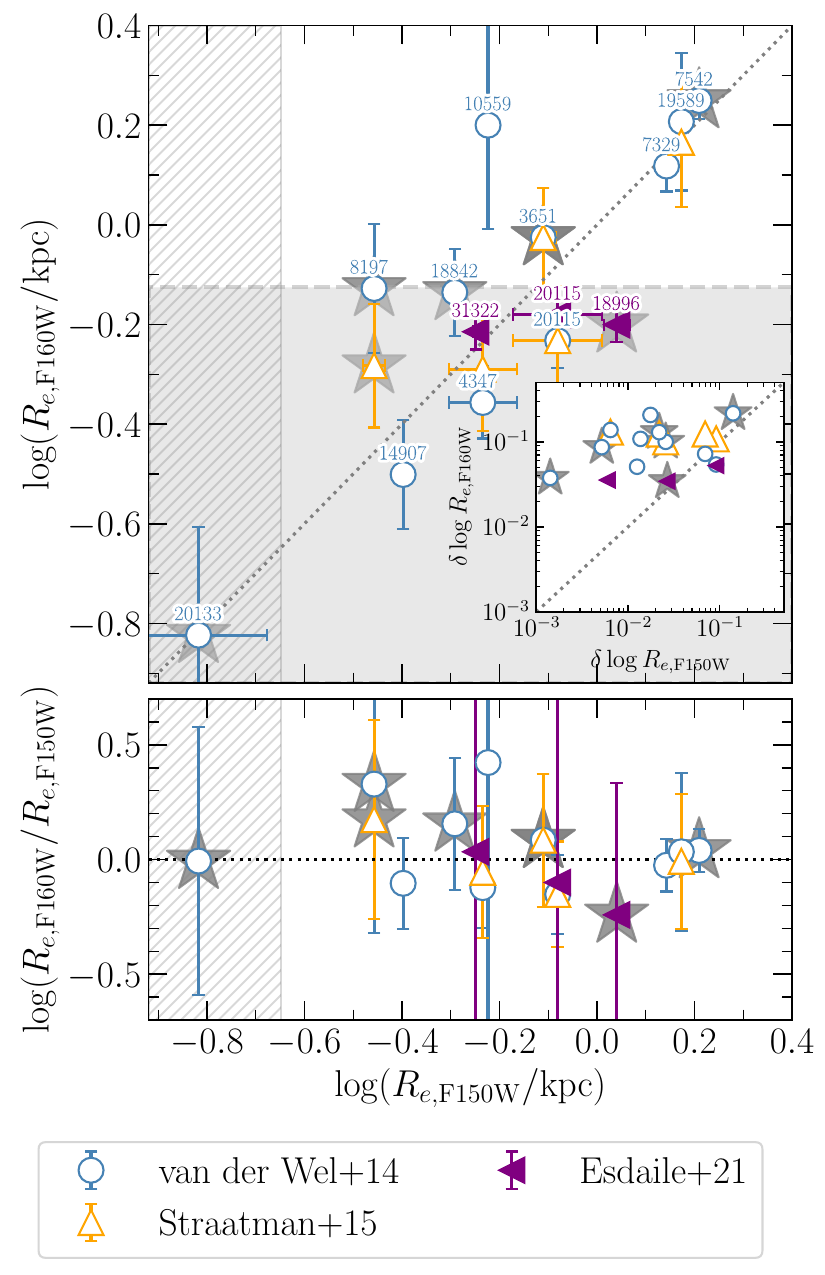}
\caption{\editone{Comparison of $1.5\mu$m sizes (NIRCam/F150W) and $1.6\mu$m sizes (WFC3/F160W) for massive quiescent galaxies at $3<z<4$. The $1.6\mu$m sizes are taken from \cite{Straatman2015} (triangle), \cite{vanderWel2014} (circle), and \cite{Esdaile2021} (left triangle). The bottom panels show the logarithmic size differences, with star markers indicating galaxies identified as potential AGN based on significant detections of prominent rest-frame optical emission lines \citep{Nanayakkara2025}. The gray-shaded region indicates sizes below half the FWHM of the F160W PSF at $z=3.5$, while the hatched region represents sizes below half the FWHM of the F150W PSF. The inset compares size uncertainties from NIRCam/F150W and WFC3/F160W.}}
\label{fig:comparesize_jwst_hst}
\end{figure}

\section{Comparison Between JWST/NIRCam and HST Size Measurements \label{sec:compare_size_nircam_wfc3}}
\par \editone{Our sample of massive quiescent galaxies is a subset of spectroscopically confirmed, $K$-selected quiescent galaxies from the ZFOURGE \citep{Straatman2016} and 3D-HST \citep{Skelton2014} surveys \citep{Schreiber2018}. \citet{Straatman2016} cross-matched the ZFOURGE photometric catalog with the structural parameters catalog from \cite{vanderWel2012}, which provides galaxy sizes and structural properties derived from S\'{e}rsic profile fits to HST/WFC3 F160W images using \textsc{Galfit} \citep{Peng2010}. Using this cross-matched catalog, we compare the $1.6\mu$m size measurements from \cite{vanderWel2012} (defined as the half-light radius along the semimajor axis) to our $1.5\mu$m size measurements for 12 galaxies in the COSMOS and UDS fields.}

\par \editone{Figure~\ref{fig:comparesize_jwst_hst} shows that the size measurements at $1.5\mu$m and $1.6\mu$m generally agree, with a median logarithmic ratio of $\log(R_{e,\mathrm{F160W}} / R_{e,\mathrm{F150W}}) = 0.04 \pm 0.05$ dex. Discrepancies arise for galaxies with sizes smaller than half the FWHM of the F160W PSF ($R_{e} < 0.5 \times \mathrm{FWHM} = 0\farcs1$, or $\sim 0.75$ kpc at $z = 3.5$), where size measurements are potentially difficult. Notably, ZF-COS-10559 shows a significant discrepancy, with $\log(R_{e,\mathrm{F160W}} / R_{e,\mathrm{F150W}} )=0.4$ dex. Its $1.6\mu$m size is $1.58 \pm 0.76$ kpc, while its $1.5\mu$m size is $0.60 \pm 0.02$ kpc.  The size difference is unlikely to result from variations in the spatial distribution of stellar populations, as the galaxy's size remains consistent across NIRCam wavelengths ($0.9\mu$m$-$4.4$\mu$m), with a median of $0.65\pm0.04$ kpc. To test whether image preprocessing or structural measurement tools (\textsc{Galight} vs. \textsc{Galfit}) contributed to the discrepancy, we reprocessed and fit the F160W image of the ZF-COS-10559 using \textsc{Galight}, following the same procedure applied to F150W images. This analysis yielded a 1.6$\mu$m size of $1.19\pm0.15$ kpc, approximately $0.3$ dex larger than the 1.5$\mu$m size. These results confirm that the size discrepancy is not caused by differences in fitting techniques or measurement tools between our work and \cite{vanderWel2012}. Instead, the discrepancy is likely attributed to the smaller PSF size of F150W (FWHM=$0\farcs06$) compared to F160W (FWHM=$0\farcs2$), which enables more precise size measurements in the F150W band, particularly important for compact and high-redshift galaxies.} 
\par In addition, \cite{Straatman2015} used \textsc{Galfit} to fit S\'{e}rsic profiles to HST/F160W images and measure the sizes (half-light radius along the semimajor axis) and structural parameters for quiescent and star-forming galaxies at $2 \le z < 4.2$ from the ZFOURGE survey. \edittwo{\citet{Straatman2015} found that their the structural measurements closely match those from \cite{vanderWel2012,vanderWel2014}, reporting a median size ratio of  $1.004\pm0.01$ and differences in S\'{e}rsic indices of $-0.012\pm0.058$. Since our 1.5$\mu$m size measurements also exhibit good agreement with those from \citet{vanderWel2012,vanderWel2014}, it follows that our results should align with \citet{Straatman2015} as well. Indeed, we find that our 1.5$\mu$m sizes are closely aligned with the measurement from \citet{Straatman2015}, with a median logarithmic size ratio of $\log(R_{e,\mathrm{F160W}} / R_{e,\mathrm{F150W}})=-0.01\pm0.08$ dex (Figure~\ref{fig:comparesize_jwst_hst}).} 


\begin{figure*}
\centering
\includegraphics[width=\textwidth]{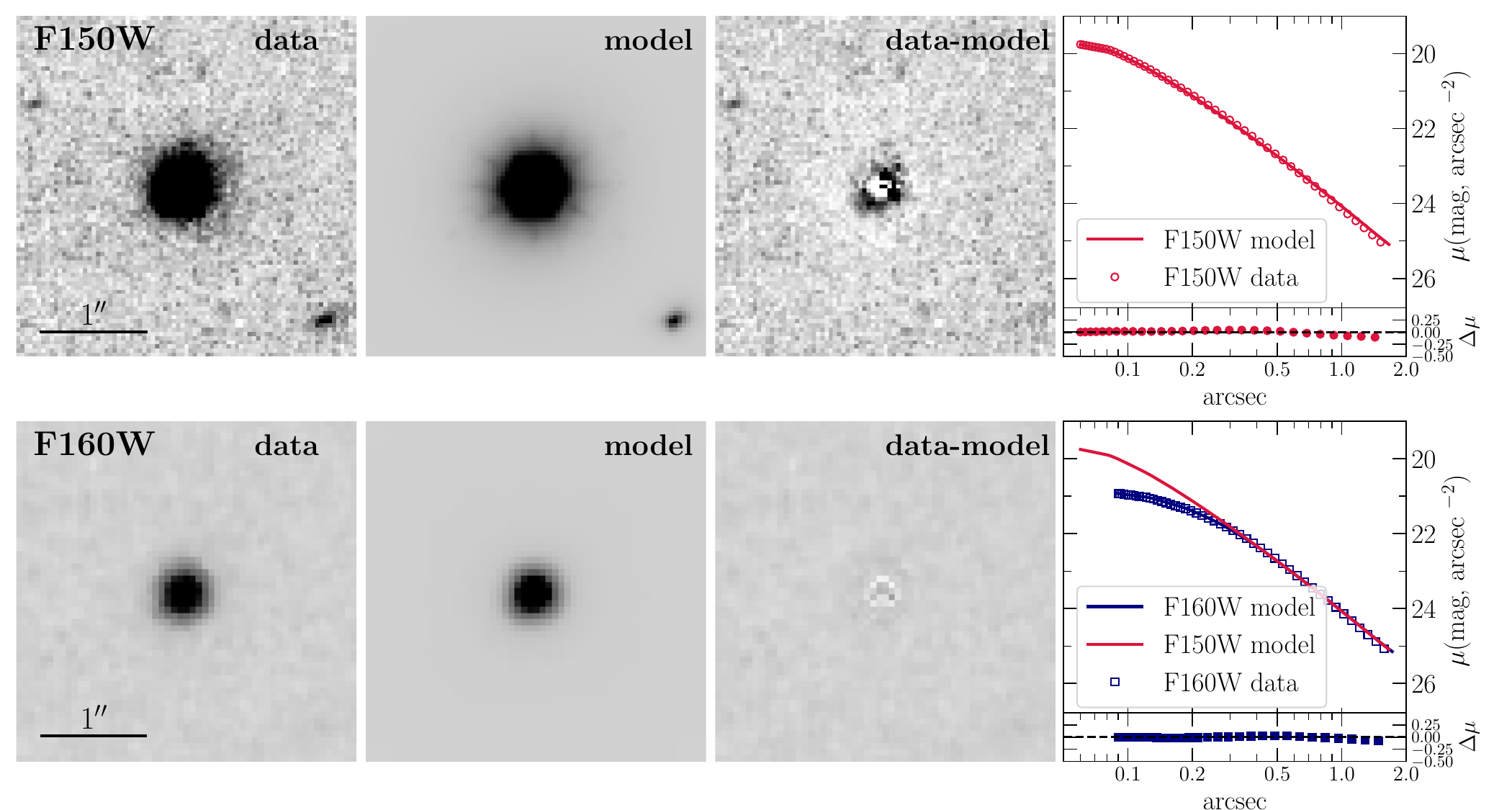}
\caption{\editone{Comparison of light profile fits for the galaxy 3D-EGS-18996 in F150W (JWST/NIRCam) and F160W (HST/WFC3). The first three columns show the data, model (PSF-convolved), and residual images (same colorscale), while the last column presents the 1D radial surface brightness profiles. \textbf{Top row:} F150W achieves $r_e = 1.07~\mathrm{kpc}$ and $n = 8.9$, capturing both compact and extended structures with minimal residuals ($\sim2\%$ of the total light). \textbf{Bottom row:} F160W shows $r_e = 0.51\pm0.06~\mathrm{kpc}$ and $n = 6.1$, with structured residuals ($\sim4\%$) and reduced sensitivity to faint outskirts. The overplotted F150W model (crimson solid line) highlights discrepancies, with F150W yielding sizes and S\'{e}rsic indices larger by factors of 2.3 and 1.3, respectively. These differences reflect JWST’s superior surface brightness sensitivity and sharper PSF ($0\farcs06$ vs. $0\farcs2$), enabling more accurate galaxy structure measurements.}}
\label{fig:comparesize_jwst_hst_3d_egs_18996}
\end{figure*}

\par \editone{Moreover, \cite{Esdaile2021} measured the sizes and structural properties of four spectroscopically-confirmed massive quiescent galaxies at $3.2 < z < 3.7$, three of which are also included in our sample: ZF-COS-20115, 3D-EGS-18996, and 3D-EGS-31322. These structural properties were measured using HST/F160W images with \textsc{Galfit}, and the authors define galaxy size using the semimajor axis, consistent with our work and previous studies \citep[e.g.,][]{vanderWel2012, Straatman2015}. For ZF-COS-20115, 3D-EGS-18996, and 3D-EGS-31322, \cite{Esdaile2021} reported $1.6 \, \mu$m sizes of $r_e = 0.66 \pm 0.08$ kpc, $0.63 \pm 0.05$ kpc, and $0.61 \pm 0.05$ kpc, respectively, corresponding to size differences of $\log(R_{e,\mathrm{F160W}} / R_{e,\mathrm{F150W}}) = -0.10$, $-0.24$, and $0.03$ dex, respectively.}

\par \editone{The significant size difference for 3D-EGS-18996, where \citeauthor{Esdaile2021} measured $r_e = 0.63 \pm 0.05$ kpc in F160W compared to our measurement of $r_e = 1.10 \pm 0.07$ kpc in F150W, is unlikely to be driven by wavelength dependence of the galaxy size. Across the observed wavelength range of $\lambda = 1.5 \, \mu$m–$2.7 \, \mu$m, the galaxy size in the JWST/NIRCam images remains consistent, as demonstrated by measurements in longer NIRCam filters: $r_e = 0.99 \pm 0.04$ kpc in F200W and $r_e = 0.81 \pm 0.06$ kpc in F277W. To investigate this discrepancy further, we reprocessed and fit the F160W image using \textsc{Galight} with the same procedure applied to F150W, utilizing the F160W PSF from the 3D-HST catalog \citep{Skelton2014}.}

\par \editone{In Figure~\ref{fig:comparesize_jwst_hst_3d_egs_18996}, we present the 2D light profiles of 3D-EGS-18996 observed in F150W and F160W. Our remeasured F160W size, $r_e = 0.51 \pm 0.06$ kpc, is in agreement with \cite{Esdaile2021} within uncertainties. The comparison between the 2D images and 1D profiles in both filters highlights key differences. The F150W image, with JWST’s superior surface brightness sensitivity and sharper PSF ($\text{FWHM} \sim 0\farcs06$), reveals faint extended structures contributing flux at larger radii and a steeper slope in the central region. This results in a larger effective radius ($r_e = 1.07 \, \mathrm{kpc}$) and a higher Sérsic index ($n = 8.9$). In contrast, the F160W image, with its lower sensitivity and broader PSF ($\text{FWHM} \sim 0\farcs2$), lacks the extended faint structures and exhibits a flatter central profile, leading to a smaller size ($r_e = 0.51 \, \mathrm{kpc}$) and lower Sérsic index ($n = 6.1$). These results underscore JWST/NIRCam’s ability to accurately capture both compact and extended components of galaxy structure.}

\par {\editone{Finally, we compare the uncertainties in the $1.5\mu$m and $1.6\mu$m size measurements, as shown in the inset of Figure~\ref{fig:comparesize_jwst_hst}. The median uncertainty for our $1.5\mu$m size measurements is $\delta \log R_{e} = 0.02 \pm 0.01$ dex, approximately $0.1$ dex lower than the uncertainties for the $1.6\mu$m measurements reported by \cite{vanderWel2014} ($\delta \log R_{e} = 0.10 \pm 0.02$ dex) and \cite{Straatman2015} ($\delta \log R_{e} = 0.12 \pm 0.01$ dex). For the three galaxies in common with \cite{Esdaile2021}, the F150W size uncertainties are comparable to or slightly higher than those in F160W. Specifically, ZF-COS-20115 has an uncertainty of $0.18$ kpc in F150W, compared to $0.08$ kpc in F160W. These results demonstrate that while JWST/NIRCam measurements generally provide higher precision due to its smaller PSF, individual uncertainties may vary depending on specific structural properties and data quality. This reduction in uncertainty again highlights the higher precision of JWST/NIRCam structural parameter measurements, enabled by its smaller PSF compared to F160W, which allows for more accurate size constraints.}}

\begin{figure*}
\centering
\includegraphics[width=\textwidth]{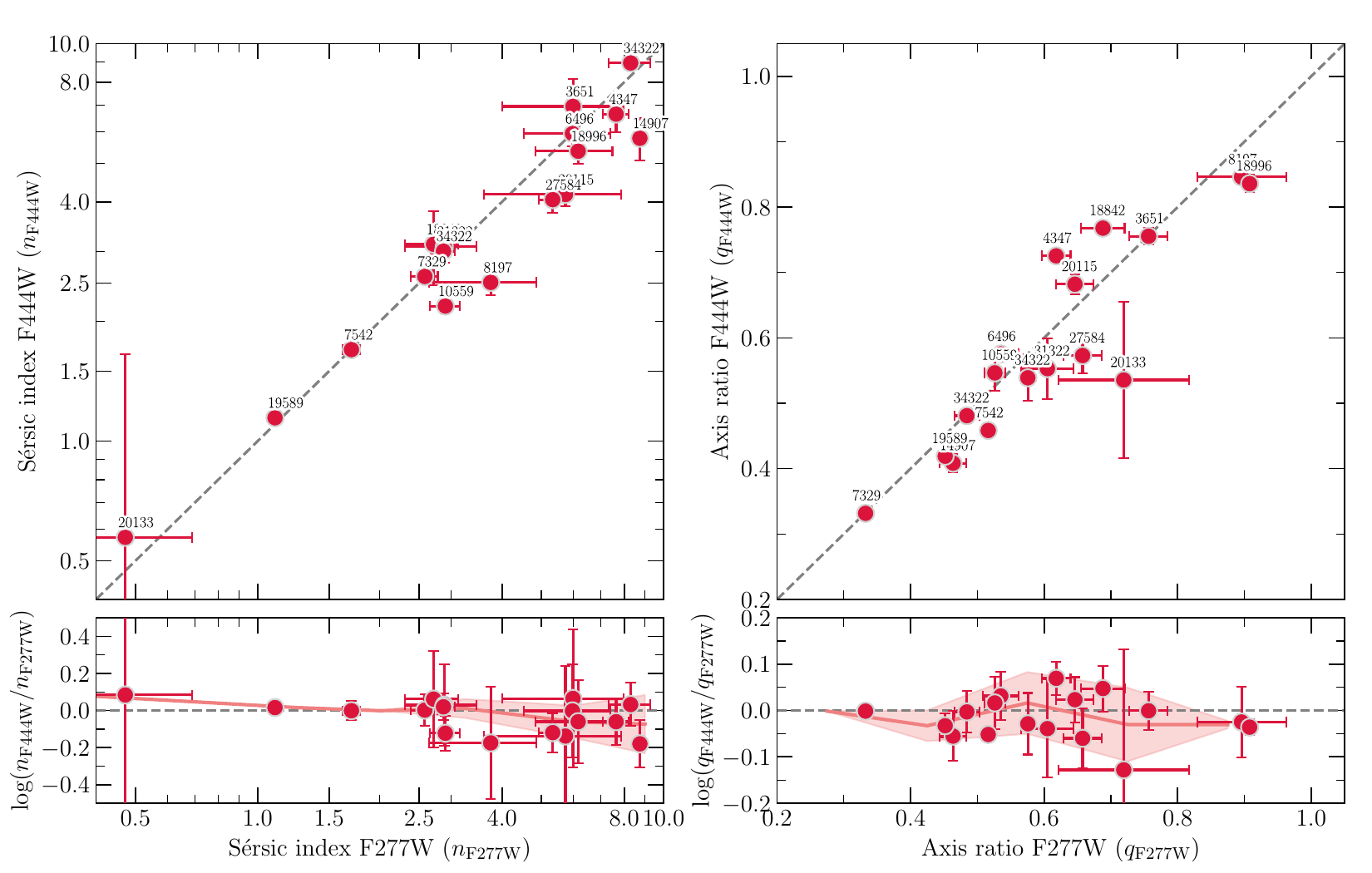}
\caption{\editthree{Comparison of Sérsic indices ($n$; left) and axis ratios ($q$; right) for massive quiescent galaxies at $z=3$–4, measured independently in the F277W ($n_{\mathrm{F277W}}$) and F444W ($n_{\mathrm{F444W}}$) filters. The top panels show direct comparisons between the two bands, while the bottom panels display the ratio of each morphological parameter ($n_{\mathrm{F444W}} / n_{\mathrm{F277W}}$ and $q_{\mathrm{F444W}} / q_{\mathrm{F277W}}$) as a function of their F277W values. Binned medians and scatter (computed via the median absolute deviation) are shown as solid lines with shaded regions. With the exception of ZF-COS-20133, which is unresolved or marginally resolved, most galaxies show statistically significant differences between the two filters. However, the typical variation is modest ($\lesssim$30\%) and the F277W and F444W measurements are strongly correlated, suggesting that the structural properties of these galaxies are largely consistent across rest-frame wavelengths.}}

\label{fig:comparesize_sersic_axisratio_f277w_f444w}
\end{figure*}

\section{Comparison of Morphological Parameters in F277W and F444W}
\label{appendix:compare_sersic_axisratio_f277w_f444w}
\editthree{In this section, we examine the wavelength dependence of S\'{e}rsic index and projected axis ratio for our sample of massive quiescent galaxies at $z=3-4$. The left panel of Figure~\ref{fig:comparesize_sersic_axisratio_f277w_f444w} compare S\'{e}rsic indices measured in F277W ($n_{\mathrm{F277W}}$) and F444W ($n_{\mathrm{F444W}}$). To quantify the variation, we compute the ratio $n_{\mathrm{F444W}} / n_{\mathrm{F277W}})$. With the exception of ZF-COS-20133, which is unresolved or marginally resolved, 16 galaxies show statistically significant differences between the two filters. However, the typical variation in $n$ and $q$ is modest ($\lesssim30\%$), and the measurements in F277W and F444W are strongly correlated, indicating largely consistent structural properties across rest-frame wavelengths.} 
\par \editthree{Among these, nine galaxies (e.g., ZF-COS-10559, ZF-8197, ZF-COS-20115, ZF-COS-14907) exhibit $n_{\mathrm{F444W}} / n_{\mathrm{F277W}} < 1$, with ratios ranging from 0.66 to 0.99. These systems are more centrally concentrated in the rest-frame optical than in the near-IR, consistent with the presence of compact, younger stellar populations or residual star formation at shorter wavelengths. Conversely, seven galaxies (e.g., ZF-COS-19589, ZF-3651, 3D-UDS-35168,3D-EGS-34322) show $n_{\mathrm{F444W}} / n_{\mathrm{F277W}} > 1$, with ratios from 1.04 to 1.15, suggesting higher concentration in the near-IR, possibly due to centrally concentrated older stellar populations.}
\par \editthree{Dust attenuation can, in principle, influence structural parameters by obscuring central light (flattening Sérsic profiles) and altering apparent shape through anisotropic extinction. For example, 3D-EGS-34322—the dustiest galaxy in the sample, with $\tau_{\mathrm{dust}} \sim 1.7$; \citep{Nanayakkara2025}—shows a mildly higher Sérsic index in F444W, consistent with the expected impact of dust flattening the optical profile. In contrast, ZF-COS-20115, a quiescent galaxy with similarly high dust content \citep[$\tau_{\mathrm{dust}} \sim 1.5$;][]{Nanayakkara2025}, exhibits a significantly higher Sérsic index in F277W than in F444W ($n_{\mathrm{F444W}} / n_{\mathrm{F277W}}=0.73$). This behavior is opposite to that expected from central dust obscuration, suggesting for this system -- and likely others in our sample -- the observed wavelength dependence is more plausibly driven by stellar population gradients or compact central star-forming regions than dust attenuation.}

\par \editthree{We also compare the projected axis ratios measured in F277W ($q_{\mathrm{F277W}}$) and F444W ($q_{\mathrm{F444W}}$), as shown in the right panel of Figure~\ref{fig:comparesize_sersic_axisratio_f277w_f444w}. Following the same approach, we compute the ratio $q_{\mathrm{F444W}} / q_{\mathrm{F277W}}$. All 16 galaxies exhibit statistically significant differences in axis ratio between the two filters. Among them, 10 galaxies (e.g., ZF-COS-19589, ZF-COS-14907, ZF-7542, 3D-EGS-27584) have $q_{\mathrm{F444W}} / q_{\mathrm{F277W}} < 1$, indicating that they appear more elongated at longer wavelengths. These systems may exhibit flatter stellar mass distributions or the emergence of disk-like components in the rest-frame near-IR. Conversely, six galaxies (e.g., ZF-6496, ZF-COS-20115, ZF-COS-18842, ZF-4347) show $q_{\mathrm{F444W}} / q_{\mathrm{F277W}} > 1$, with ratios up to 1.17, suggesting rounder morphologies in the near-IR or a clearer view of spheroid-dominated structures at longer wavelengths.}

\par \editthree{Together, these results demonstrate that while the structural parameters of massive quiescent galaxies at $z=3$–4 are strongly correlated across rest-frame wavelengths, modest but statistically significant differences in Sérsic index and axis ratio are common. These differences are most consistent with underlying gradients in stellar population age and structure—such as centrally concentrated star-forming regions or older stellar bulges—rather than dust obscuration. This highlights the importance of accounting for rest-frame wavelength when interpreting morphological measurements at high redshift, particularly when comparing to lower-redshift samples or inferring intrinsic structural properties.}

\begin{deluxetable*}{lcccc|cccc}
\tablecaption{Environmental Density Measurements of Massive Quiescent Galaxies at $z=3-4$ derived using the Traditional $N$th Nearest Neighbors ($\log(1+\delta)_N$) and Bayesian-motivated $N$th nearest neighbors ($\log(1+\delta^{\prime} )_N$) density estimators \label{tab:overdensity_measurements}}
\tablehead{
\colhead{ID} & 
\multicolumn{4}{c}{ $\log(1+\delta)_N$} & 
\multicolumn{4}{c}{$\log(1+\delta^{\prime}_{N})$} \\ 
\cline{2-9} 
 & $N=3$ & $N=5$ & $N=7$ & $N=10$ & $N=3$ & $N=5$ & $N=7$ & $N=10$}
\startdata
ZF-UDS-3651 & $1.90$ & $1.16$ & $0.74$ & $0.65$ & $1.83$ & $1.42$ & $1.06$ & $0.84$ \\
ZF-UDS-4347 & $0.35$ & $0.11$ & $0.19$ & $0.30$ & $0.31$ & $0.15$ & $0.16$ & $0.22$ \\
ZF-UDS-6496 & $1.25$ & $1.09$ & $1.18$ & $1.15$ & $1.19$ & $1.20$ & $1.19$ & $1.18$ \\
ZF-UDS-7329 & $0.24$ & $0.28$ & $0.21$ & $0.32$ & $0.32$ & $0.32$ & $0.27$ & $0.29$ \\
ZF-UDS-7542 & $0.18$ & $0.15$ & $0.29$ & $0.09$ & $0.13$ & $0.13$ & $0.19$ & $0.20$ \\
ZF-UDS-8197 & $-0.07$ & $0.09$ & $0.08$ & $0.11$ & $-0.07$ & $0.01$ & $0.04$ & $0.07$ \\
ZF-COS-10559 & $-0.01$ & $-0.16$ & $-0.06$ & $-0.12$ & $-0.06$ & $-0.14$ & $-0.11$ & $-0.12$ \\
ZF-COS-14907 & $0.31$ & $0.22$ & $0.28$ & $0.24$ & $0.52$ & $0.33$ & $0.31$ & $0.28$ \\
ZF-COS-18842 & $0.51$ & $0.38$ & $0.36$ & $0.12$ & $0.63$ & $0.46$ & $0.41$ & $0.31$ \\
ZF-COS-20133 & $0.52$ & $0.68$ & $0.71$ & $0.77$ & $0.73$ & $0.68$ & $0.68$ & $0.72$ \\
3D-EGS-18996 & $0.29$ & $-0.08$ & $0.07$ & $0.16$ & $0.23$ & $0.01$ & $0.02$ & $0.08$ \\
3D-EGS-27584 & $0.28$ & $0.27$ & $0.36$ & $0.34$ & $0.47$ & $0.34$ & $0.34$ & $0.34$ \\
3D-EGS-31322 & $0.80$ & $0.50$ & $0.51$ & $0.09$ & $0.95$ & $0.74$ & $0.60$ & $0.34$ \\
3D-EGS-34322 & $0.71$ & $0.83$ & $0.51$ & $0.48$ & $0.72$ & $0.77$ & $0.71$ & $0.59$ \\
3D-UDS-35168 & $-0.10$ & $0.01$ & $0.02$ & $0.02$ & $-0.11$ & $-0.05$ & $-0.02$ & $0.01$ \\
GS-9209 & $-0.04$ & $-0.05$ & $-0.06$ & $-0.01$ & $0.13$ & $0.02$ & $0.00$ & $-0.01$ \\
RUBIES-EGS-QG1 & $1.16$ & $0.87$ & $0.86$ & $0.84$ & $1.15$ & $1.07$ & $0.95$ & $0.90$ \\
\enddata
\tablecomments{Note: $\log(1+\delta)_N$ is derived using the traditional $N$th nearest neighbor estimator (Equation~\ref{eq:traditional_sigmaN}), while $\log(1+\delta^{\prime}_{N})$ is based on the Bayesian-motivated estimator (Equation~\ref{eq:bayesian_density_cowan}) for the local surface number density of galaxies. ZF-COS-20115 and ZF-COS-19589 are located at the edge of the PRIMER COSMOS mosaic image, preventing reliable measurement of their $N$th nearest neighbor distances and environmental densities. Consequently, these galaxies are excluded from the analysis of environmental effects on morphological properties. GS-9209 is a spectroscopically confirmed massive quiescent galaxy with $\log(M_{\ast}/M_{\odot}) = 10.58$ at $z = 4.658$, reported by \citealt{Carnall2023Nature}. Similarly, RUBIES-EGS-QG1 is also a spectroscopically confirmed massive quiescent galaxy with $\log(M_{\ast}/M_{\odot}) = 10.9, z = 4.886$, reported by \citealt{deGraaff2024}.}
\end{deluxetable*}

\begin{figure*}
    \centering
     \includegraphics[width=1\textwidth]{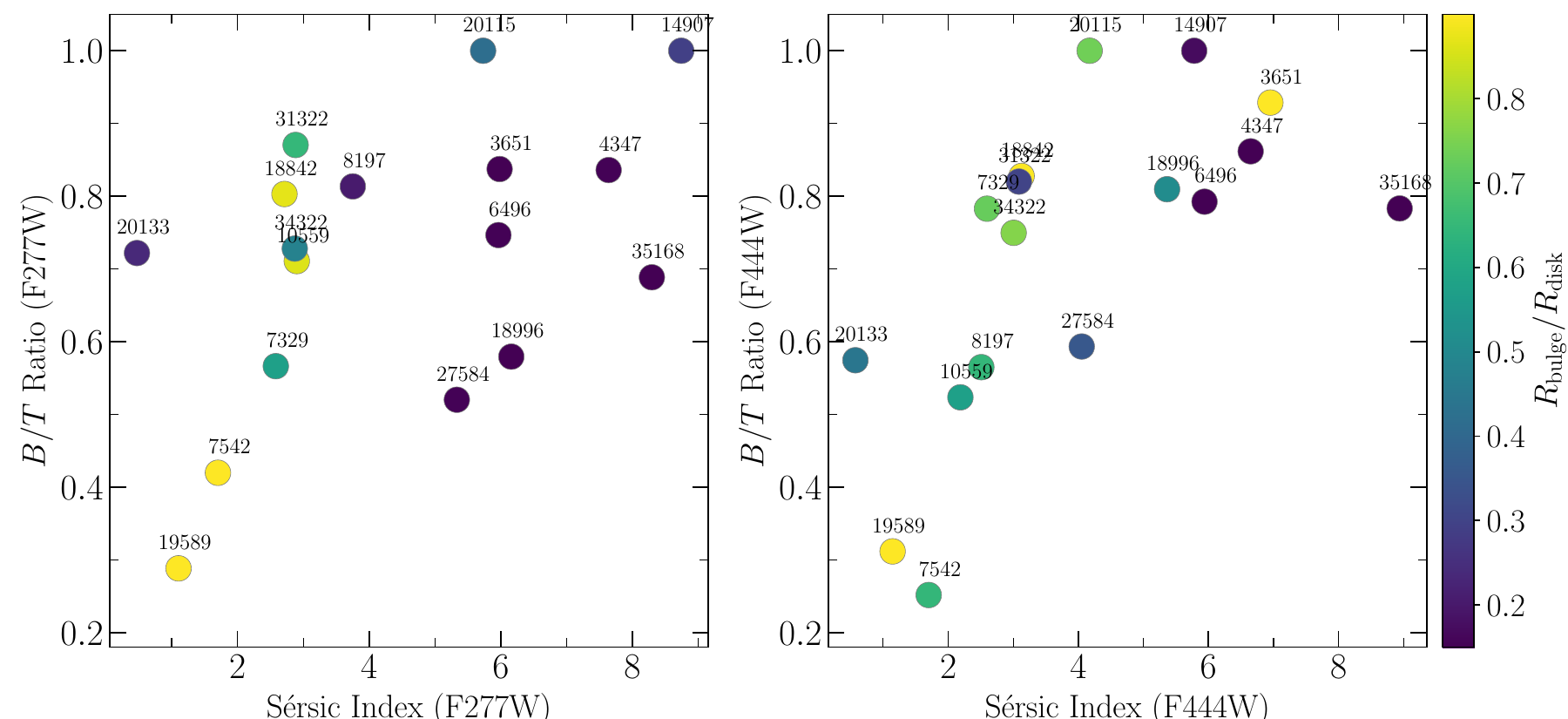}
\caption{Bulge-to-total flux ratio ($B/T$) from the two-component bulge–disk decomposition versus Sérsic index $n$ from the \emph{single-component} fit for massive quiescent galaxies at $z=3$–4, shown for the F277W (left) and F444W (right) bandpasses. In the two-component fits, the bulge Sérsic index is fixed to $n_{\mathrm{bulge}} = 4$. Points are color-coded by the ratio of bulge to disk size ($R_{\mathrm{bulge}}/R_{\mathrm{disk}}$), with more extended disks (lower ratios) appearing darker. \editfour{An overall positive correlation between $B/T$ and $n$ is evident, particularly at $n \lesssim 4$. However, substantial scatter persists across the full range of $n$, reflecting structural differences not captured by a single-parameter fit. For example, ZF-COS-18842 and ZF-6496 both have similarly high $B/T$ values of $\sim$0.8 in F444W, yet differ in $n$ (3.13 vs.\ 5.94) and bulge-to-disk size ratio ($(R_{\mathrm{bulge}}/R_{\mathrm{disk}})_{F444W}=0.90$ vs.\ 0.14). Such cases illustrate how extended disks can increase $n$ even at fixed $B/T$. These findings highlight the non-unique mapping between $n$ and bulge prominence, underscoring the value of two-component decompositions for robust structural characterization.}}
    \label{fig:bt_vs_sersicindex} 
\end{figure*}

\begin{figure*}
    \centering
    \includegraphics[width=0.9\textwidth]{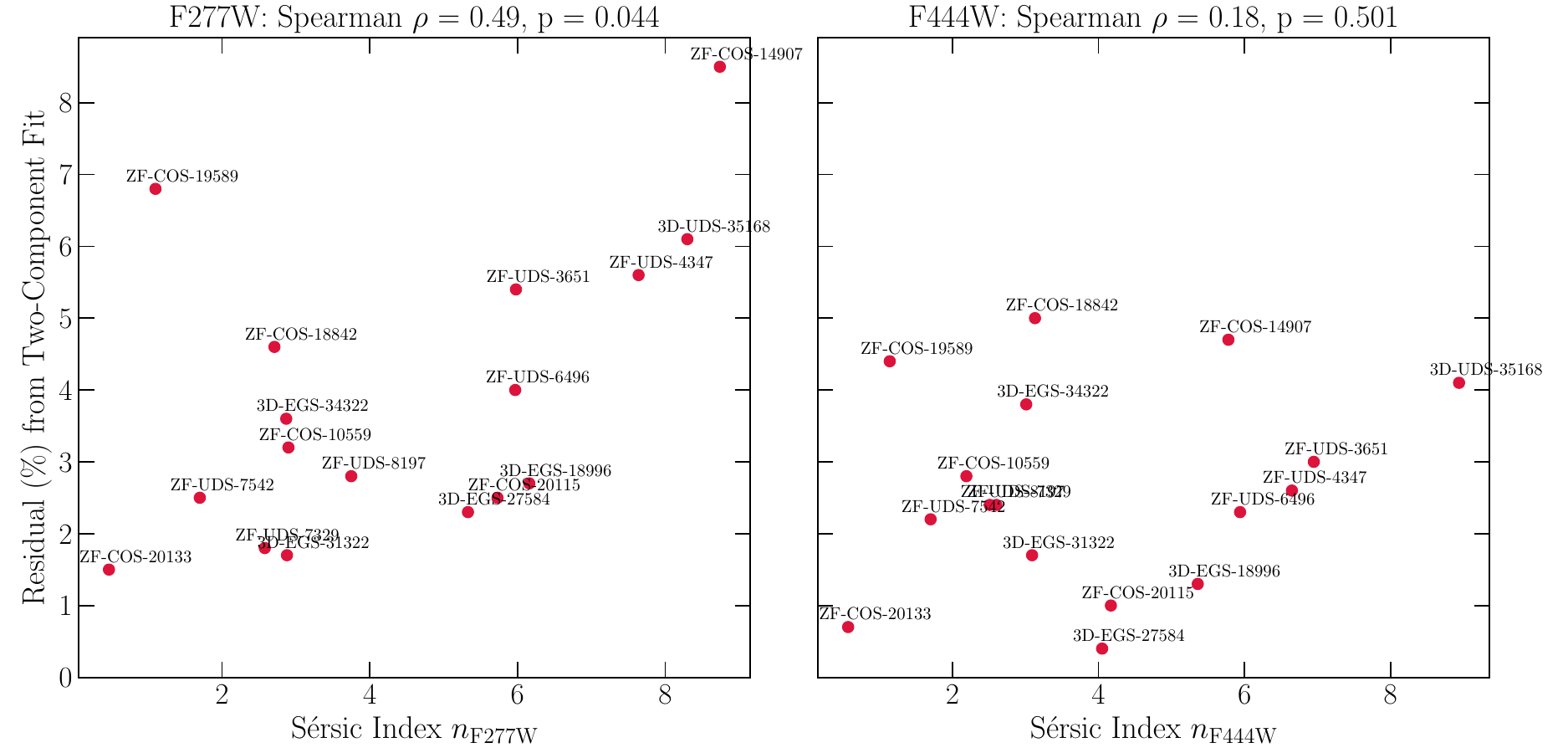}
    \caption{\editfour{Residual percentage from the two-component (bulge–disk) fit as a function of Sérsic index $n$ from the \emph{single-component} fit, for F277W (left) and F444W (right). In the two-component fits, the bulge Sérsic index is fixed to $n_{\mathrm{bulge}} = 4$. Each point represents a galaxy in our sample, with its ID labeled. The Spearman rank correlation coefficients ($\rho$) and associated $p$-values are indicated. A moderate positive correlation is present in F277W, where galaxies with $n_{F277W} > 6$ tend to have $\sim$5–10\% higher residuals, while no significant correlation is seen in F444W.}}
    \label{fig:residual_vs_sersicindex}
\end{figure*}

\section{Comparison Between $B/T$ and S\'{e}rsic Index\label{appendix:compare_bt_sersic}}

\par To evaluate the reliability of single-component S\'ersic fits as a proxy for galaxy structure, we compare the S\'ersic index ($n$) from the \emph{single-component} fit with the bulge-to-total flux ratio ($B/T$) derived from two-component bulge+disk decompositions. In the two-component fits, the bulge S\'ersic index is fixed to $n_{\mathrm{bulge}} = 4$. While a general positive correlation between $B/T$ and $n$ is expected—and indeed observed, particularly at $n \lesssim 4$—the relationship is not one-to-one. A high S\'ersic index can result either from a more dominant bulge (i.e., higher $B/T$) or from a more extended disk surrounding a compact bulge, which steepens the overall light profile without significantly altering the flux ratio \citep[e.g.,][]{Lang2014}. To test this, we examine the distribution of $B/T$ versus $n$ in both F277W and F444W, with points color-coded by the ratio of bulge to disk sizes ($R_{\mathrm{bulge}}/R_{\mathrm{disk}}$), as shown in Figure~\ref{fig:bt_vs_sersicindex}.

\par This structural degeneracy is illustrated by galaxies such as ZF-COS-18842 and ZF-6496, both of which have high $B/T$ values of $\sim$0.8 in F444W. However, ZF-COS-18842 has a single-component S\'ersic index of 3.13 and $R_{\mathrm{bulge}}/R_{\mathrm{disk}} = 0.90$, while ZF-6496 has a much higher single-component index of 5.94 but a significantly smaller size ratio of $R_{\mathrm{bulge}}/R_{\mathrm{disk}} = 0.14$. The higher $n$ of ZF-6496 likely reflects its extended disk rather than a more prominent bulge. These cases illustrate that even among galaxies with similar $B/T$, differences in structural configuration can lead to a broad range in $n$. Conversely, galaxies with similar $n$ may have different $B/T$ values depending on how the light is spatially distributed between components.

\par While part of the observed scatter in $B/T$ at fixed $n$ may stem from model assumptions—e.g., fixing the bulge component to $n_{\mathrm{bulge}} = 4$—the presence of this scatter at intermediate $n$ values ($n \sim 3$--4) even in robust fits suggests that it is not solely driven by fitting constraints. Overall, this highlights the importance of two-component decomposition in capturing structural diversity not fully encoded in single-component S\'ersic fits.

\par \editfour{In addition, we assess whether residuals from the two-component fits are correlated with the single-component S\'ersic index (Figure~\ref{fig:residual_vs_sersicindex}). We find a moderate positive correlation in F277W ($\rho = 0.49$, $p = 0.044$), indicating that galaxies with higher single-component S\'ersic indices tend to show systematically larger residuals—by $\sim$5--10\% on average—after bulge–disk fitting. No significant correlation is present in F444W ($\rho = 0.18$, $p = 0.501$). This supports the interpretation that fixed-component bulge–disk models are less effective at modeling strongly peaked or extended profiles, which are more prominent in the rest-frame optical morphologies traced by F277W.}

\begin{figure*}
    \centering
     \includegraphics[width=1\textwidth]{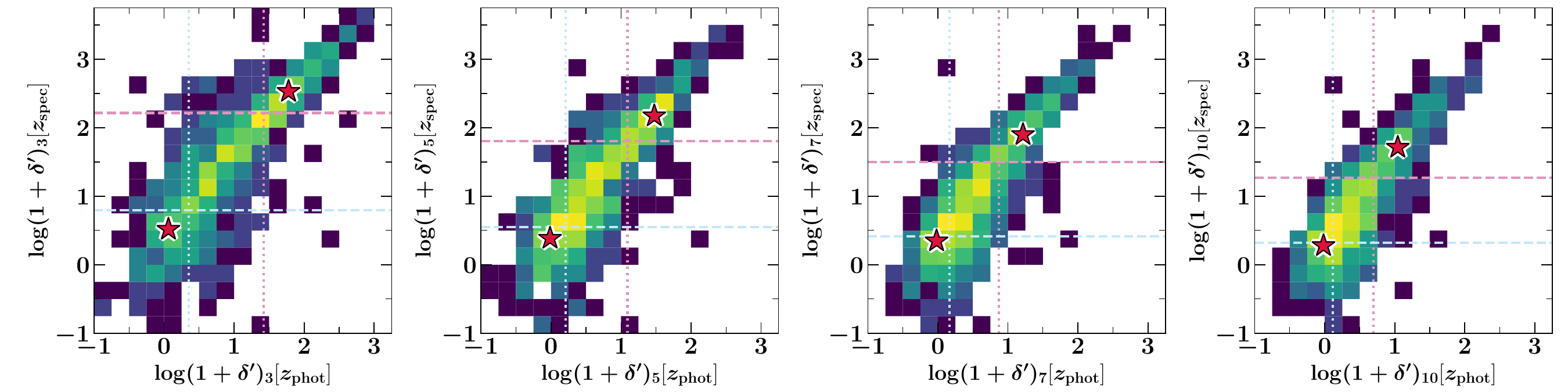}
    \caption{Measurements of galaxy overdensities for massive galaxies with $\log(M_{\ast}/M_{\odot})>10.2$ at $3<z<4.5$, selected from UniverseMachine mock galaxy catalog for JWST \citep{Behroozi2020}. The overdensities of galaxies are computed using both photometric redshifts, ($\log(1+\delta^{\prime}_{N}) [z_{\mathrm{phot}}];~\Delta z=0.02(1+z)$, $x$-axis), and spectroscopic redshifts ($\log(1+\delta^{\prime}_{N}) [z_{\mathrm{spec}}];~\Delta v=2100~\mathrm{km~s}^{-1}$; $y$-axis). The panels show results for the Bayesian-motivated estimator of the local surface density of galaxies for $N=3, 5, 7,$ and 10 (from the left to the right). The light blue and red vertical dotted lines indicate the 25th and 75th percentiles of $\log(1+\delta^{\prime}_{N}) [z_{\mathrm{phot}}]$ distribution of the $z=3-4$ massive galaxies in the mock, marking the lowest- and highest-density environments defined using $z_{\mathrm{phot}}$. Similarly, the light blue and red horizontal dash lines represent the 25th and 75th percentiles of the $\log(1+\delta^{\prime}_{N}) [z_{\mathrm{spec}}]$ distribution. The red star show the median values of  $(\log(1+\delta^{\prime}_{N}) [z_{\mathrm{spec}}]$ for galaxies identified as residing in the low- and high-density environments using $z_{\mathrm{phot}}$. A strong correlation is observed between the density estimates from  $z_{\mathrm{phot}}$ and $z_{\mathrm{spec}}$. This indicates that, with the photometric redshift precision achieved in the JWST/PRIMER and CEERS surveys, galaxies in the highest- (lowest-) density quartiles identified using $z_{\mathrm{spec}}$ over this redshift range can be reliably recovered using $z_{\mathrm{phot}}$.}
    \label{fig:log1pd_specz_photoz_mock} 
\end{figure*}
\section{Measuring Environmental Densities with Photometric Redshifts\label{appendix:measure_overdensity}}
In this work, we quantify the environmental density of massive quiescent galaxies at $z=3-4$ using the (projected) distance to the $N$th nearest-neighbor, $d_{N}$ with $N$ chosen within typical range of 3 to 10 in literature \citep[e.g.,][]{Dressler1980, Baldry2006, Muldrew2012, Kawinwanichakij2017,Papovich2018, Gu2021,Chang2022}. One of the advantages of this estimator is that it provides a reliable measure of ``local density'' on scales internal to galaxy group halos, particularly when $N$ is small \citep{Muldrew2012}. The local surface density of a galaxy based on the distance to the $N$th nearest neighbor is given by 
\begin{equation}
\Sigma_{N} = N/(\pi d^{2}_{N})
\label{eq:traditional_sigmaN}
\end{equation}
\noindent where $d_{N}$ is the projected distance to the $N$th nearest neighbor.
Following \citet{Kawinwanichakij2017,Papovich2018, Gu2021,Chang2022}, we improve our measurement of environmental density by adopting a Bayesian-motivated estimator of the local surface density of galaxies, which uses the distances to all $N$th nearest neighbors to improve accuracy and mitigate projection effects compared to the traditional $N$th nearest-neighbor estimator (Equation~\ref{eq:traditional_sigmaN}) as demonstrated by  \citet{Ivezic2005} and \citet{Cowan2008} (see Appendix B of \cite{Ivezic2005} and also Appendix A of \cite{Kawinwanichakij2017}). Specifically, we use the Bayesian-motivated estimator of environmental density given by \cite{Cowan2008}: 
\begin{equation}
\Sigma^{\prime}_{N} \propto \frac{1}{\sum_{i=1}^{N}  d_{i}^{2}},
\label{eq:bayesian_density_cowan}
\end{equation}
\noindent where $d_{i}$ is the projected distance to the $i$th nearest neighbor.

\par Following \cite{Gu2021} \citep[and see also][]{Chang2022}, we define the environment of a galaxy in terms of the dimensionless overdensity as: 
\begin{equation}
(1+\delta^{\prime}_{N}) =  \frac{\Sigma^{\prime}_{N}}{ \left<\Sigma^{\prime}_{N} \right>_{\mathrm{uniform}}},
\label{eq:onepdelta}
\end{equation}
\noindent where $\left<\Sigma^{\prime}_{N} \right>_{\mathrm{uniform}}$ is the standard value of Bayesian-motivated density estimator when galaxies are distributed in the uniform environment.  Specifically, we estimate $\left<\Sigma^{\prime}_{N} \right>_{\mathrm{uniform}}$ using the following procedure. First, we create a uniform map by dropping galaxies into the real sky coverage, such that they are uniformly distributed. These are galaxies in the same redshift range as the galaxy sample we aim to measure the environmental density and with stellar masses down to the stellar mass completeness limit adopted at a given redshift range (see below). We then compute the Bayesian density estimator, $\Sigma^{\prime}_{N}$ for each target galaxy (e.g., a sample of massive quiescent galaxies) in the simulated uniform map. In practice, we generate multiple ($\gtrsim5000$) uniform maps of galaxies. We then adopt the median value of $\Sigma^{\prime}_{N}$ from all uniform maps for each galaxy as a typical value of the overall density estimator in the uniform condition,  $\left<\Sigma^{\prime}_{N} \right>_{\mathrm{uniform}}$. Thus, $\log(1+\delta^{\prime}_{N})>0$ indicates that the environmental density of a galaxy exceeds the density standard in the uniform condition and vice versa.

\par In Table~\ref{tab:overdensity_measurements}, we present our measurements of the environmental densities of massive quiescent galaxies at $z=3-4$. We note that ZF-COS-20115 and ZF-COS-19589 are located at the edge of the PRIMER COSMOS mosaic image, were excluded from the analysis of environmental effects on morphological properties due to unreliable nearest-neighbor distances and overdensity measurements. Additionally, we include our environmental density measurements for RUBIES-EGS-QG1, a spectroscopically confirmed massive quiescent galaxy with $\log(M_{\ast}/M_{\odot}) = 10.9, z = 4.886$ \citep{deGraaff2024}, and GS-9209, another spectroscopically confirmed quiescent galaxy with $\log(M_{\ast}/M_{\odot}) = 10.58$ at $z = 4.658$ \citep{Carnall2023Nature}.
\par Our measured environmental density of RUBIES-EGS-QG1 indicates that it resides in overdense environments, with $\log(1+\delta^{\prime}_{3})=1.15$, qualitatively consistent with \cite{deGraaff2024}, who used galaxies at $4<z<6$ with robust spectroscopic redshifts to show that RUBIES-EGS-QG1 at $z=4.9$ resides in a significantly overdense environment ($3\sigma$ above the background level) within a velocity range of 4000 km s$^{-1}$. The agreement between our density estimator, derived from the distance to all $N$th nearest neighbors using photometric redshifts, and those of \citeauthor{deGraaff2024} strengthens our confidence in the reliability of our density estimator at this redshift range.


\subsection{Validation of Environmental Densities with Photometric Redshifts Using Mock Catalogs\label{appendix:mocktest_overdensity}}

\par We evalute how well our environmental densities ($\log(1+\delta^{\prime}_{N})$), derived using the Bayesian-motivated estimator of local surface density, $\Sigma^{\prime}_{N}$, reproduce the ``true" densities measured from spectroscopic redshift surveys. To test this, we utilize a mock galaxy catalog for JWST based on UniverseMachine \citep{Behroozi2020}. This catalog includes galaxies with ``true" redshifts (incorporating both cosmological expansion and peculiar velocity) down to the stellar mass completeness limit of the PRIMER and CEERS photometric catalogs at $3<z<6$ \citep[$\log(M_{\ast}/M_{\odot})=8.5$;][see their Figure C1]{Valentino2023}. We perturb the redshifts in the mock catalog by a random offset drawn from a normal distribution with $\sigma_{z} = 0.02(1+z)$, reflecting typical JWST photometric redshift uncertainties.
\par Using this mock data, we calculate distances to the $N$th nearest neighbors ($N=3, 5, 7, 10$) in two ways:
\begin{enumerate}
\item Spectroscopic redshifts: Distances are calculated using ``true" redshifts, within a cylindrical volume with a radial length corresponding to $dv = 2100~\mathrm{km~s}^{-1}$, simulating density estimates from spectroscopic surveys.
\item Photometric redshifts: Distances are calculated using perturbed redshifts, simulating JWST-like photometric redshift data.
\end{enumerate}

\par For both cases, we compute two estimates of the local surface density, $\Sigma^{\prime}_{N}$, using the Bayesian-motivated estimator defined in Equation~\ref{eq:bayesian_density_cowan}. The overdensity, $(1+\delta^{\prime}_{N})$, is then derived using Equation~\ref{eq:onepdelta}.

\par We define $\log(1+\delta^{\prime}_{N}) [z_{\mathrm{phot}}]$ as the density measured using photometric redshift uncertainties and $\log(1+\delta^{\prime}_{N}) [z_{\mathrm{spec}}]$ as the density measured using ``true" redshifts from the mock catalog. Figure~\ref{fig:log1pd_specz_photoz_mock} compares $\log(1+\delta^{\prime}_{N}) [z_{\mathrm{spec}}]$ to $\log(1+\delta^{\prime}_{N}) [z_{\mathrm{phot}}]$ for the Bayesian 3rd nearest neighbors estimator. The comparison shows a clear correlation: galaxies identified in low and high overdensity environments using spectroscopic redshifts exhibit similar trends when measured using photometric redshifts. The Spearman's correlation coefficients between $\log(1+\delta^{\prime}_{N}) [z_{\mathrm{phot}}]$ and $\log(1+\delta^{\prime}_{N}) [z_{\mathrm{spec}}]$ are 0.82, 0.82, 0.77, and 0.74 for $N=3$, $5$, $7$, and $10$, respectively. The corresponding $p$-values are effectively zero, indicating a very strong correlation across all $N$th nearest neighbors.

\begin{figure*}
    \centering
     \includegraphics[width=1\textwidth]{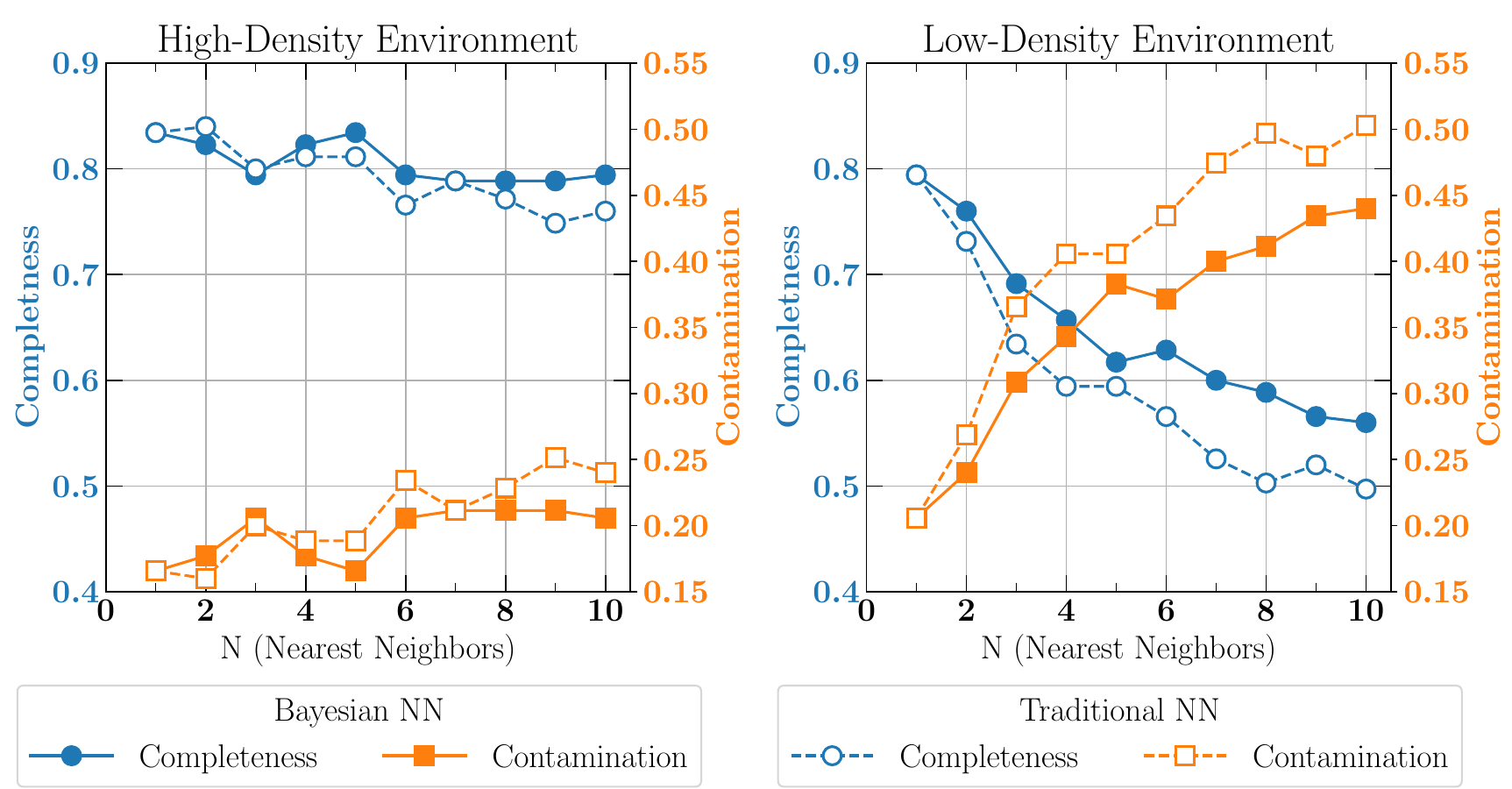}
    \caption{
Completeness and contamination as a function of $N$ (nearest neighbors) for galaxies in the stellar mass range $\log(M_{\ast}/M_{\odot}) > 10.2$ and $3<z<4.3$, derived from a mock JWST galaxy catalog based on UniverseMachine \citep{Behroozi2020}. The left panel shows results for selecting galaxies in the highest density quartile in $\log(1+\delta^{\prime}_{N}) [z_{\mathrm{phot}}]$ using $N=1-10$, while the right panel shows results for lowest density quartile. Blue lines represent completeness, and orange lines represent contamination fraction. The selections based on the Bayesian density estimator, $\log(1+\delta^{\prime}_{N}) [z_{\mathrm{phot}}]$, and traditional estimator, $\log(1+\delta_{N}) [z_{\mathrm{phot}}]$ $N$-th nearest neighbors are indicated as solid curve with filled symbol and dashed curve with open symbol, respectively. As $N$ increases, completeness slightly decreases due to smoothing over larger scales, while contamination shows a modest increase, The Bayesian nearest neighbor method demonstrates higher completeness and lower contamination compared to the traditional estimator, highlighting its improved performance in identifying environmental densities.}
\label{fig:completeness_contam_mock} 
\end{figure*}

\par Second, we estimate the completeness and contamination in low- and high-density environments derived using JWST-like photometric redshifts. Our goal is to robustly identify galaxies in low-density (high-density) environments with minimal contamination, i.e., a low fraction of galaxies misclassified between density environments. Specifically, we define samples of galaxies in high-density and low-density environments based on the rank quartiles, consistent with our analysis of the massive quiescent galaxy sample.

\par For galaxies in the highest density quartile in $\log(1+\delta^{\prime}_{N}) [z_{\mathrm{phot}}]$ using $N=2-10$, we recover $\sim80\%$ of galaxies that are also in the highest density quartile in $\log(1+\delta^{\prime}_{N}) [z_{\mathrm{spec}}]$ (Figure~\ref{fig:completeness_contam_mock}). The completeness decreases slightly (by $\lesssim5\%$) with increasing $N$. Importantly, the contamination fraction—defined as galaxies in the highest density quartile in $\log(1+\delta^{\prime}_{N}) [z_{\mathrm{phot}}]$ that belong to the medium or lowest density quartiles in $\log(1+\delta^{\prime}_{N}) [z_{\mathrm{spec}}]$—remains consistently low at $\sim20\%$ across the range of $N$. These results indicate that we achieve a relatively pure sample of galaxies in high-density environments. In other words, while we lose $\sim20\%$ of galaxies that should belong to the high-density quartile, the vast majority of galaxies identified in this quartile are truly in high-density environments as measured by $\log(1+\delta^{\prime}_{N}) [z_{\mathrm{spec}}]$. This demonstrates the low impact of chance alignments compared to physical associations. Using the traditional $N$th nearest neighbor estimator, $\log(1+\delta_{N}) [z_{\mathrm{phot}}]$, we find a $\sim2-5\%$ increase in completeness and a $\sim10-15\%$ reduction in contamination when using the Bayesian-motivated estimator, highlighting the superiority of the Bayesian method.

\par For galaxies in the lowest density quartile in $\log(1+\delta^{\prime}_{N}) [z_{\mathrm{phot}}]$, we recover $\sim70\%$ of galaxies also in the lowest density quartile in $\log(1+\delta^{\prime}_{N}) [z_{\mathrm{spec}}]$ for $N=3$. The completeness decreases to $\sim60\%$ for $N=10$ (Figure~\ref{fig:log1pd_specz_photoz_mock}). Contamination for this sample remains low, increasing from $\sim30\%$ at $N=3$ to $\sim40\%$ at $N=10$, with most contamination arising from galaxies in the medium-density quartile of $\log(1+\delta^{\prime}_{N}) [z_{\mathrm{spec}}]$. Notably, there is almost no contamination from galaxies in the highest-density quartile. This suggests that $\sim70\%$ of galaxies identified in the lowest-density quartile are genuinely in low-density environments. The Bayesian estimator further improves performance in this regime, with completeness fractions that are $\sim5-15\%$ higher than the traditional estimator, while contamination fractions are reduced by $10-20\%$. These improvements are most evident at higher values of $N$.

\par Collectively, these results provide strong justification for the use of the Bayesian-motivated estimator, which demonstrates superior recovery of galaxies in both high- and low-density environments with strong correlations between measured and true densities. For this study, we adopt $N=3$ nearest neighbors as our standard for its balance of high completeness ($\sim80\%$) and low contamination ($\sim20\%$). Additionally, the 3rd nearest neighbors Bayesian estimator is also notable for its faithful measurement of local environments on the scales of galaxy and galaxy group halos, as highlighted by \citet{Muldrew2012}.

\begin{figure*}
   \centering
   \includegraphics[width=\textwidth]{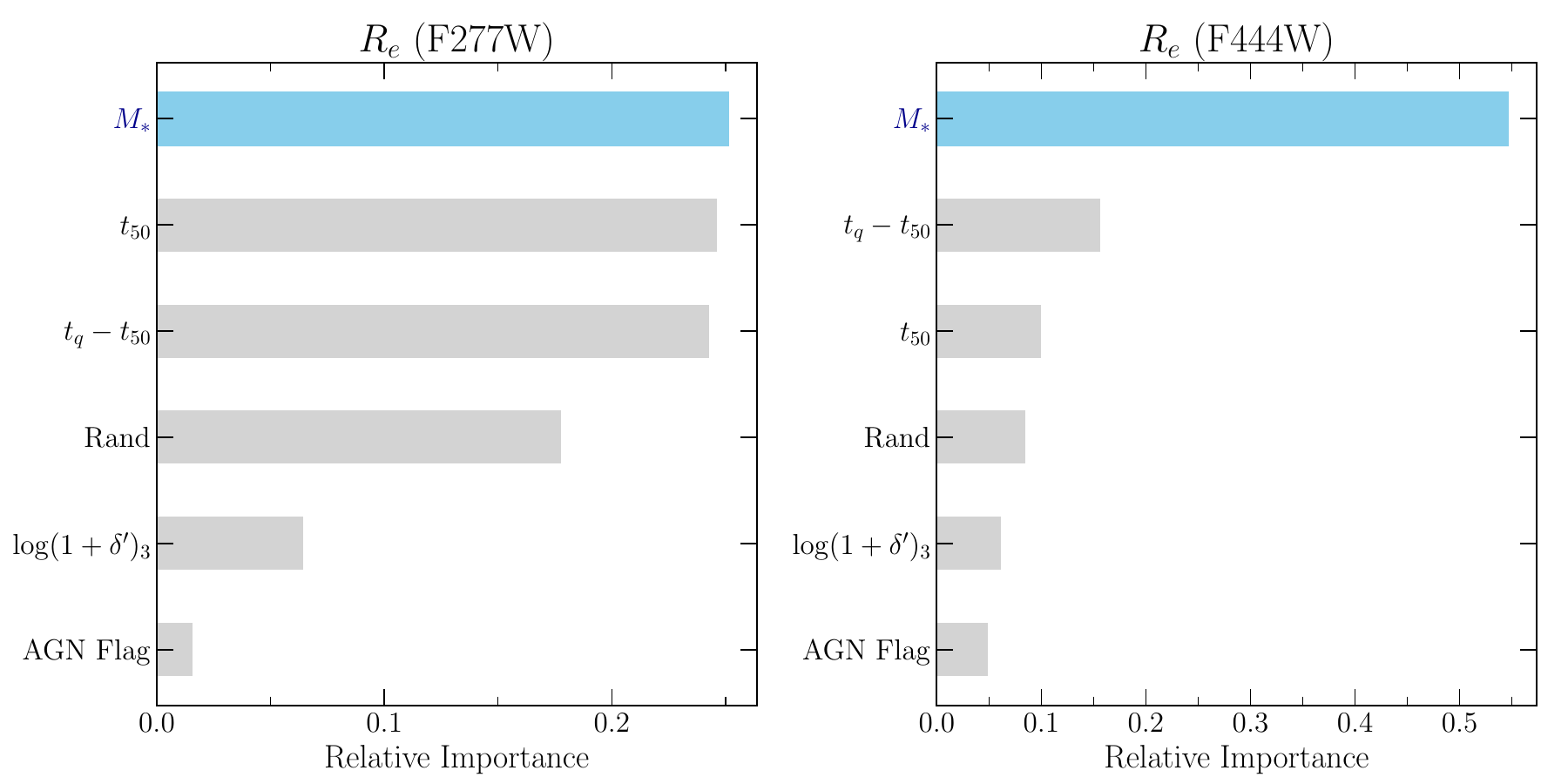}
    \caption{Feature importance from random forest regression models predicting the effective radius ($R_{e}$) of massive quiescent galaxies at $z=3-4$, measured in the F277W (left) and F444W (right) filters. In both cases, stellar mass ($M_{\ast}$) is the most important predictor, consistent with the well-established size-mass relation. For F277W sizes, which trace rest-frame optical light, the second and third most important parameters are the stellar formation time ($t_{50}$) and quenching timescale ($ t_q - t_{50} $), suggesting sensitivity to recent star formation history. In contrast, F444W sizes, which better reflect the rest-frame near-infrared and thus the underlying stellar mass distribution, are more strongly influenced by the quenching timescale than by  $t_{50}$. This implies that the structural imprint of quenching plays a key role in shaping mass-weighted sizes, while the formation epoch primarily affects light-weighted measurements. The minimal importance of environmental density ($\log(1 + \delta^{\prime}_{3})$) and AGN flag in both filters suggests these factors contribute little to size variation at fixed mass.}
   \label{fig:RF_size_importance}
\end{figure*}

\section{Predicting Galaxy Size with Random Forest Regression\label{appendix:RF_size}}

\par \editthree{To investigate the physical drivers of galaxy size, we apply the same random forest (RF) regression method described in the main text (Section~\ref{sec:randomforest_analysis})—previously used to predict $B/T$ ratios—to predict the effective radii ($R_e$) of massive quiescent galaxies at $z = 3\text{--}4$, measured in the F277W and F444W filters. The aim is to quantify the relative importance of various galaxy properties in determining observed size in each band. The results are shown in Figure~\ref{fig:RF_size_importance}.}

\par \editthree{In both filters, stellar mass ($M_{\ast}$) is identified as the most important predictor of galaxy size, with feature importance scores of 0.25 and 0.55 for F277W and F444W, respectively. This is consistent with the well-established size–mass relation observed across cosmic time. However, the importance of other parameters varies between the two filters, reflecting their differing sensitivity to stellar population age and star formation history.}

\par \editthree{For F277W sizes, which trace rest-frame optical light, the second and third most important features are the stellar formation time ($t_{50}$) and the quenching timescale ($t_q - t_{50}$), both with feature importance around 0.25. This indicates that F277W sizes are moderately sensitive to the star formation history. These results are consistent with our earlier finding that galaxies with more recent formation times tend to be more compact, exhibiting sizes up to $\sim 0.2$ dex smaller than predicted by the median relation (Figure~\ref{fig:logsizeoffset_tform_tq}).}

\par \editthree{In contrast, F444W sizes—tracing the rest-frame near-infrared and thus more closely reflecting the underlying stellar mass distribution—are most strongly influenced by stellar mass, with a high feature importance of 0.55. The quenching timescale ($t_q - t_{50}$) emerges as the second most important parameter (0.16), while $t_{50}$ has a lower feature importance of 0.10. This suggests that the structural imprint left by the quenching process—such as compaction or inside-out cessation of star formation—has a stronger effect on mass-weighted sizes than the formation epoch alone. The random number control variable ranks higher than both the environmental density contrast ($\log(1 + \delta^{\prime}_{3})$) and the AGN flag in both filters, indicating that these two factors contribute little to the variation in galaxy size at fixed stellar mass within this sample.}

\par \editfour{Although the lack of a significant Spearman correlation between size offset and quenching timescale (see Section~\ref{sec:result_smr_sfh}) may initially appear inconsistent with the Random Forest regression results presented here—where $\Delta t_q$ ranks as the third most important predictor of F277W size and the second most important for F444W—this discrepancy is expected. The Random Forest model captures nonlinear relationships and interactions among variables such as stellar mass and formation time, whereas the Spearman test assesses only monotonic, one-dimensional trends. As such, $\Delta t_q$ may still meaningfully influence galaxy size in combination with other physical parameters, even if it shows no simple trend in isolation.}

\par \editthree{Overall, these results support a scenario in which galaxy structure, particularly as traced by near-infrared sizes, is primarily shaped by stellar mass assembly and the details of quenching, while the roles of large-scale environment and AGN activity are subdominant at this redshift and mass scale.}

\begin{figure*}
  \centering
  \includegraphics[width=\textwidth]{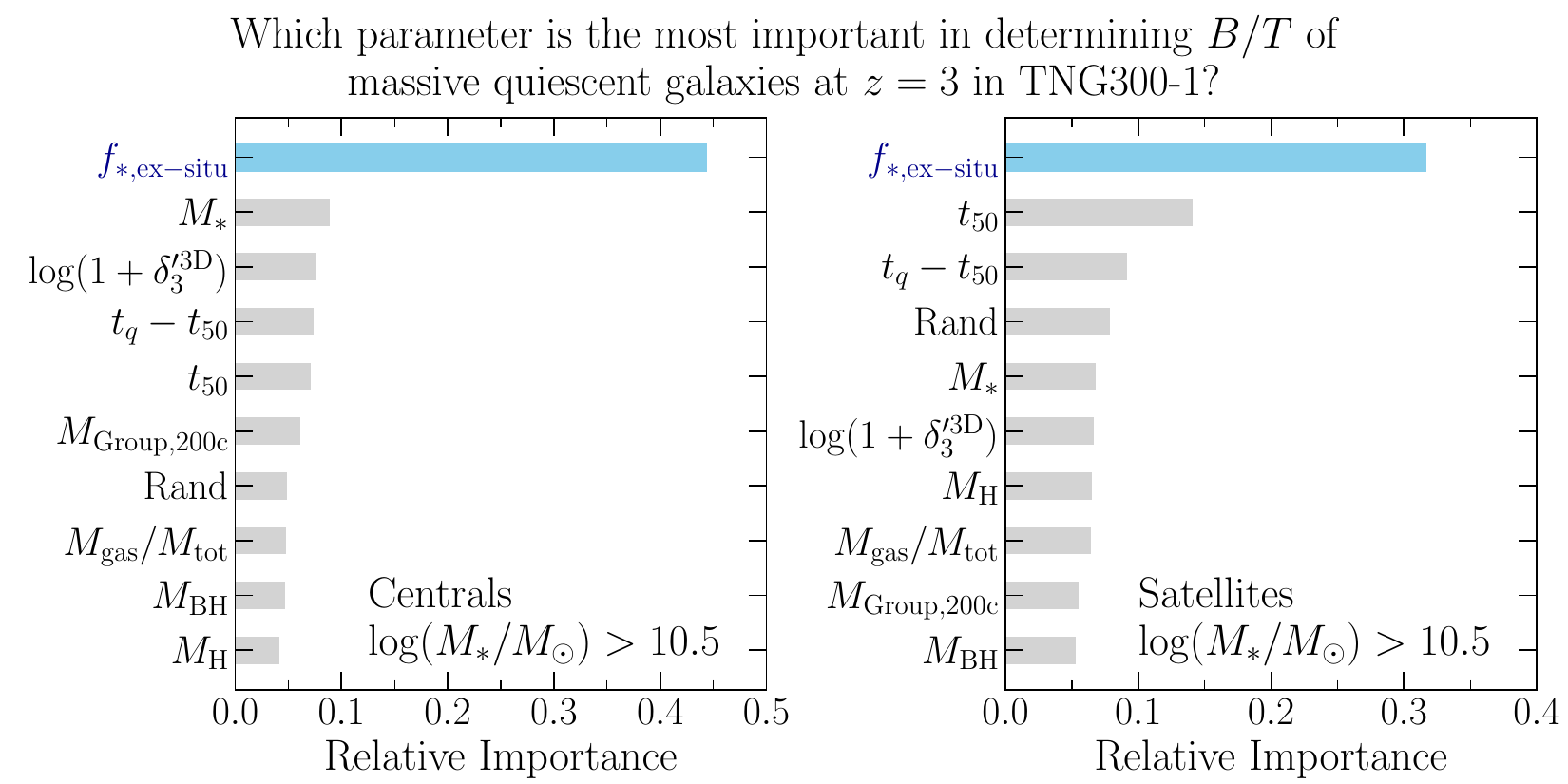}
 \caption{Results from a Random Forest (RF) regression analysis to predict the bulge-to-total ratio ($B/T$) of massive quiescent galaxies ($\log(M_{\ast}/M_{\odot}) > 10.5$, $z=3$, $\mathrm{sSFR} < 1.5 \times 10^{-10} \, \mathrm{yr}^{-1}$) selected from the TNG300-1 simulation, including total stellar mass ($M_{\ast}$) in the input feature set. The $x$-axis shows the relative importance of each parameter (listed on the $y$-axis) for predicting $B/T$. The fraction of stellar mass formed ex-situ ($f_{\ast,\mathrm{ex-situ}}$) is the most important parameter for both centrals (left) and satellites (right), reinforcing the dominant role of mergers and stellar accretion—especially prior to infall—in building bulge-dominated systems. For centrals, total stellar mass ($M_\ast$) becomes the second most important feature, consistent with its correlation with $f_{\ast,\mathrm{ex-situ}}$. Environmental metrics such as group halo mass ($M_{\mathrm{group,200c}}$) and local overdensity ($\log(1+\delta^{\prime\mathrm{3D}}_3)$) also retain some predictive power, potentially reflecting environmental modulation of merger and interaction rates within group-scale halos. \editthree{For satellites, formation history parameters—namely the formation time ($t_{50}$) and the quenching timescale ($t_q - t_{50}$)—emerge as the second and third most important predictors of $B/T$, ahead of $M_\ast$ and all environmental variables. These features likely trace internal processes such as early gas collapse, violent disk instabilities, and rapid self-regulated quenching, which govern bulge formation before infall. Together with $f_{\ast,\mathrm{ex-situ}}$, which encodes pre-infall merger activity, these results suggest that the structural evolution of satellites is primarily driven by internal and early-time processes, rather than by their present-day environment.}
}

\label{fig:tng_RF_btratio_Mstar}
\end{figure*}

\section{Sensitivity of Random Forest Results to Parameter Choices in TNG300-1 Simulations \label{appendix:RF_tests_TNG}}

\subsection{ Impact of Including Total Stellar Mass in RF Analysis in TNG300-1 Simulations \label{appendix:Mstar_RF_test}}
\par \edittwo{To further examine the role of total stellar mass ($M_{\ast}$) in predicting $B/T$ of massive quiescent galaxies in the IllustrisTNG 300-1, we perform an additional RF analysis by introducing $M_{\ast}$ into the feature set. This allows us to assess whether its inclusion alters the relative importance of environmental parameters, particularly group halo mass ($M_{\mathrm{group,200c}}$) and local overdensity ($\log(1+\delta^{\prime\mathrm{3D}}_{3})$).}

\par \edittwo{For centrals, we find that $M_{\ast}$ emerges as the second most important predictor of $B/T$ (Figure~\ref{fig:tng_RF_btratio_Mstar}), which is expected given its strong correlation with $f_{\ast,\mathrm{ex-situ}}$. In the simulation, $M_{\ast}$ is explicitly defined as the sum of in-situ and ex-situ stellar mass. Since galaxies that experience more mergers tend to be more massive and have larger ex-situ fractions, this reinforces the predictive overlap between $M_{\ast}$ and $f_{\ast,\mathrm{ex-situ}}$.}

\par \edittwo{However, the inclusion of $M_{\ast}$ in the RF model leads to a reduction in the relative importance of $M_{\mathrm{group,200c}}$ and $\log(1+\delta^{\prime\mathrm{3D}}_{3})$, potentially obscuring environmental effects. This is likely because $M_{\ast}$ already encodes much of the merger-driven stellar mass growth that may itself be environmentally regulated. As a result, disentangling the direct influence of halo environment on bulge formation becomes more challenging when $M_{\ast}$ is included.}

\par \edittwo{For satellites, the addition of $M_{\ast}$ does not significantly affect the ranking of predictive features. The fraction of ex-situ stellar mass ($f_{\ast,\mathrm{ex-situ}}$), the formation time ($t_{50}$), and the quenching timescale ($t_q - t_{50}$) remain the top three predictors of $B/T$, while $M_{\ast}$ and environmental parameters have relatively low importance. This supports the conclusion that bulge growth in satellites is primarily governed by internal processes and pre-infall formation histories—such as mergers, early star formation, and rapid quenching—rather than by present-day environment or stellar mass.}

\begin{figure*}
  \centering
 \includegraphics[width=\textwidth]{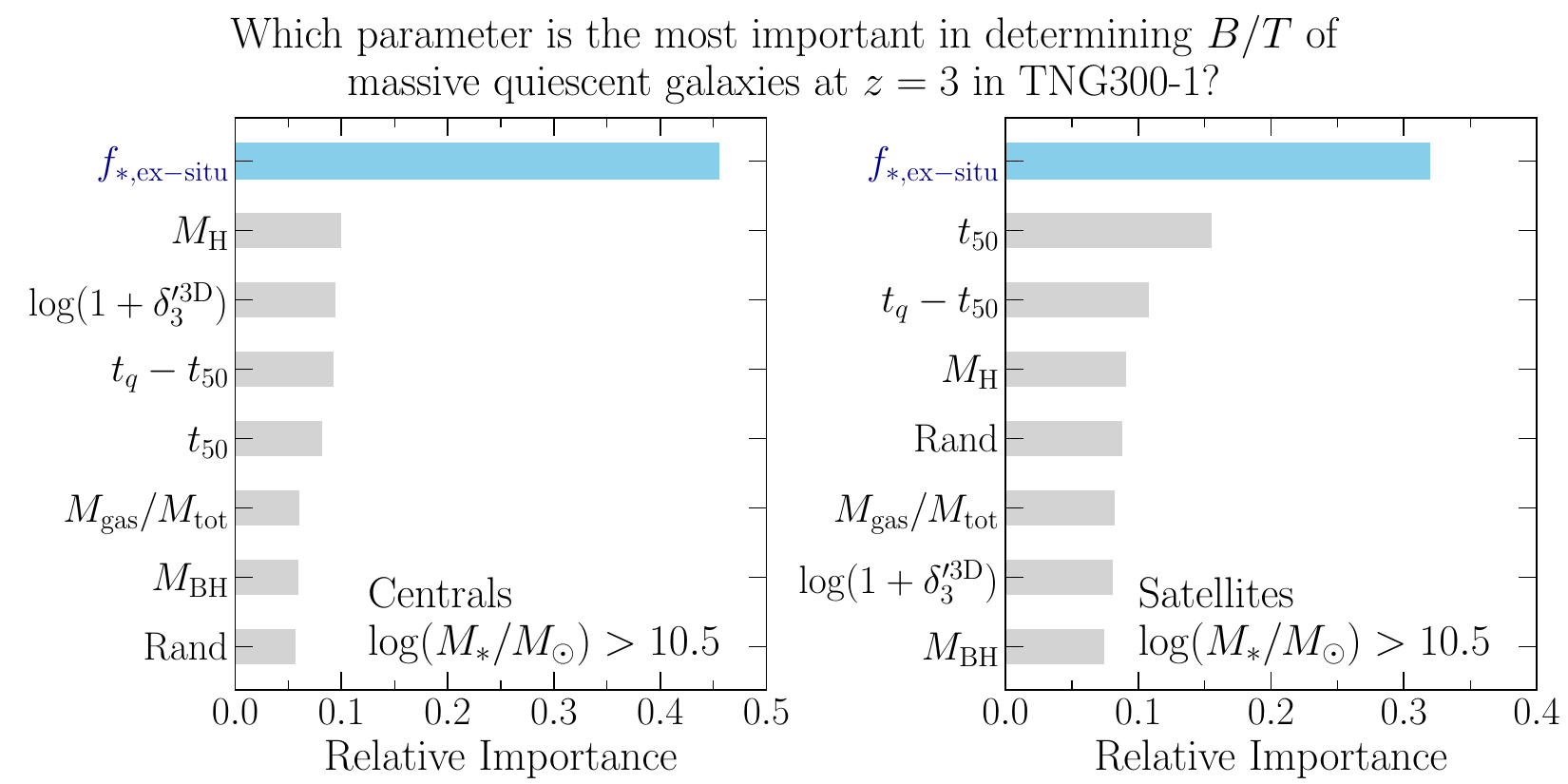}
\caption{\editthree{Results from a Random Forest (RF) regression analysis to predict the bulge-to-total ratio ($B/T$) of massive quiescent galaxies ($\log(M_{\ast}/M_{\odot}) > 10.5$, $z=3$, $\mathrm{sSFR} < 1.5 \times 10^{-10} \, \mathrm{yr}^{-1}$) selected from the TNG300-1 simulation, after excluding group halo mass ($M_{\mathrm{group,200c}}$) from the feature set. The $x$-axis shows the relative importance of each parameter (listed on the $y$-axis) for predicting $B/T$. For central galaxies (left), the local environmental density at $z=3$—quantified by $\log(1+\delta^{\prime\mathrm{3D}}_3)$ measured at $z=3$—becomes the third most important predictor after $f_{\ast,\mathrm{ex-situ}}$ and subhalo mass, suggesting that small-scale environment contributes independently to bulge growth, even in the absence of group halo mass. In contrast, for satellites (right), the removal of $M_{\mathrm{group,200c}}$ does not noticeably elevate the ranking of $\log(1+\delta^{\prime\mathrm{3D}}_3)$, which remains a low-importance feature. This indicates that the local overdensity at $z=3$ provides little additional predictive power for $B/T$ of satellites once intrinsic formation history—captured by $f_{\ast,\mathrm{ex-situ}}$, $t_{50}$, and $t_q - t_{50}$—is taken into account. These results highlight that bulge growth in centrals is more sensitive to their environment at $z=3$, while satellites retain structural imprints primarily from earlier evolutionary stages.}}

\label{fig:tng_RF_btratio_nomgroup}
\end{figure*}

\subsection{Impact of Excluding Group Halo Mass \texorpdfstring{$M_{\mathrm{group,200c}}$}{Mgroup,200c}}

\par \editthree{We also assess whether the predictive power of environmental overdensity ($\log(1+\delta^{\prime\mathrm{3D}}_{3})$) is influenced by the inclusion of group halo mass ($M_{\mathrm{group,200c}}$), given the known correlation between these two quantities \citep[e.g.,][]{Chiang2013}. To this end, we rerun the Random Forest (RF) analysis for both central and satellite galaxies, excluding $M_{\mathrm{group,200c}}$ from the feature set.}

\par \editthree{The resulting feature importance rankings are shown in Figure~\ref{fig:tng_RF_btratio_nomgroup}. For central galaxies, the overdensity parameter $\log(1+\delta^{\prime\mathrm{3D}}_{3})$ becomes the third most important predictor of $B/T$, with relative importance comparable to that of subhalo mass in the original RF analysis. This suggests that local density contributes additional predictive power beyond group halo mass—likely by tracing small-scale environmental effects such as local interactions or enhanced merger rates within the group potential.}

\par \editthree{In contrast, for satellites, the removal of $M_{\mathrm{group,200c}}$ does not noticeably elevate the ranking of $\log(1+\delta^{\prime\mathrm{3D}}_{3})$, which remains a low-importance feature. This implies that, for satellites, the current local overdensity at $z=3$ does not encode significant additional information about bulge growth once the effects of internal evolution and pre-infall history—captured by $f_{\ast,\mathrm{ex-situ}}$, $t_{50}$, and $t_q - t_{50}$—are accounted for. These findings reinforce the idea that the role of local environment in shaping $B/T$ is more relevant for centrals than for satellites, and that bulge structure in satellites is primarily set by processes operating prior to group infall.}

\bibliography{references}{}
\bibliographystyle{aasjournal}



\end{document}